\documentclass[a4paper,11pt]{article}
\usepackage{jinstpub} 
\usepackage{orcidlink}
\usepackage{subfigure, dcolumn}
\usepackage{booktabs}

\title{\boldmath Understanding Energy Dependent Hadronic Calorimeter Response from a Machine Learning Perspective}

\author[a]{Shuai-Chun Wang\orcidlink{0009-0000-3536-3408}}
\author[a]{Huang-Ran Shen\orcidlink{0009-0003-4743-4164}}
\author[a,b,1]{Wan-Bing He\orcidlink{0000-0002-3854-4965}\note{Corresponding author.}}
\author[a,b]{Wei-Hu Ma\orcidlink{0000-0003-4125-937X}}
\author[a,b]{Peng-Jie Li\orcidlink{0000-0001-6542-4697}}
\author[a,b,1]{De-Qing Fang\orcidlink{0000-0002-6123-3014}}
\author[a,b,c,1]{Yu-Gang Ma\orcidlink{0000-0002-0233-9900}}

\affiliation[a]{Key Laboratory of Nuclear Physics and Ion-beam Application (MOE), Institute of Modern Physics, Fudan University, Shanghai, 200433, China}
\affiliation[b]{Shanghai Research Center for Theoretical Nuclear Physics, NSFC and Fudan
University, Shanghai, 200438, China}
\affiliation[c]{School of Physics, East China Normal University, Shanghai, 200062, China}
\emailAdd{hewanbing@fudan.edu.cn}
\emailAdd{dqfang@fudan.edu.cn}
\emailAdd{mayugang@fudan.edu.cn}

\abstract{
To meet the precision requirements of future high-energy physics experiments, improving the energy resolution of hadronic calorimeters remains a critical challenge. This work presents a systematic investigation of hadronic energy reconstruction using machine learning, highlighting the roles of various signal channels, including scintillation light, Cherenkov light, charged particles, and the full three-dimensional topology of hadronic showers in the energy range up to 10 GeV. Throughout this study, detector effects are not taken into account. Under these conditions, the intrinsic resolution of hadronic showers reaches approximately $(10.8\pm0.3)\% / \sqrt{E/GeV}$ when all signal channels and the full 3D shower information are fully utilized. Compared with the traditional signal-summing approach, machine-learning-based reconstruction can significantly improve energy resolution, even under a limited sampling fraction of 10\%, enhancing it from $(57.6\pm3.7)\%/\sqrt{E/GeV}$ to $(34.1\pm2.8)\%/\sqrt{E/GeV}$. These results highlight the critical importance of both multi-channel information and detailed spatial shower features in hadronic energy reconstruction, and demonstrate the substantial potential of combining high-granularity and dual-readout calorimeter designs with machine-learning-based reconstruction techniques for future experiments.
}


\begin{document}
\maketitle
\flushbottom

\section{Introduction}
\label{sec:introduction}
In modern high-energy physics experiments, calorimeters are indispensable detector systems for particle energy measurement and play a central role in both offline reconstruction and online triggering~\cite{Fernow_2023,RevModPhys.75.1243}. In particular, hadronic calorimeters are critical for jet energy reconstruction, neutral-hadron measurements, and missing transverse energy determination, thereby directly impacting the physics sensitivity of current and future collider experiments~\cite{Accardi_2016,Benedikt_2025}. Despite extensive technological development, hadronic energy measurement remains intrinsically more challenging than electromagnetic calorimetry, and the achievable energy resolution of hadronic calorimeters is generally significantly inferior~\cite{Brau:2010zz, WangXY2024}.

This limitation originates from the complex and stochastic nature of hadronic shower development. Unlike electromagnetic showers, which are governed by well-understood bremsstrahlung and pair-production processes, hadronic showers are initiated by inelastic hadron–nucleus interactions and involve a wide range of secondary processes, including electromagnetic subshowers, nuclear breakup, slow-neutron production, and nuclear excitation~\cite{WELLISCH200165,GROOM2007633,Wang2024NST,Zhao2025NST}. A substantial fraction of the incident particle energy is dissipated through invisible channels, such as nuclear binding energy losses and low-energy neutrons, which are not directly detectable by conventional active media. These effects introduce large event-by-event fluctuations in the calorimeter response and represent a fundamental limitation of hadronic calorimetry.

In addition to invisible energy losses, the relative fractions of electromagnetic and non-electromagnetic components in hadronic showers fluctuate significantly from event to event. Since most calorimeter technologies exhibit different responses to these components, such fluctuations further degrade energy resolution. Moreover, hadronic showers exhibit complex three-dimensional spatial structures, with a longitudinal development governed by the nuclear interaction length and a nontrivial transverse morphology. As a result, the calorimeter response to a hadronic shower is determined not only by its total deposited energy, but also by the detailed spatial and compositional characteristics of the shower.

Traditional approaches to hadronic energy reconstruction largely rely on averaged detector responses and low-dimensional observables, such as total deposited energy or simple longitudinal weighting schemes. To mitigate the resulting performance degradation, several detector concepts and correction strategies have been developed, including compensating calorimeters~\cite{HILGER1987488,CELLETTI1984493}, dual-readout calorimetry~\cite{RevModPhys.90.025002,instruments6030036,Pareti_2024,WIGMANS2013475}, and highly granular imaging calorimeters~\cite{Amendola_2025,Zubankov2025NST}. Compensation-based designs aim to equalize the detector response to electromagnetic and hadronic shower components, while dual-readout calorimeters exploit the simultaneous detection of scintillation and Cherenkov light to perform event-by-event corrections. Highly granular calorimeters, motivated by particle-flow reconstruction~\cite{Thomson_2011,THOMSON200925}, emphasize fine transverse and longitudinal segmentation to enable detailed imaging of shower development.

Historically, most hadronic calorimeters have adopted sampling architectures as a practical compromise among energy resolution, radiation hardness, and engineering feasibility~\cite{Yu2024NST}. Representative implementations include the brass–scintillator hadronic calorimeter of CMS~\cite{particles9010001}, optimized for operation in high-luminosity hadron-collider environments, and the ATLAS hadronic calorimeter system, which combines steel–scintillator and copper–liquid-argon technologies to achieve fine longitudinal segmentation and long-term stability~\cite{Solovyanov_2009}. The ZEUS uranium–scintillator calorimeter demonstrated the intrinsic advantages of compensation-based designs by achieving excellent linearity and hadronic energy resolution through suppression of electromagnetic fraction fluctuations~\cite{HILGER1987488}. However, uranium-based calorimeters have largely been phased out due to safety, response-time, and engineering constraints.

As collider experiments progress toward higher luminosities and increased event complexity, calorimeter development has shifted toward highly granular, information-rich detector systems. Examples include the CMS High-Granularity Calorimeter (HGCAL)~\cite{Amendola_2025}, designed to operate under extreme pile-up conditions at the HL-LHC, and the ALICE Forward Calorimeter (FoCal)~\cite{VANDERKOLK2020162059}, which employs ultra-high spatial resolution to address forward-physics challenges. Future detector concepts, such as the EPIC detector at the Electron–Ion Collider~\cite{ABDULKHALEK2022122447}, further emphasize fine segmentation, precise timing, and close integration with tracking detectors.

Beyond detector hardware innovations, these developments share a common implication: modern calorimeters provide access to high-dimensional, highly correlated observables encoding detailed shower information~\cite{Benedikt_2025, achasov_2023}. In this context, hadronic energy reconstruction can be naturally reformulated as a high-dimensional regression problem, in which the goal is to infer the incident particle energy from a complex set of calorimeter signals that reflect both visible and invisible components of the shower. This perspective highlights a key limitation of traditional reconstruction approaches, which are often unable to fully exploit the available information content due to their reliance on simplified, low-dimensional models.

Machine learning, and deep learning in particular, offers a powerful framework for addressing this challenge. By construction, deep neural networks are capable of learning nonlinear mappings and complex correlations in high-dimensional data, making them well suited for extracting subtle features related to shower topology, signal correlations, and event-by-event fluctuations. Recent studies have demonstrated the potential of machine-learning-based approaches to improve energy resolution, particle identification, and pile-up mitigation in highly granular calorimeter systems~\cite{Lai_2024,Lai_2024_1,Aamir_2024,Fei2025NST,He2023NST,Liao2024,Giannelli2024,Wu2024NST}.

Nevertheless, an important and largely unresolved question remains: which information is most relevant for hadronic energy reconstruction, and how does its relative importance depend on the shower energy? This issue is particularly critical in the low-energy regime, where nuclear breakup processes, slow-neutron energy dissipation, and large fluctuations in the electromagnetic shower fraction become increasingly dominant. In this regime, conventional calibration, compensation, or weighting schemes often exhibit strong nonlinearities and degraded performance, even in advanced calorimeter designs.

Motivated by these considerations, this work adopts a machine-learning-centric approach to systematically investigate hadronic energy reconstruction in information-rich calorimeter systems. By employing deep learning models to analyze multiple signal channels and three-dimensional shower information, we quantitatively assess the relative contributions of different observables to reconstruction performance across a wide energy range, with particular emphasis on low-energy hadronic showers. The results are expected to provide guidance for future calorimeter design choices, readout strategies, and reconstruction methodologies in next-generation collider experiments.

This paper is organized as follows. Section~\ref{sec:sim} describes the Geant4-based simulation framework and the construction of the datasets used in this study. Section~\ref{sec:ml_method} introduces the machine-learning methodology, including the three-dimensional ResNet architecture employed for hadronic energy reconstruction. In Section~\ref{sec:energy_resolution}, the procedures for evaluating the energy resolution using both conventional reconstruction algorithms and machine-learning methods are presented. To account for realistic detector responses, Section~\ref{sec:digitization} investigates the impacts of the photon detection processes on the overall energy resolution. The results are discussed in Section~\ref{sec:results}, where the homogeneous calorimeter is used to establish the physical limit of hadronic energy resolution, followed by a detailed investigation of sampling calorimeters and the impact of sampling structure on resolution performance. Finally, Section~\ref{sec:summary} summarizes the main findings and presents the conclusions of this work.

\section{Simulation and dataset}
\label{sec:sim}

The dataset used in this work is entirely based on Geant4 simulations of hadronic shower development induced by $\pi^-$ mesons in a homogeneous $\mathrm{PbWO_4}$ calorimetric medium. $\mathrm{PbWO_4}$ is a well-established calorimeter material owing to its high density and compactness, characterized by a short radiation length ($X_0 \approx 0.89$~cm) and a relatively small nuclear interaction length ($\lambda_{\mathrm{int}} \approx 20$~cm). In addition, $\mathrm{PbWO_4}$ simultaneously produces scintillation light and Cherenkov radiation, making it particularly suitable for studies of dual-readout techniques and for investigating the complementarity of different signal channels~\cite{AKCHURIN2008273}.

Hadronic interactions and shower development are simulated using the FTFP-BERT physics list in Geant4 version 11.3.0. This physics list provides a reliable description of hadron--nucleus interactions in the considered energy range and is widely adopted in calorimeter simulation studies. In all simulations, the underlying shower physics remains unchanged; different datasets are constructed exclusively through alternative signal extraction and sampling strategies applied to the same simulated showers.

The calorimeter volume is defined as a $\mathrm{PbWO_4}$ crystal with dimensions of $1~\mathrm{m} \times 1~\mathrm{m} \times 2~\mathrm{m}$, sufficient to contain both the longitudinal and transverse development of hadronic showers in the studied energy range. The incident $\pi^-$ mesons enter the detector from the center of one side surface. To preserve detailed spatial information, the crystal is segmented into a $100 \times 100 \times 100$ three-dimensional grid, corresponding to voxel dimensions of $1~\mathrm{cm} \times 1~\mathrm{cm} \times 2~\mathrm{cm}$. Regarding the photon transport process, given the high-granularity calorimeter structure, the optical paths are sufficiently short that absorption losses are considered secondary. In terms of detection, wavelength-dependent efficiency factors were applied to simulate SiPM response. It is worth noting that these optical processes can introduce additional fluctuations, which may affect the overall energy resolution. Their impact is evaluated and discussed in detail in Section \ref{sec:digitization}. The results indicate that, within the studied energy range, their contribution is limited compared to the intrinsic stochastic fluctuations of hadronic showers, thereby supporting the treatment adopted in this work.

Regarding the energy containment of this geometry, the total average energy leakage of the detector is 5.9\%, with lateral leakage averaging 3.9\% and longitudinal leakage averaging 2.0\%. Consequently, the energy leakage remains within an acceptable range for the analysis of the intrinsic shower properties.

\begin{figure}[htbp]
\subfigure[]{
\label{ScintPhoton}
\includegraphics[width=0.4\textwidth]{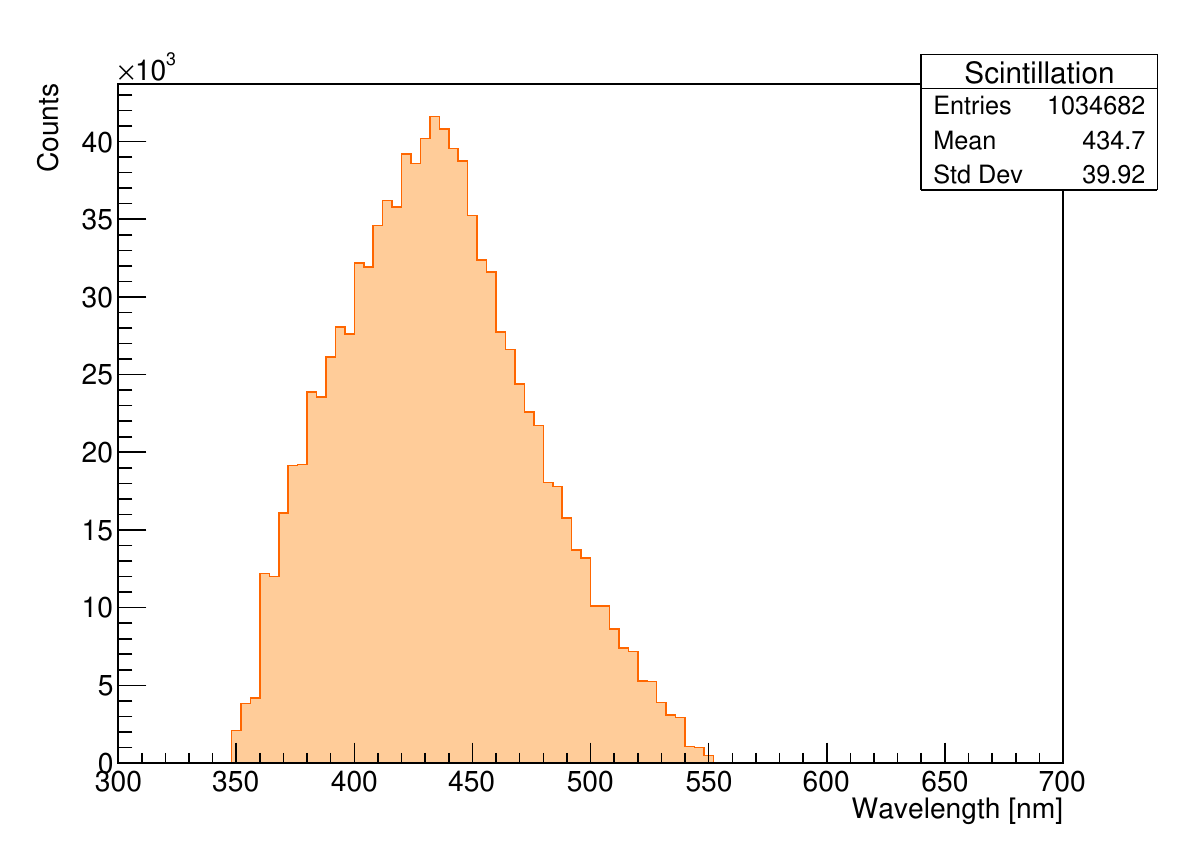}}
\subfigure[]{
\label{CerenPhoton}
\includegraphics[width=0.4\textwidth]{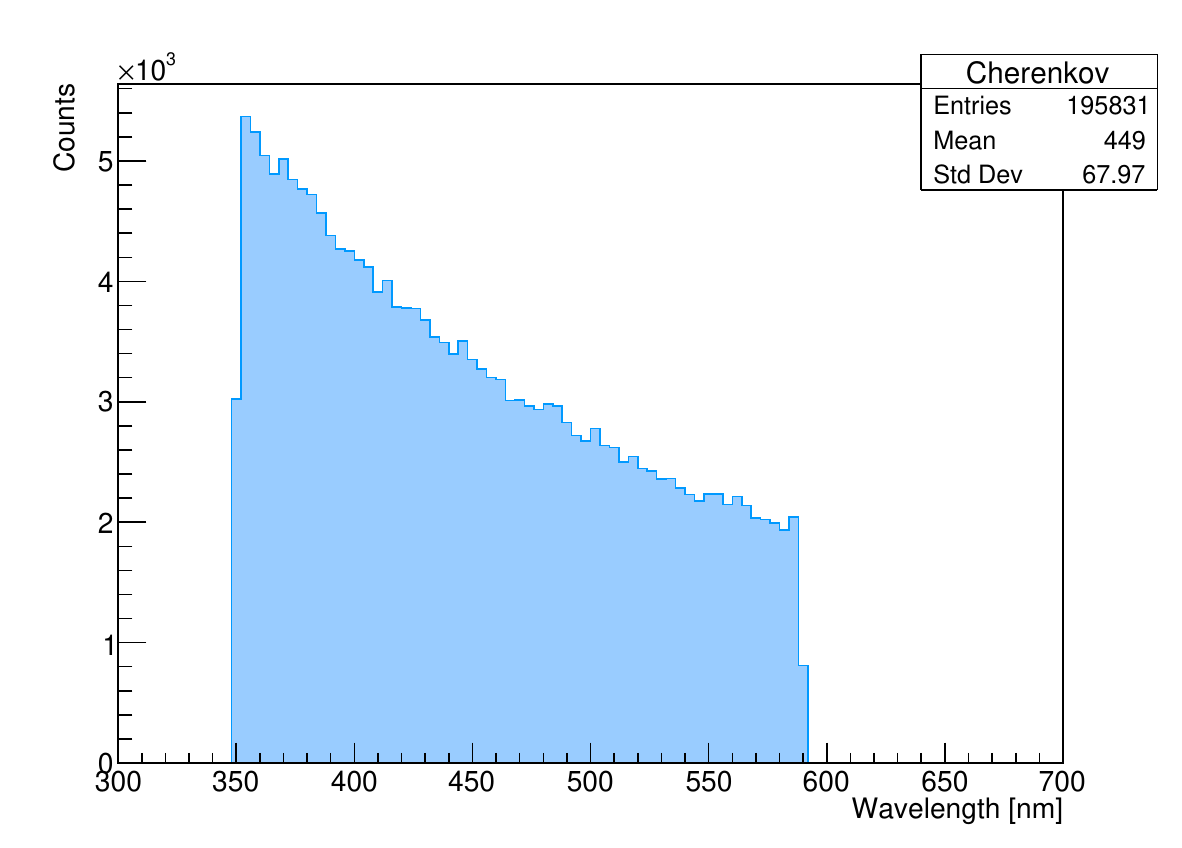}}
\caption{Photon energy spectra within the assumed acceptance of a photodetector for 10 GeV $\pi^-$ mesons incident on the calorimeter: (a) scintillation photon emission spectrum, (b) Cherenkov photon emission spectrum.} 
\label{Pure}
\end{figure}

Three voxel-level signal channels are constructed. Scintillation and Cherenkov photons are distinguished directly by their production processes in Geant4, forming two independent optical channels. These photon spectra, shown in figs.\ref{ScintPhoton} and \ref{CerenPhoton}, are derived from the shower results of a single 10 GeV incident $\pi^-$, where the number of scintillation photons is generally an order of magnitude larger than that of Cherenkov photons. To account for the spectral response of the EQR20\cite{MELIKYAN2025170604} series Silicon Photomultiplier (SiPM), the effective wavelength of both photon sources is constrained to the 350 nm – 550 nm interval. In addition, the number of charged particles traversing each voxel is recorded as a third channel, representing the charged component of the hadronic shower. These channels constitute the input representation used for subsequent energy reconstruction studies.

The design of the homogeneous option follows the fine voxel-based transverse and longitudinal segmentation described above, where the individual voxel readout is assumed. Building upon this configuration, to emulate the response of sampling calorimeters commonly employed in collider experiments, sampling effects are introduced at the signal-processing level rather than by altering the shower simulation. The detector response is treated as a homogeneous energy deposition in Geant4, without imposing any segmentation during the simulation stage. Longitudinal sampling is then emulated in a post-processing step by dividing the shower axis into predefined depth intervals and selectively retaining photons in designated intervals, while discarding the others.

In this study, the sampling segmentation is defined as the total thickness of one unit, comprising both active and absorber layers, while the sampling ratio is defined as the ratio of the active layer thickness to the total thickness of one unit. By selectively retaining energy deposits in designated active layers and discarding others, $10^5$ statistically independent hadronic showers are analyzed. Datasets corresponding to different sampling segmentations and sampling ratios are constructed by applying various sampling configurations to the same ensemble of simulated showers. This procedure ensures that the impact of sampling-induced information loss on the energy resolution can be studied in a controlled and unbiased manner, without conflating it with shower-to-shower fluctuations.

\section{Machine-Learning Method}
\label{sec:ml_method}

For the reconstruction of hadronic energy from three-dimensional calorimeter data, a three-dimensional residual network (3D-ResNet) architecture is employed. The network is based on ResNet-10, in which the residual blocks reformulate the learning objective from $H(x)$ to $F(x)=H(x)-x$, facilitating optimization. The architecture comprises an input convolutional layer, four residual modules, a global average-pooling layer, and a fully connected layer.

The input layer performs initial 3D convolution and batch normalization (BN) on the voxelized calorimeter data. The outputs then pass sequentially through four residual modules, each containing two 3D convolutional layers followed by BN layers. ReLU activations are applied after the initial convolution and within each residual module. Global average pooling is applied before the fully connected layer to produce the final prediction.

Since the total number of photons is an important feature for energy reconstruction, global quantities corresponding to the total counts of each channel (scintillation, Cherenkov, and charged particles) are concatenated with the flattened convolutional features prior to the fully connected layer. To further mitigate overfitting, a dropout layer with a rate of 30\% is applied before the fully connected layer.

For the regression task, the Huber loss function is adopted to improve robustness against outliers. For a single sample, it is defined as
\begin{equation}
L_\delta(y, \hat{y})=
\begin{cases}
\frac{1}{2}(\hat{y}-y)^2 & \mathrm{if}\ |\hat{y}-y|\leq\delta \\
\delta \left(|\hat{y}-y|-\frac{1}{2}\delta \right) & \mathrm{if}\ |\hat{y}-y|>\delta
\end{cases}
\end{equation}

where $y$ and $\hat{y}$ represent the true hadron energy and the model prediction, and $|\hat{y}-y|$ denotes the absolute residual. The parameter $\delta$ serves as a threshold that determines the transition between quadratic and linear loss behaviors.

For a dataset containing $n$ samples, the overall loss is
\begin{equation}
\mathrm{Huber\ Loss} = \frac{1}{n}\sum_{i=1}^{n} L_\delta(y_i, \hat{y}_i).
\end{equation}

To improve generalization and prevent overfitting, L2 regularization (weight decay) is incorporated into the loss function,
\begin{equation}
\mathcal{L}_\mathrm{total} = \mathcal{L}_\mathrm{task} + \frac{\lambda}{2}\|\theta\|^2,
\end{equation}
where $\theta$ represents the trainable parameters and $\lambda$ is the regularization coefficient. In this study, $\lambda = 10^{-5}$ is chosen to balance convergence stability and effective regularization.

The network is trained using $10^5$ simulated events with $\pi^-$ energies uniformly distributed from 0 to 10~$\mathrm{GeV}$. An independent set of $10^4$ events is used for testing. To further reduce overfitting, 20\% of the training events are reserved as a validation set, and early stopping is applied when the validation loss does not decrease for five consecutive epochs. Optimization is performed with the AdamW algorithm, which provides adaptive learning-rate adjustment and decoupled weight decay. The initial learning rate is set to $2\times 10^{-4}$, and a dynamic learning-rate schedule based on validation loss is applied using the \texttt{ReduceLROnPlateau} strategy: if the validation loss does not improve for three consecutive epochs, the learning rate is reduced by 50\%, down to a minimum of $1\times 10^{-6}$. The batch size is set to 16, providing a balance between computational efficiency and gradient stability. This choice also favors generalization and stable convergence in regression tasks involving the calorimeter energy distributions.

For machine-learning (ML) approaches, different sets of input variable were used to account for different reconstruction options. Hereinafter, the following legend is applied: ML1 for scintillation only, ML2 for Cherenkov only, ML3 for charged particle only, ML12 for scintillation and Cherenkov, ML13 for scintillation and charged particle, and ML123 for scintillation, Cherenkov and charged particle.

\section{Energy Resolution Evaluation Methods}
\label{sec:energy_resolution}

To enable a quantitative comparison with the machine-learning-based energy reconstruction, the energy resolution obtained using conventional single-channel and dual-readout reconstruction methods is evaluated using identical simulated datasets.

\begin{figure}[htbp]
\centering
\subfigure[]{
\label{two_fits.1}
\includegraphics[width=0.4\textwidth]{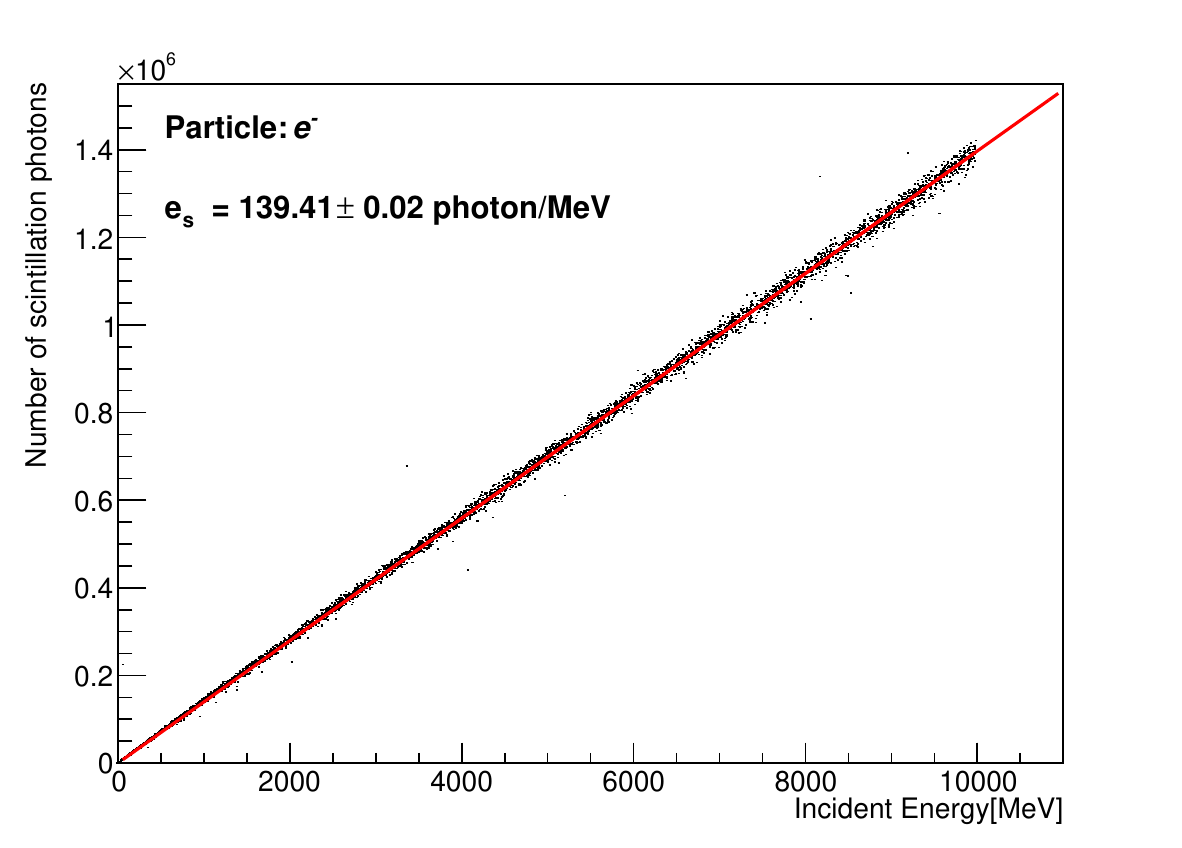}}
\subfigure[]{
\label{two_fits.2}
\includegraphics[width=0.4\textwidth]{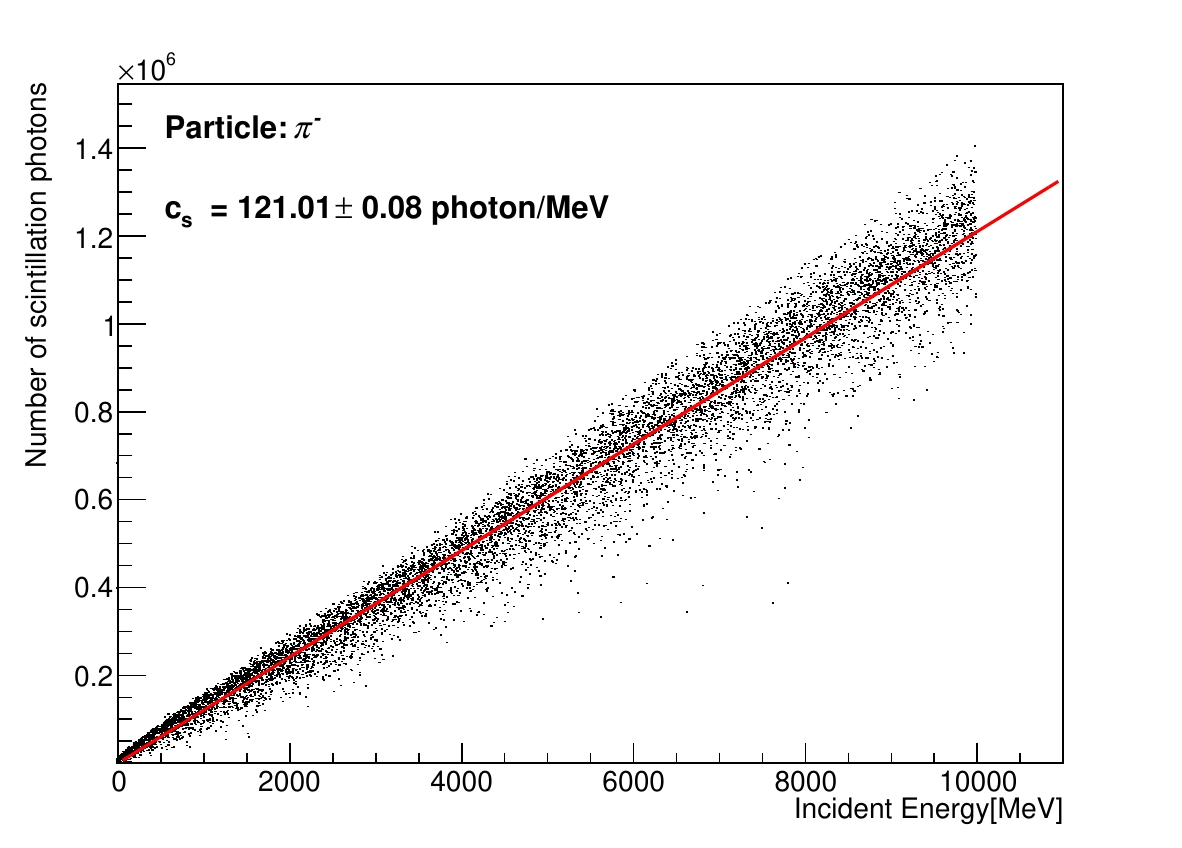}}
\caption{Response coefficients between the incident particle energy $E_{\text{in}}$ from $0$ to $10000 \text{ MeV}$ and the number of produced photons in the homogeneous calorimeter. Plot (a) shows the response coefficient obtained from incident $e^{-}$, and Plot (b) shows the response coefficient obtained from incident $\pi^{-}$. The red line is linear fit.} 
\label{fig:two_fits}
\end{figure}

For the single-channel reconstruction, the reconstructed energy is derived from a linear fit between the incident hadron energy and the number of detected photons. We define $c_s$ and $c_c$ as the hadronic response coefficients for the scintillation and Cherenkov channels. Since the photon count must be zero when the incident particle energy is zero, we force the linear fit to pass through the origin $(0,0)$. The specific fitting results are shown in fig.~\ref{two_fits.2} for scintillation photons. These coefficients $c_s$ and $c_c$ represent the response of each individual channel when the detector is treated as a conventional single-readout calorimeter. In general, the energy resolution of the scintillation channel is better than that of the Cherenkov channel due to its higher light yield. Therefore, the scintillation response is used as the single-channel for subsequent performance comparisons. For the calculation, one subset of the data is utilized to determine these coefficients, while the remaining subset is used for energy reconstruction and performance evaluation.

The parameters utilized for the single-channel algorithm across different configurations are summarized in the Tables \ref{layer_coeff}, \ref{Sample_coeff}, \ref{compare_coeff_s}, and \ref{compare_coeff_c} provided in the Addendum.

In the dual-readout calculation, a different set of calibration constants is required. We first determine the electromagnetic (EM) response coefficients,  denoted as $e_s$ for the scintillation channel and $e_c$ for the Cherenkov channel. These coefficients are extracted from linear fits of the detected photon counts against the incident energy of electrons, as illustrated in fig.~\ref{two_fits.1} . These EM coefficients $e_s$ and $e_c$ are essential for the dual-readout formula. They are used to calibrate the raw scintillation and Cherenkov signals before we combine them to reconstruct the final hadron energy.

\begin{figure}[htbp]
\centering
\includegraphics[width=0.5\textwidth]{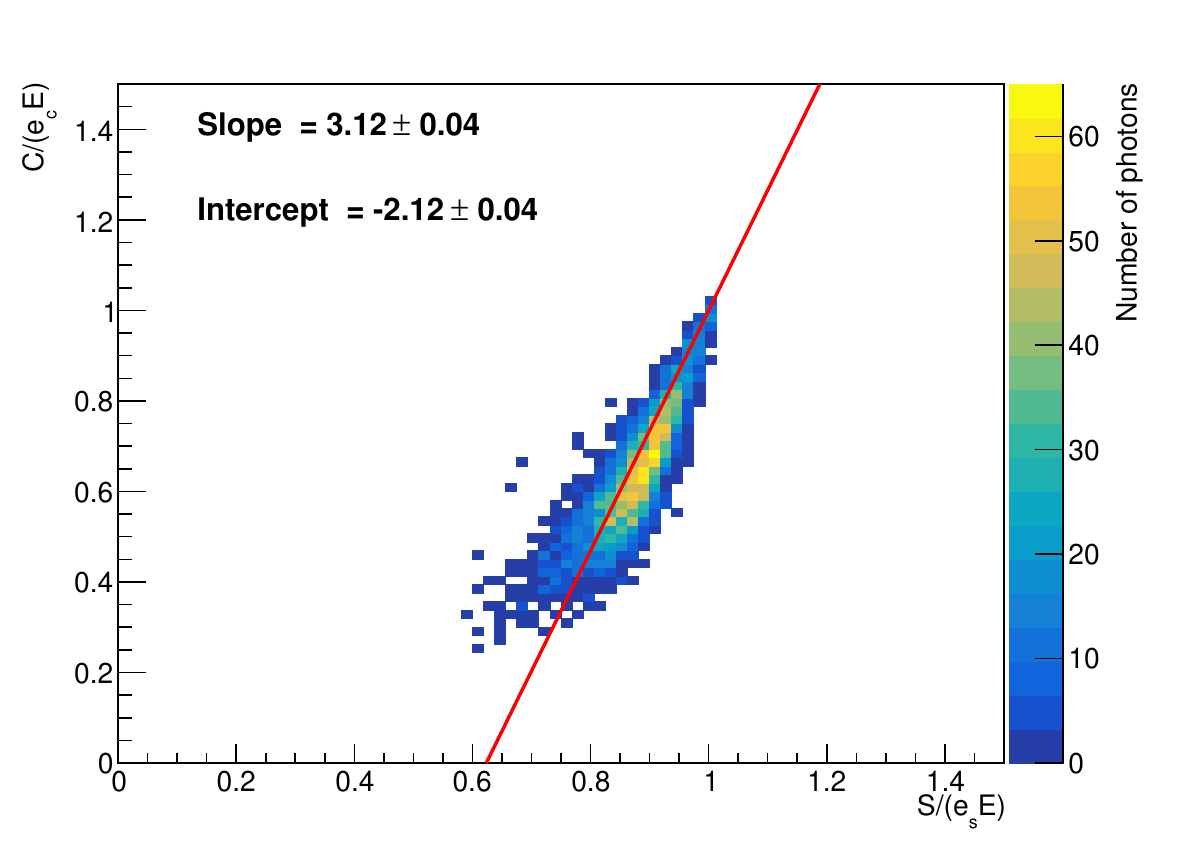}
\caption{Two-dimensional distribution of calibrated scintillation and Cherenkov photon yields for incident $\pi^-$ mesons with $E=10\;\mathrm{GeV}$ in the homogeneous calorimeter. The fitted linear correlation is used to extract the dual-readout response parameters. The red line is linear fit.}
\label{LINE_fit}
\end{figure}

Assuming the total photon counts for the scintillation and Cherenkov signals in a hadronic shower are $S$ and $C$, we calculate the corrected single-channel energy ratios, $S/(e_sE_{\text{in}})$ and $C/(e_cE_{\text{in}})$, based on the EM response coefficients. It is assumed that the slope $k$ of the dependence $C / (e_{c} E_{\text{in}})$ versus $S / (e_{s} E_{\text{in}})$ represents an intrinsic constant characteristic of the detector geometry and material, and for the fitting procedure, the $10 \text{ GeV } \pi^{-}$ events are selected, which are simulated in the homogeneous calorimeter. These results are then plotted in a two-dimensional coordinate system, as shown in fig.~\ref{LINE_fit}, where the slope is determined through a linear fit.

Finally, after obtaining the slope from the fitting process, we apply the following equation to calculate the reconstructed energy.

\begin{equation}
    E_{rec} = \frac{Sk/e_s - C/e_c }{k - 1}
\end{equation}

where $k$ denotes the fitted slope and $E_{rec}$ represents the event reconstructed energy.

The EM response coefficients $e_s$ and $e_c$, along with the slope $k$ utilized in the dual-readout calculation process, are summarized in the Tables \ref{dual_layer}, \ref{dual_sample}, \ref{compare_es}, \ref{compare_ec}, and \ref{compare_k} provided in the Addendum.

It is worth noting that the scintillation--Cherenkov correlation exhibits a clear deviation from linearity even at 10~$\mathrm{GeV}$, and this non-linear distortion becomes increasingly pronounced as the incident particle energy decreases. Such behavior reflects the growing impact of fluctuations in the electromagnetic shower fraction, nuclear breakup processes, and invisible energy losses in low-energy hadronic showers. As a consequence, the assumption of a linear $S$--$C$ correlation underlying the conventional dual-readout correction gradually breaks down, leading to a degradation of the reconstructed energy resolution and, eventually, to the failure of the dual-readout correction in the low-energy regime.

Since both the training and test samples cover a continuous range of incident energies, the energy resolution is evaluated in discrete energy intervals. The full energy range is divided into bins of equal width. For each bin, the difference between the reconstructed energy and the incident energy is calculated on an event-by-event basis. The energy resolution $\sigma_E$ is obtained from the standard deviation of this distribution, and the relative energy resolution is then calculated by dividing $\sigma_E$ by the mean reconstructed energy $\langle E_{rec} \rangle$. The bin center is adopted as the incident energy $E_{in}$ for each interval.

\begin{figure}[htbp]
\centering
\subfigure[]{
\label{contiousenergy_4}
\includegraphics[width=0.2\textwidth]{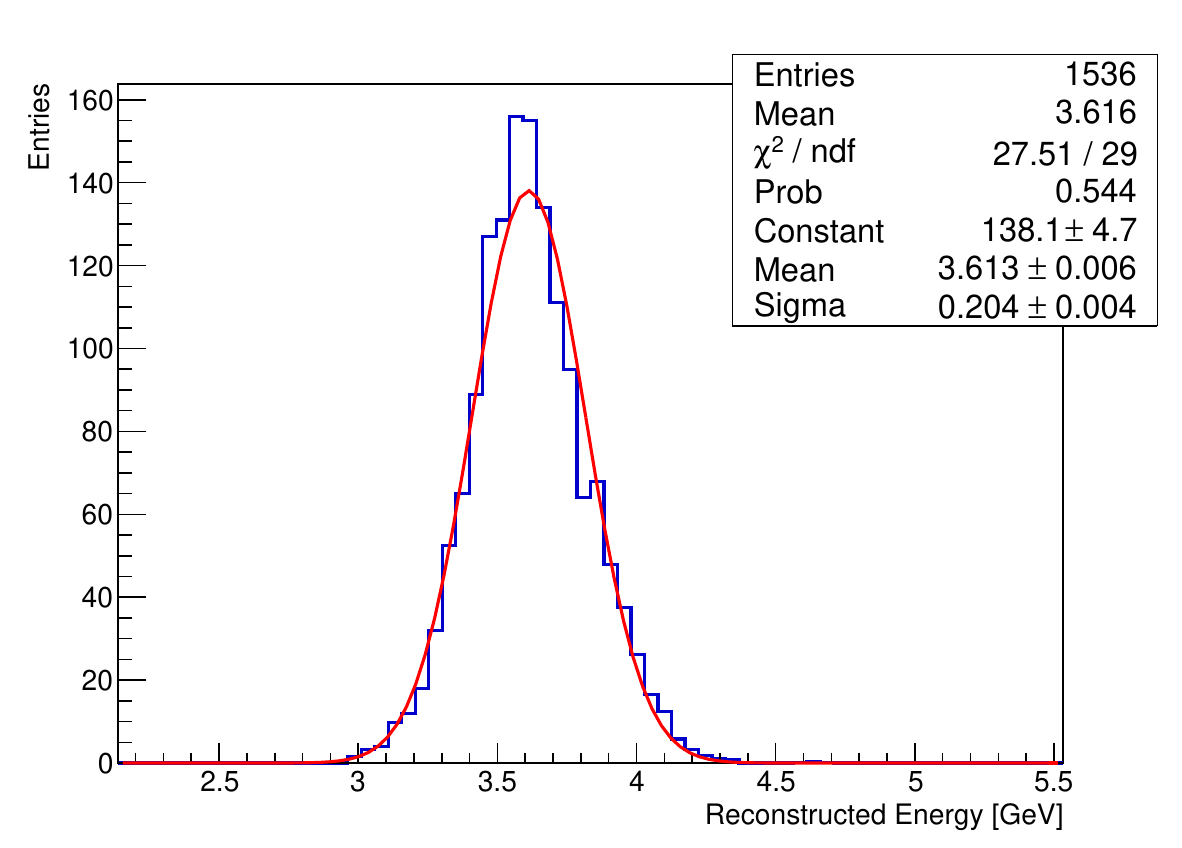}}
\subfigure[]{
\label{contiousenergy_6}
\includegraphics[width=0.2\textwidth]{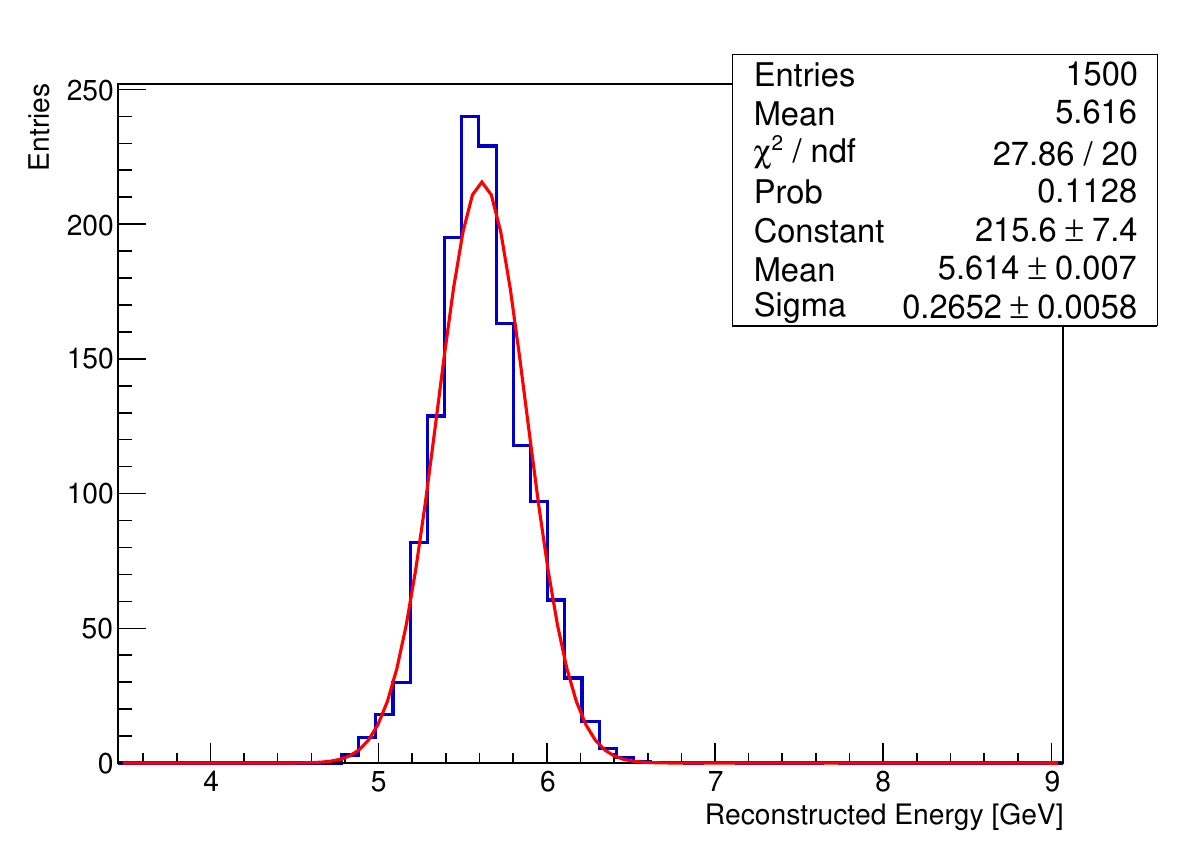}}
\subfigure[]{
\label{contiousenergy_8}
\includegraphics[width=0.2\textwidth]{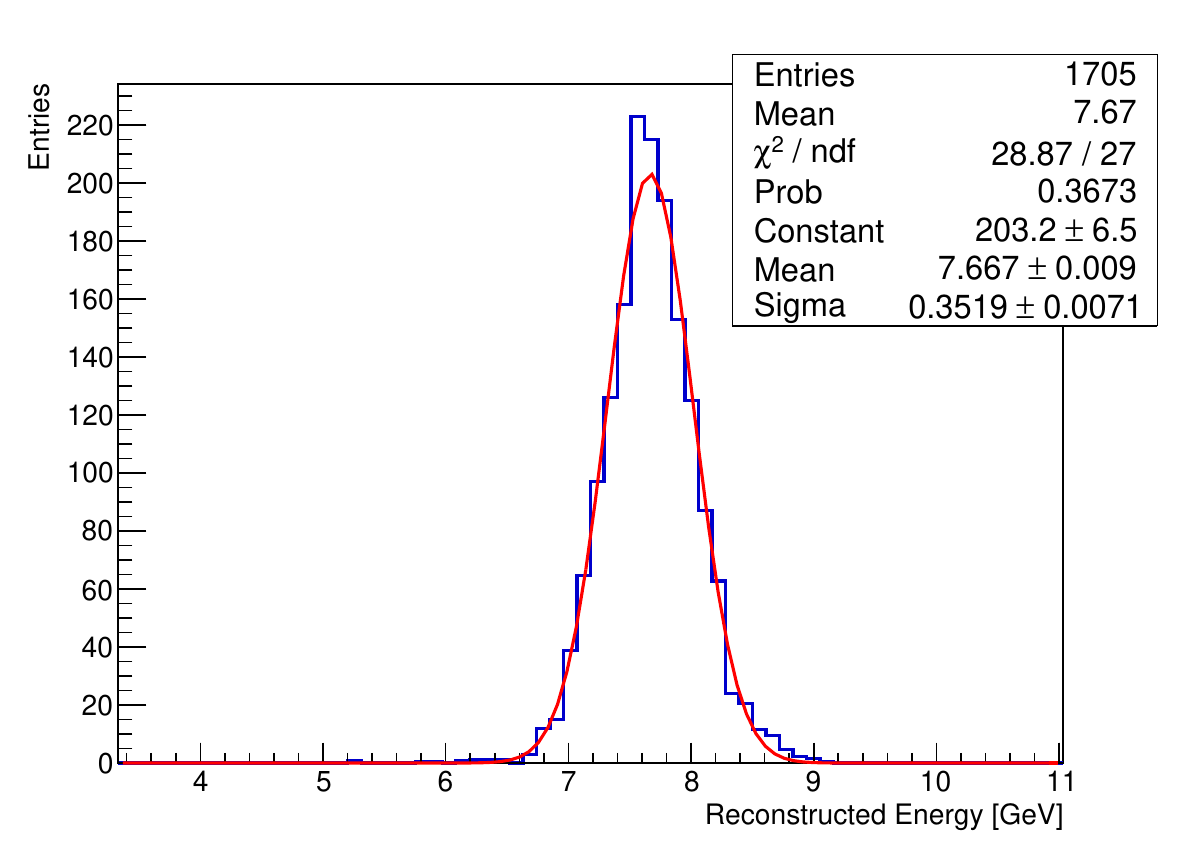}}
\subfigure[]{
\label{contiousenergy_10}
\includegraphics[width=0.2\textwidth]{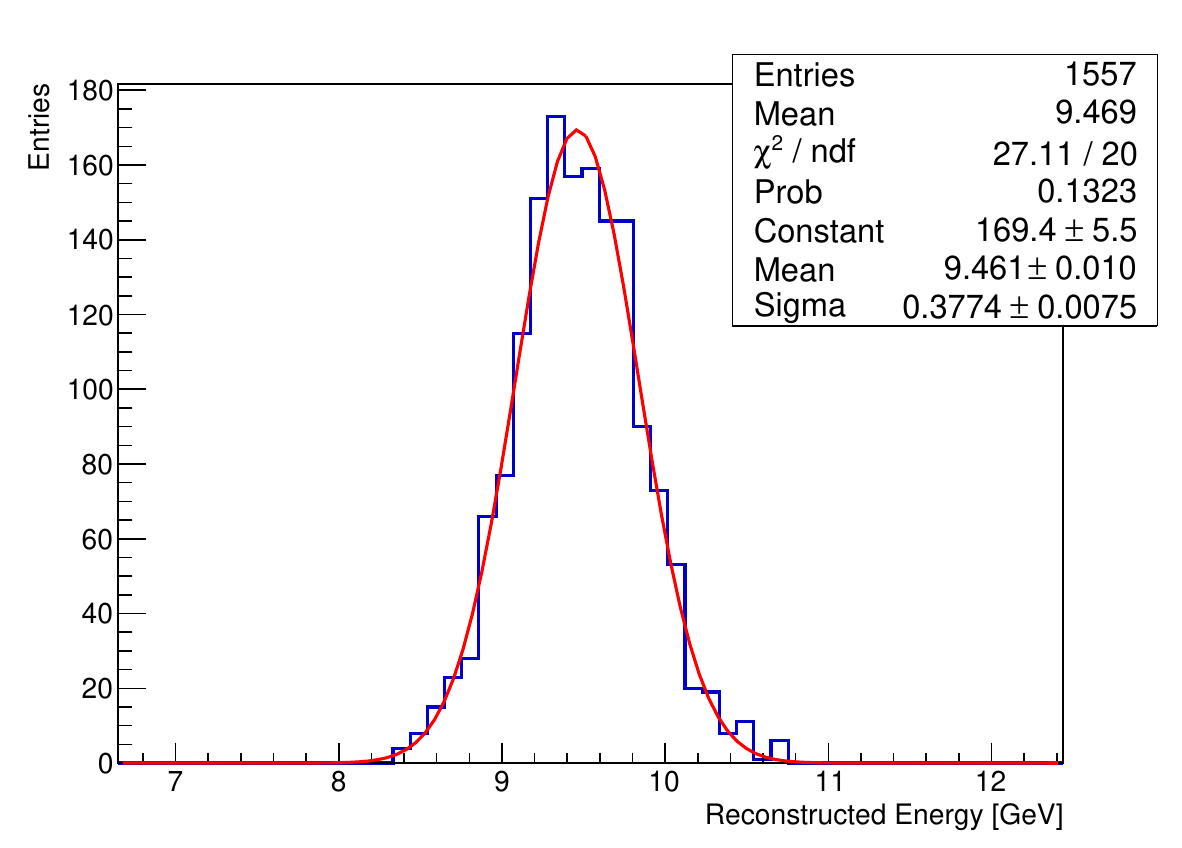}}
\caption{Distributions of reconstructed energies for continuous energy bins $3$–$4 \text{ GeV}$ (a), $5$–$6 \text{ GeV}$ (b), $7$–$8 \text{ GeV}$ (c), and $9$–$10 \text{ GeV}$ (d) in the homogeneous calorimeter. The red curves show Gaussian fits.} 
\label{contiousenergy}
\end{figure}

To demonstrate the energy reconstruction performance, we provide the distributions for several energy bins. These distributions are presented by adding the corresponding bin center value to the event-by-event residuals, as illustrated in fig.~\ref{contiousenergy}.

After obtaining the energy resolution in each energy interval, the results are fitted using the commonly adopted parametrization.
\begin{equation}
    \frac{\sigma_E}{\langle E_{rec} \rangle} = \frac{a}{\sqrt{E_{in}}} \oplus b
    \label{fiteq}
\end{equation}
where $\sigma_E$ represents the standard deviation of the reconstructed energy distribution, and $\langle E_{rec} \rangle$ denotes the mean reconstructed energy. $E_{in}$ is the incident particle energy, while the parameter $a$ represents the stochastic term of the energy resolution and $b$ accounts for the constant contribution.

\begin{figure}[htbp]
\centering
\subfigure[]{
\label{Pure_ML}
\includegraphics[width=0.2\textwidth]{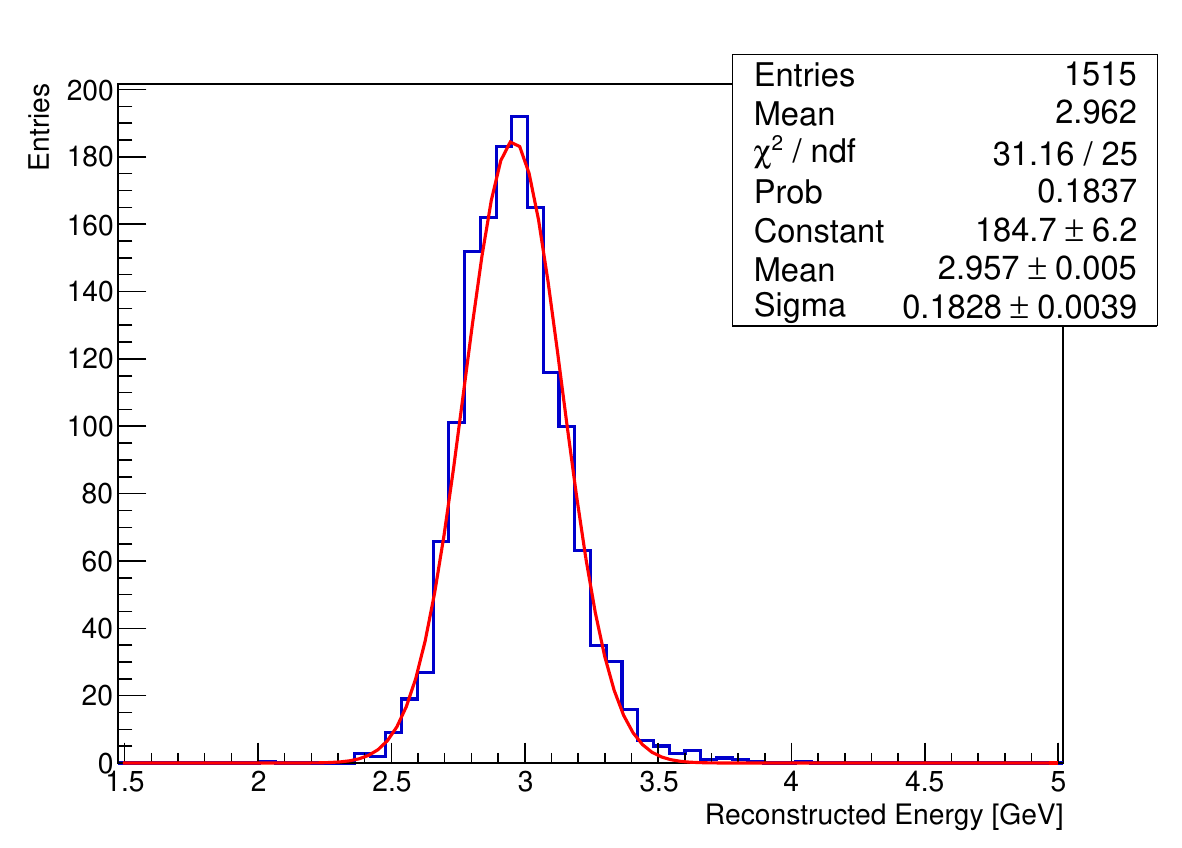}}
\subfigure[]{
\label{Pure_ML}
\includegraphics[width=0.2\textwidth]{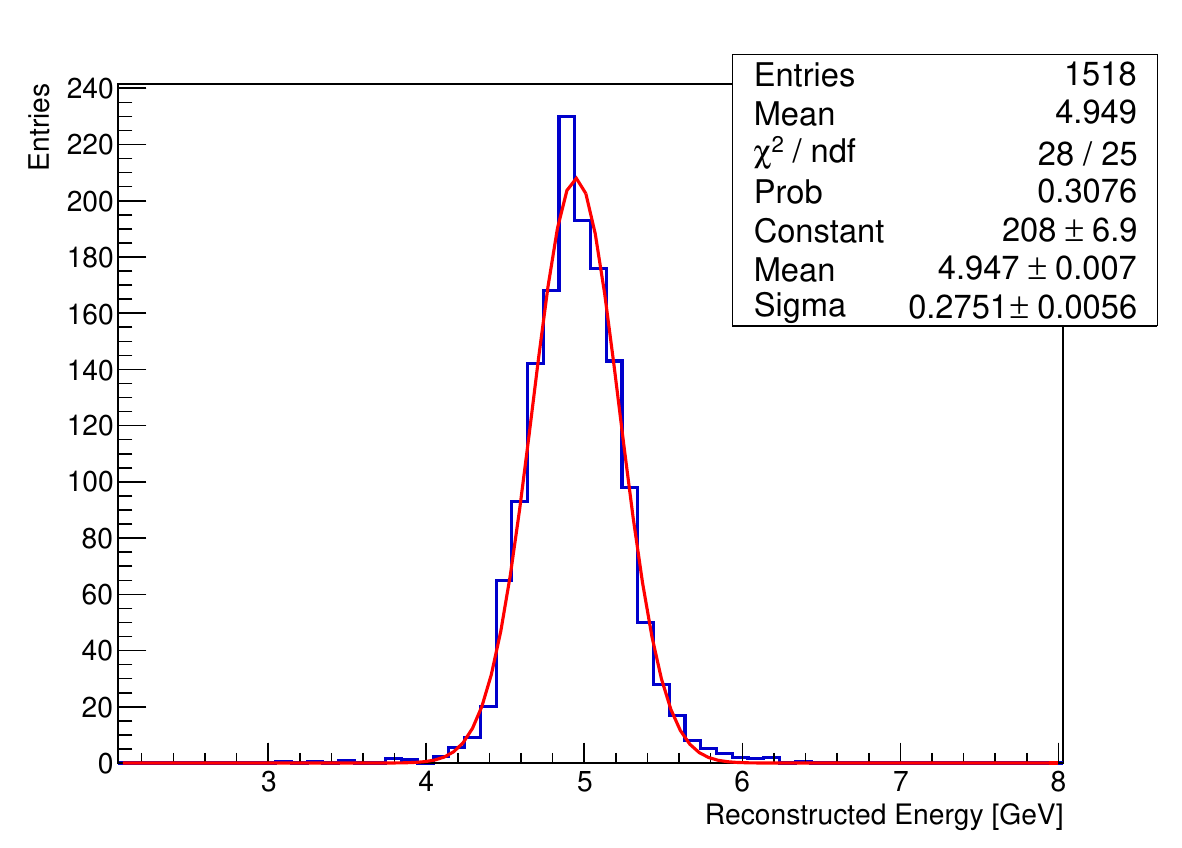}}
\subfigure[]{
\label{Pure_ML}
\includegraphics[width=0.2\textwidth]{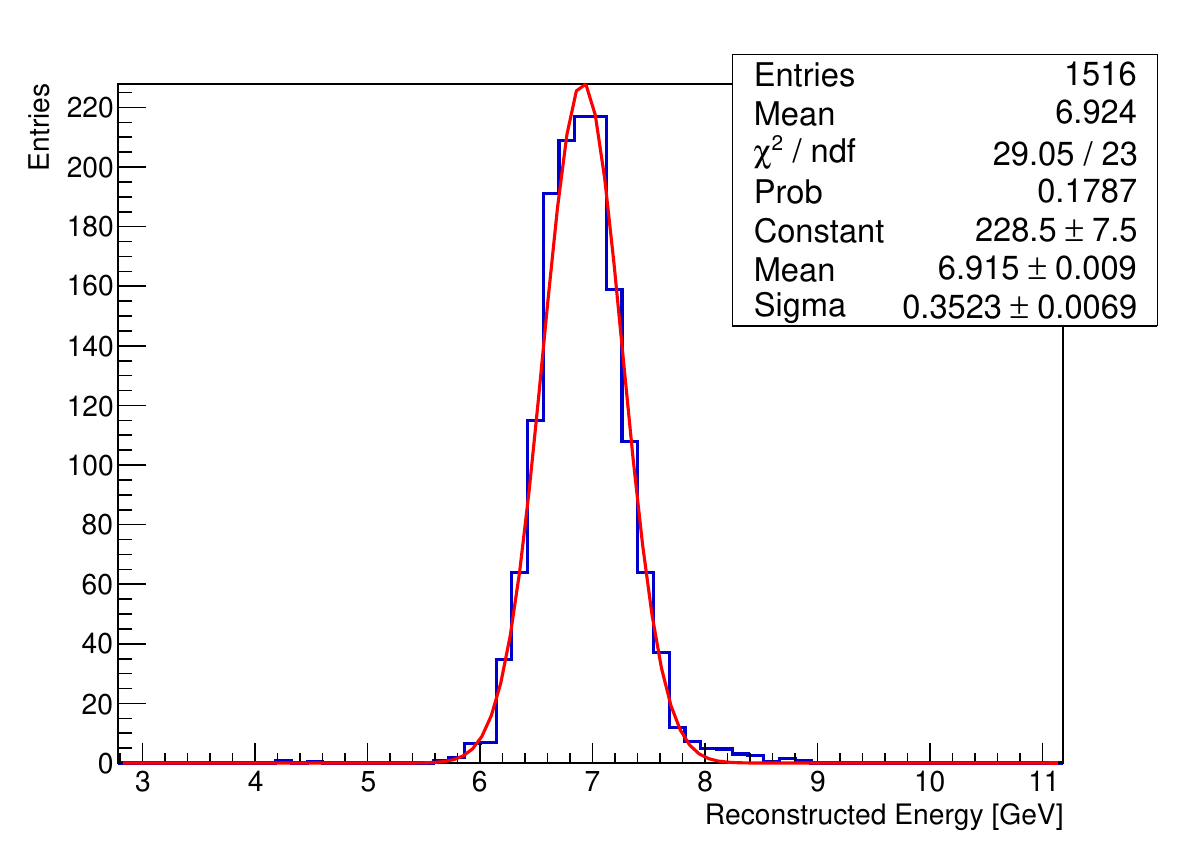}}
\subfigure[]{
\label{Pure_ML}
\includegraphics[width=0.2\textwidth]{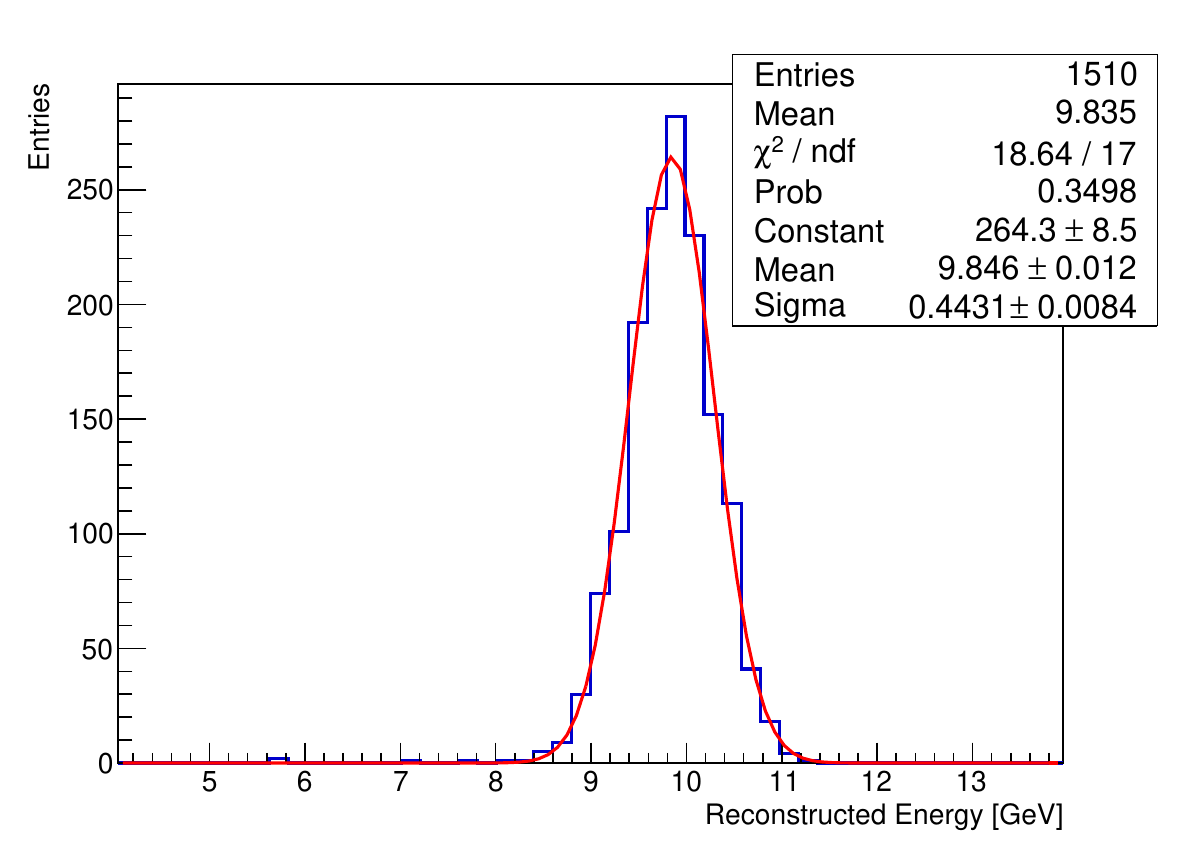}}
\caption{Distributions of reconstructed energy predicted by machine learning option
ML12 for mono-energetic points 3 GeV (a), 5 GeV (b), 7 GeV (c), and 10 GeV
(d) in the homogeneous calorimeter. The red curves show Gaussian fits.} 
\label{monoenergy}
\end{figure}

For the optimal of machine learning, uniform energy distribution were employed for both training and testing. To ensure the robustness of the results, we validated the energy resolution by comparing the continuous spectrum approach with tests performed at discrete mono-energetic points at 3, 5, 7, and 10 $\text{GeV}$ and utilized the previously trained weights to predict the energy values for these mono-energetic points. The mean value and resolution of $E_{\text{rec}}$ were then obtained through Gaussian fitting, as illustrated in fig.~\ref{monoenergy}.

\begin{figure}[htbp]
\centering
\subfigure[]{
\label{Pure_ML}
\includegraphics[width=0.4\textwidth]{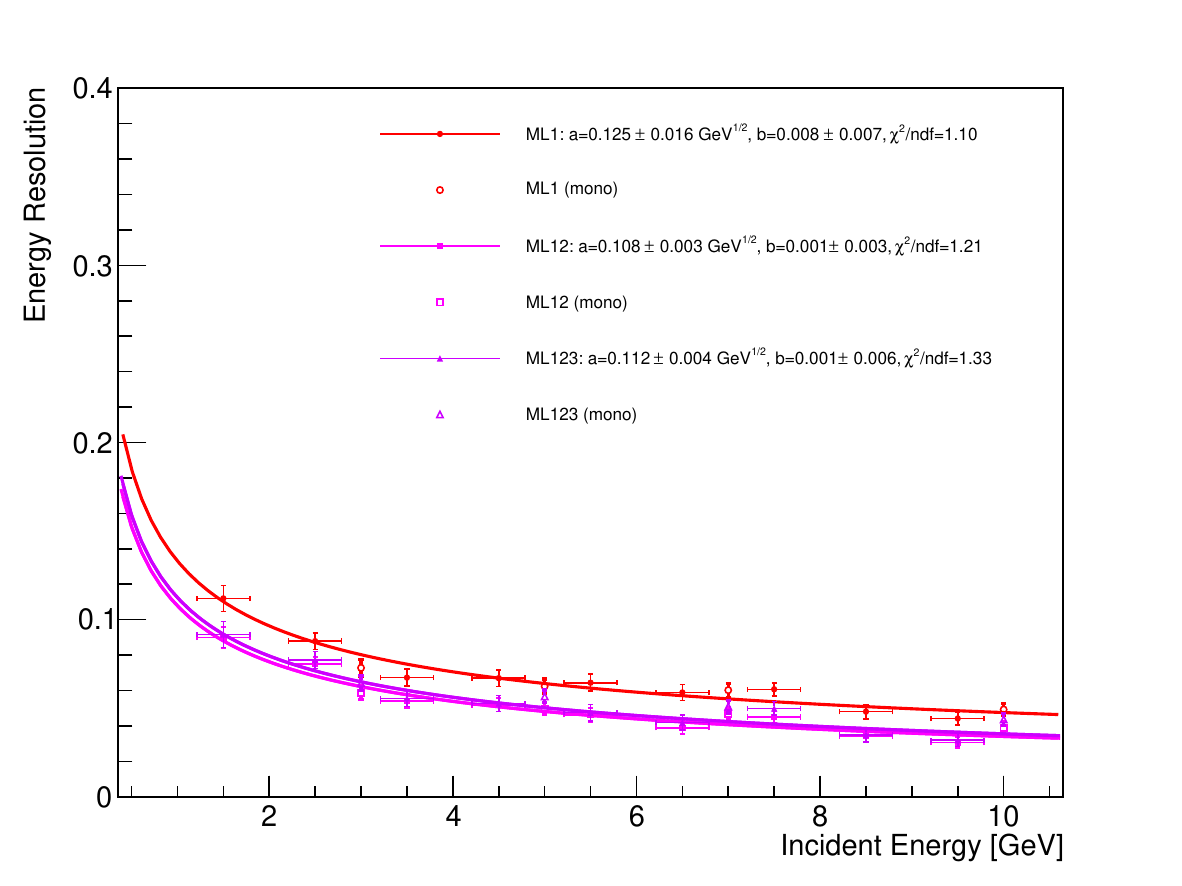}}
\subfigure[]{
\label{Pure_ML}
\includegraphics[width=0.4\textwidth]{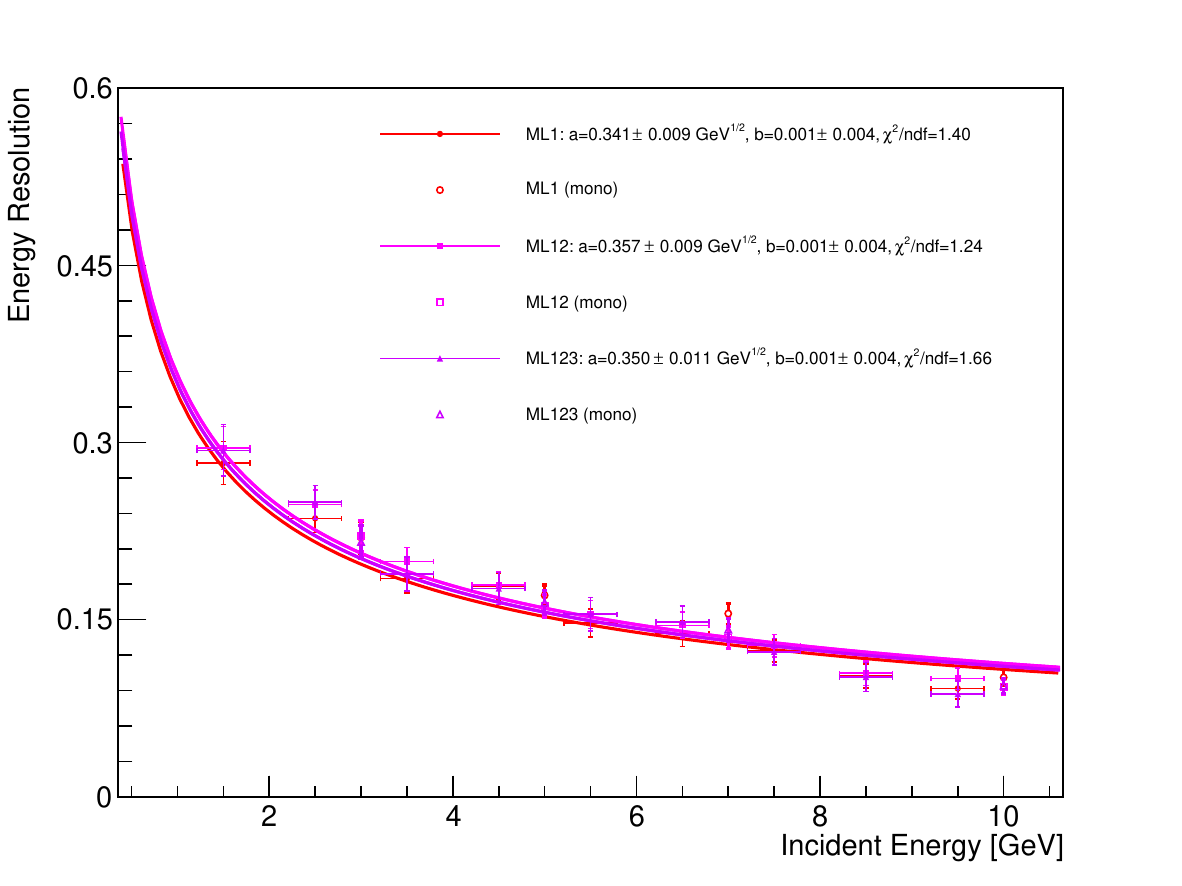}}
\caption{Comparison of relative energy resolution between mono-energetic points and continuous energy for ML1, ML12, and ML123 configurations. Plots (a) and (b) correspond to the homogeneous calorimeter and the sampling calorimeter, see text for detail. Solid markers represent results from continuous energy, while open markers denote mono-energetic data points. The fitting is performed in the range of $1.75$–$9.75 \text{ GeV}$.} 
\label{Comparison}
\end{figure}

For the monoenergetic option, the energy resolution is calculated using the ratio $\sigma/\mu$, where $\mu$ and $\sigma$ are the mean value and resolution obtained from the Gaussian fit. We then compared this relative energy resolution with the results previously obtained from continuous energy distribution, as shown in fig.~\ref{Comparison}.  We considered two distinct configurations: homogeneous and low sampling ratio (sampling segmentation of 50 mm with a sampling ratio of 10\%). It can be observed that for both homogeneous and sampling calorimeters, the results for the mono-energetic points are consistent with those from the continuous energy.

\begin{figure}[htbp]
\centering
\subfigure[]{
\label{Pure_ML}
\includegraphics[width=0.3\textwidth]{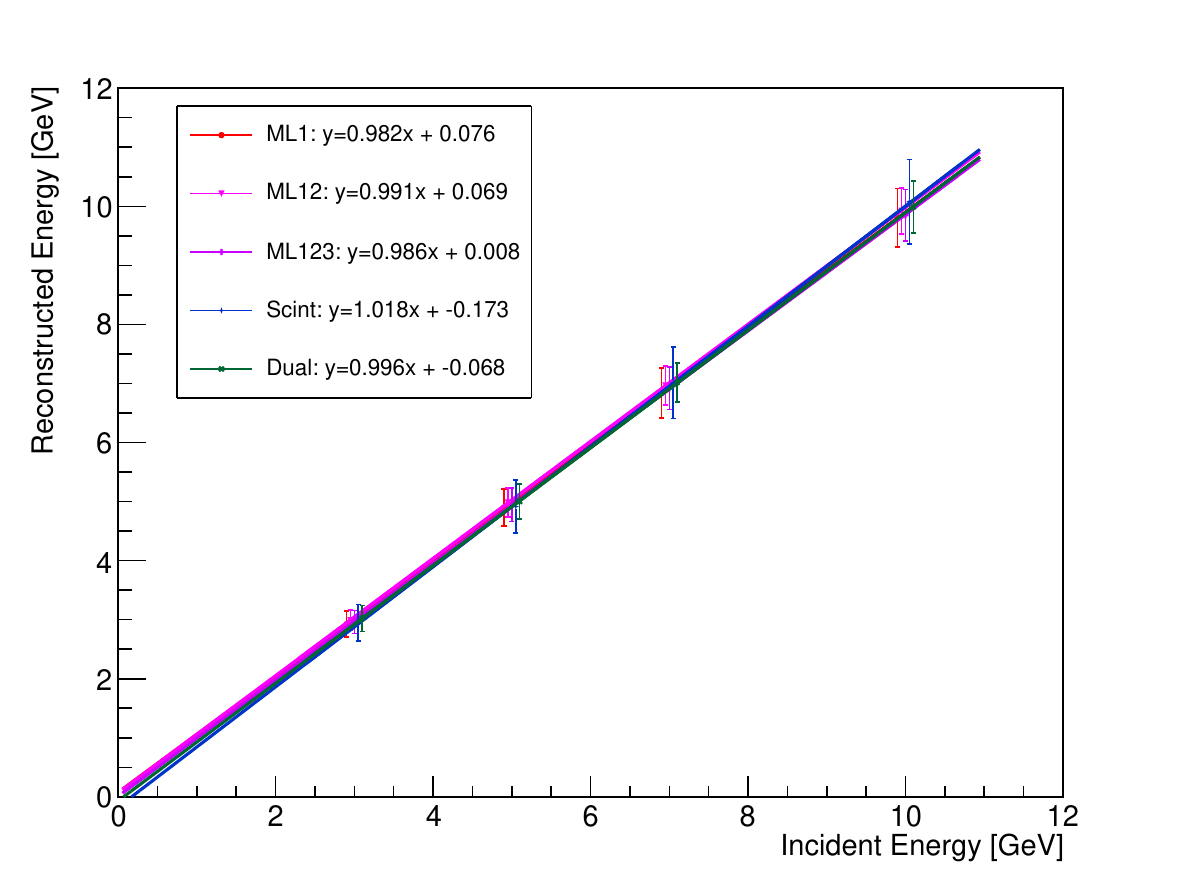}}
\subfigure[]{
\label{Pure_ML}
\includegraphics[width=0.3\textwidth]{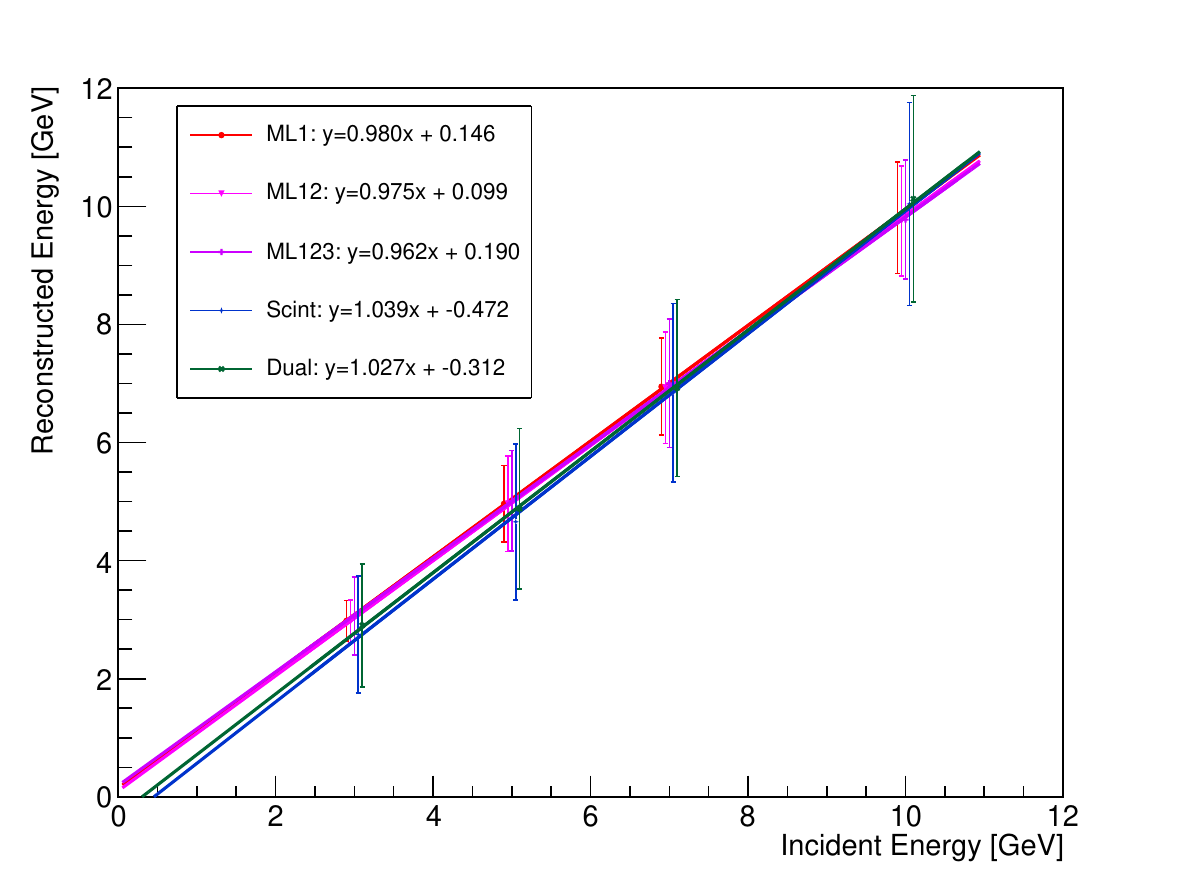}}
\caption{Linear fit of the mean reconstructed energy versus the incident particle energy, where the horizontal axis represents the incident particle energy and the vertical axis represents the mean value of the reconstructed energy. Plot (a) shows the results for a homogeneous calorimeter, while plot (b) shows the results for a sampling calorimeter. See text for details.} 
\label{Linearity}
\end{figure}

\begin{figure}[htbp]
\centering
\subfigure[]{
\label{Pure_ML}
\includegraphics[width=0.4\textwidth]{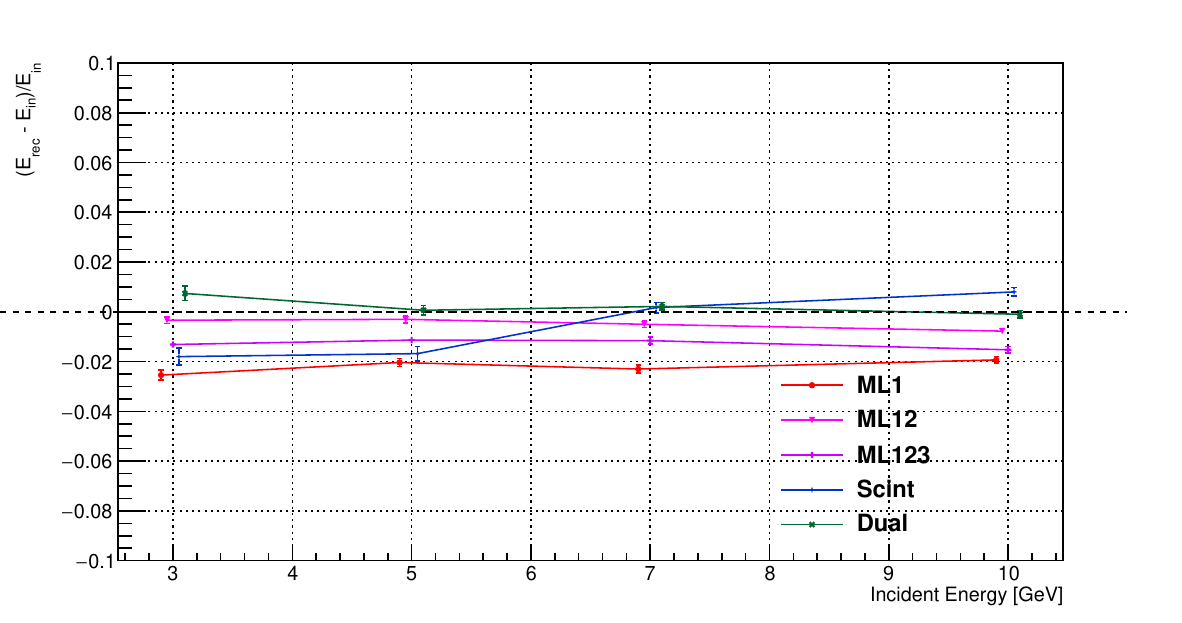}}
\subfigure[]{
\label{Pure_ML}
\includegraphics[width=0.4\textwidth]{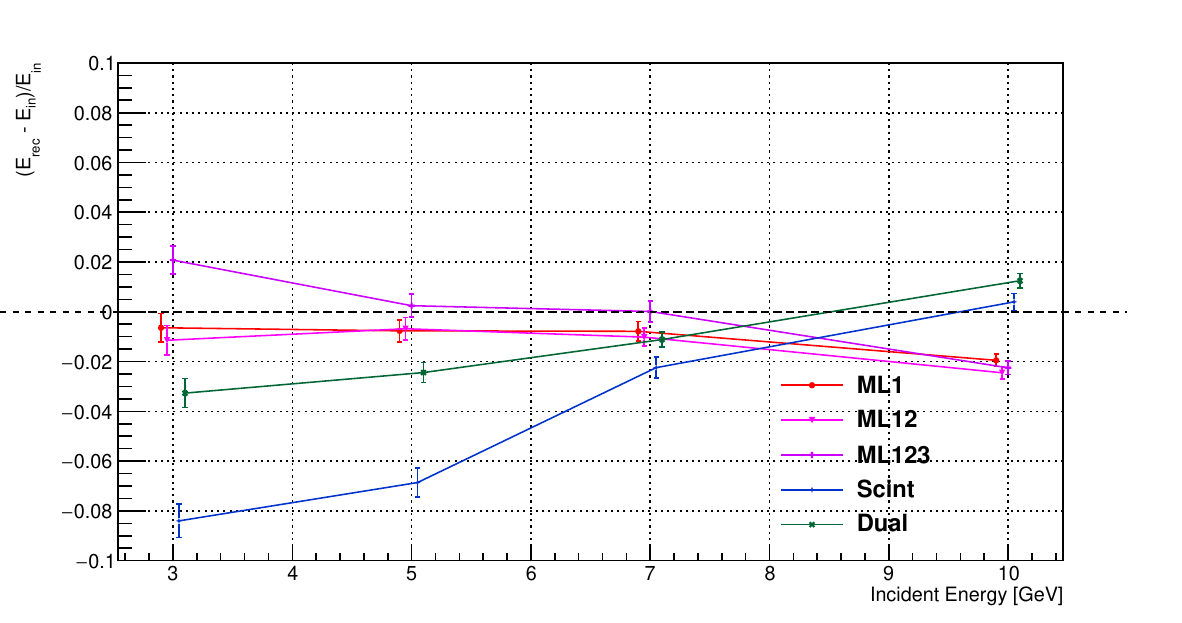}}
\caption{Residual relationship between the mean reconstructed energy and the incident particle energy, where the horizontal axis represents the incident particle energy and the vertical axis represents the residuals.  Plot (a) displays the results for a homogeneous calorimeter, and plot (b) displays the results for a sampling calorimeter. See text for details.} 
\label{Residuals}
\end{figure}

To compare the linearity between the machine learning approach and the dual-readout algorithm, the relationship between the incident energy and the mean reconstructed energy for the scint, dual, and machine learning methods is shown in fig.~\ref{Linearity}. Two representative calorimeter configurations were considered in this comparison: a homogeneous calorimeter and a low sampling-ratio calorimeter with a sampling segmentation of 50~mm and a sampling ratio of 10\%. To determine the mean reconstructed energy, the same Gaussian fitting parameters as described in fig.~\ref{monoenergy} were applied. Additionally, the data points are slightly offset along the $x$-axis to avoid overlap and ensure a clearer visualization of the error bars. To further evaluate the statistical consistency of the reconstruction across the entire energy range, the residuals, defined as $(\langle E_{\text{rec}} \rangle - E_{\text{in}})/E_{\text{in}}$, were calculated and are displayed in fig.~\ref{Residuals}. The uncertainties for these data points are defined as $\sigma / E_{\text{in}}$, where $\sigma$ is the uncertainty of the mean reconstructed energy extracted directly from the Gaussian fit.

The linearity results from Figure 8 show that neither structure achieves a perfect linear response, with the majority of options tending to underestimate the reconstructed energy. The homogeneous structure provides better linearity than the presented sampling structure, with 4\% deviation compared to 8\% for the latter. For the homogeneous option, the dual-readout choice excels while ML1 and ML123 models underestimate the energy by about 2\%. Conversely, for the sampling option, ML models show a trend toward overcompensation, though the dual-readout option still outperforms the single scintillation option as expected.

\section{Digitization Effect}
\label{sec:digitization}

\begin{figure}[htbp]
\centering
\subfigure[]{
\label{GeneScint}
\includegraphics[width=0.4\textwidth]{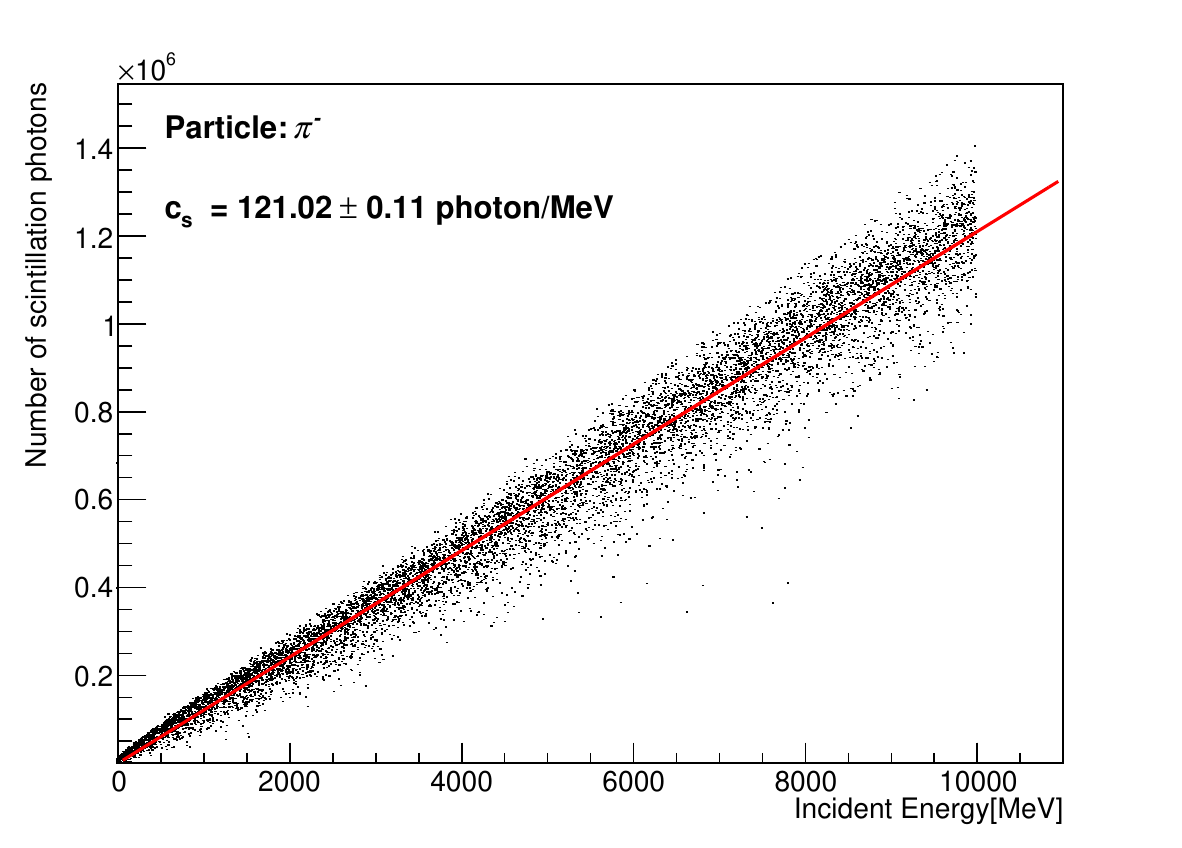}}
\subfigure[]{
\label{OutputScint}
\includegraphics[width=0.4\textwidth]{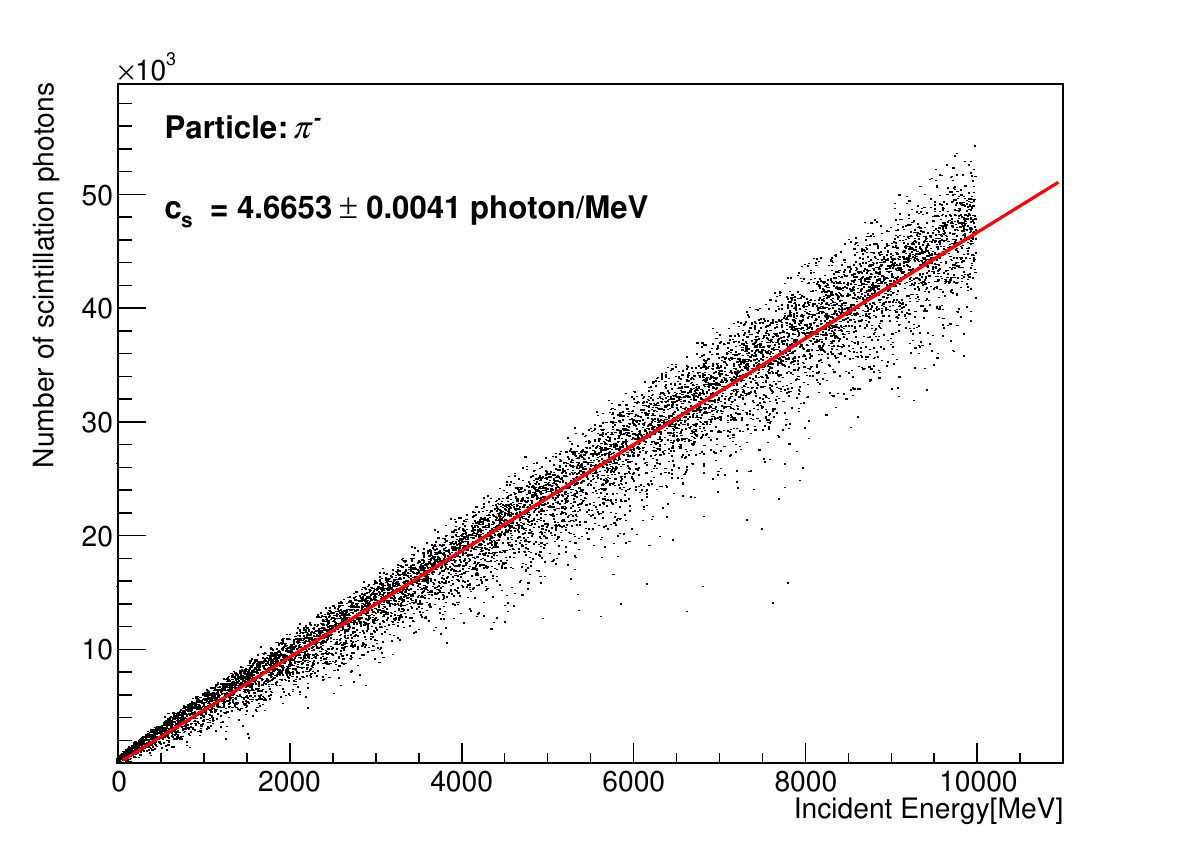}}
\caption{Comparison of scintillation photon counts for incident $\pi^{-}$ mesons with $E_{\text{in}}$ from $0$ to $10000 \text{ MeV}$ in the homogeneous calorimeter before and after the detection process, where (a) represents the initial generated photons and (b) represents the final detected photons. The red line is the linear fit.} 
\label{PWO_Scint}
\end{figure}

\begin{figure}[htbp]
\centering
\subfigure[]{
\label{PWO_Generate}
\includegraphics[width=0.4\textwidth]{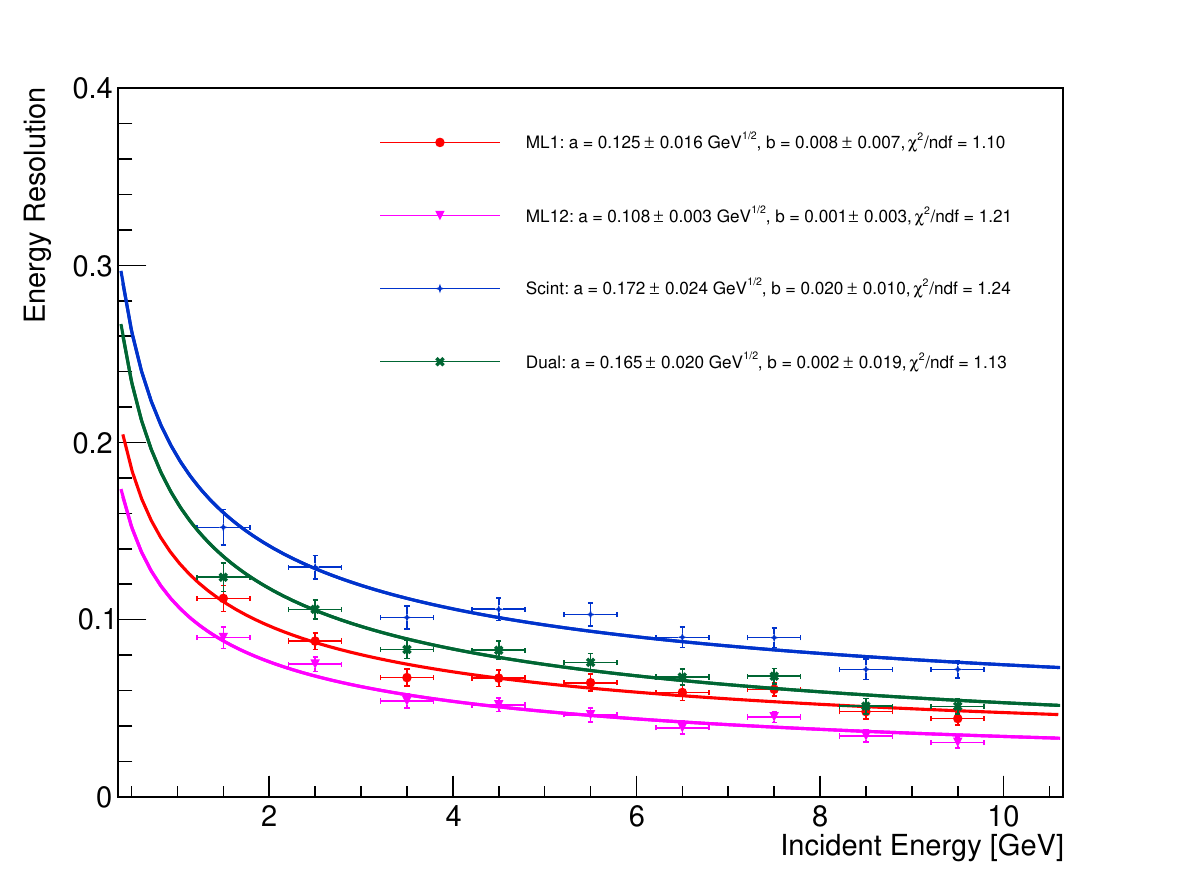}}
\subfigure[]{
\label{PWO_Detected}
\includegraphics[width=0.4\textwidth]{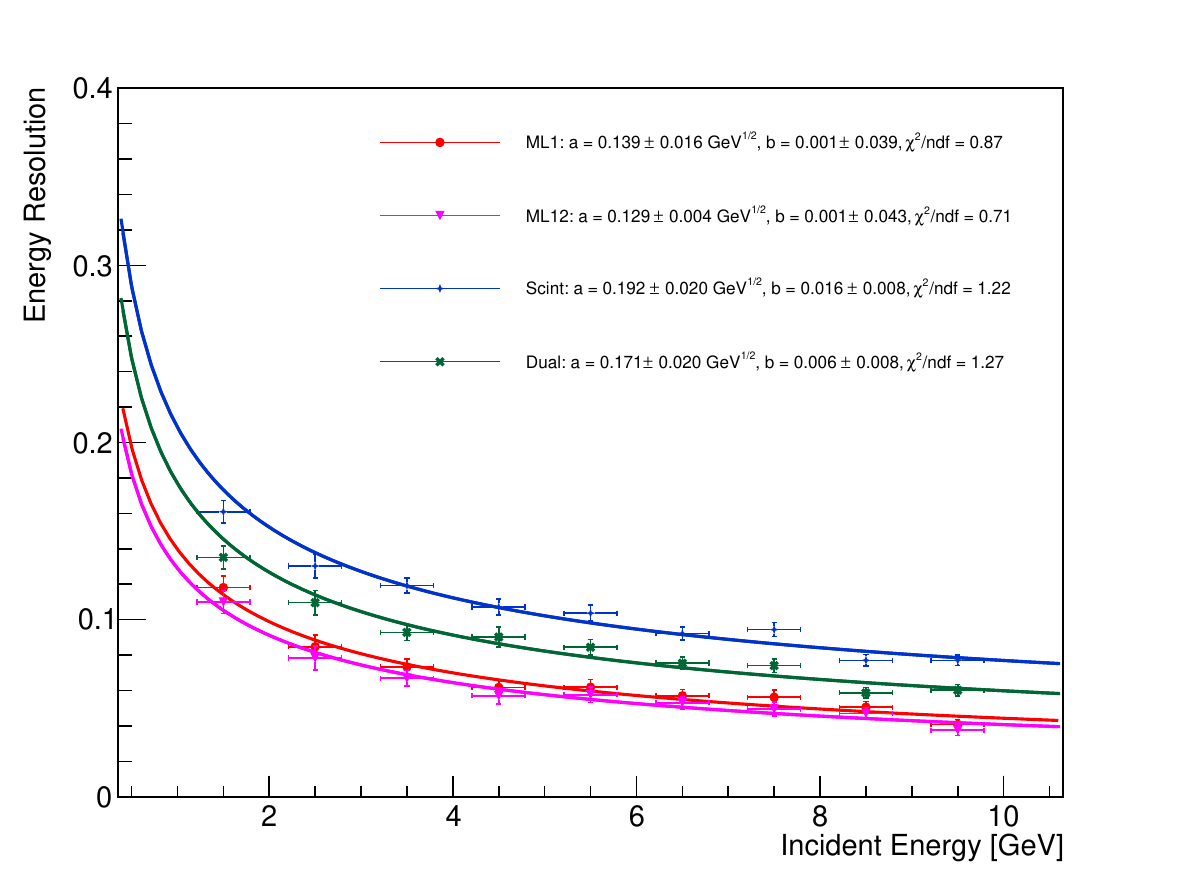}}
\caption{Comparison of energy resolution in the homogeneous calorimeter before and after the detection process, where (a) represents the resolution obtained using initial generated photons and (b) represents the resolution obtained using final detected photons. The fitting is performed in the range of $1.75$–$9.75 \text{ GeV}$.} 
\label{PWO_De}
\end{figure}

To investigate the impact of detection efficiency on energy resolution, additional simulations were performed. While maintaining the total volume of the $\mathrm{PbWO_4}$ crystal at $1 \times 1 \times 2~\mathrm{m^3}$, the detector was divided into a $100 \times 100 \times 100$ pixel array. Reflective films with an average reflectivity of 96\% were applied to all surfaces of each voxel except for one end-face, from which the scintillation and Cherenkov photons were independently collected. Furthermore, for the photons successfully escaping the crystal, the actual detection was calculated by applying the energy-dependent photon detection efficiency of the EQR-20 series SiPMs. For a homogeneous calorimeter where signals from all pixel units are read out, the resulting variations in photon counts and energy resolutions are shown in fig.~\ref{PWO_Scint} and fig.~\ref{PWO_De}, respectively. 

\begin{figure}[htbp]
\centering
\subfigure[]{
\label{SamplingScint}
\includegraphics[width=0.43\textwidth]{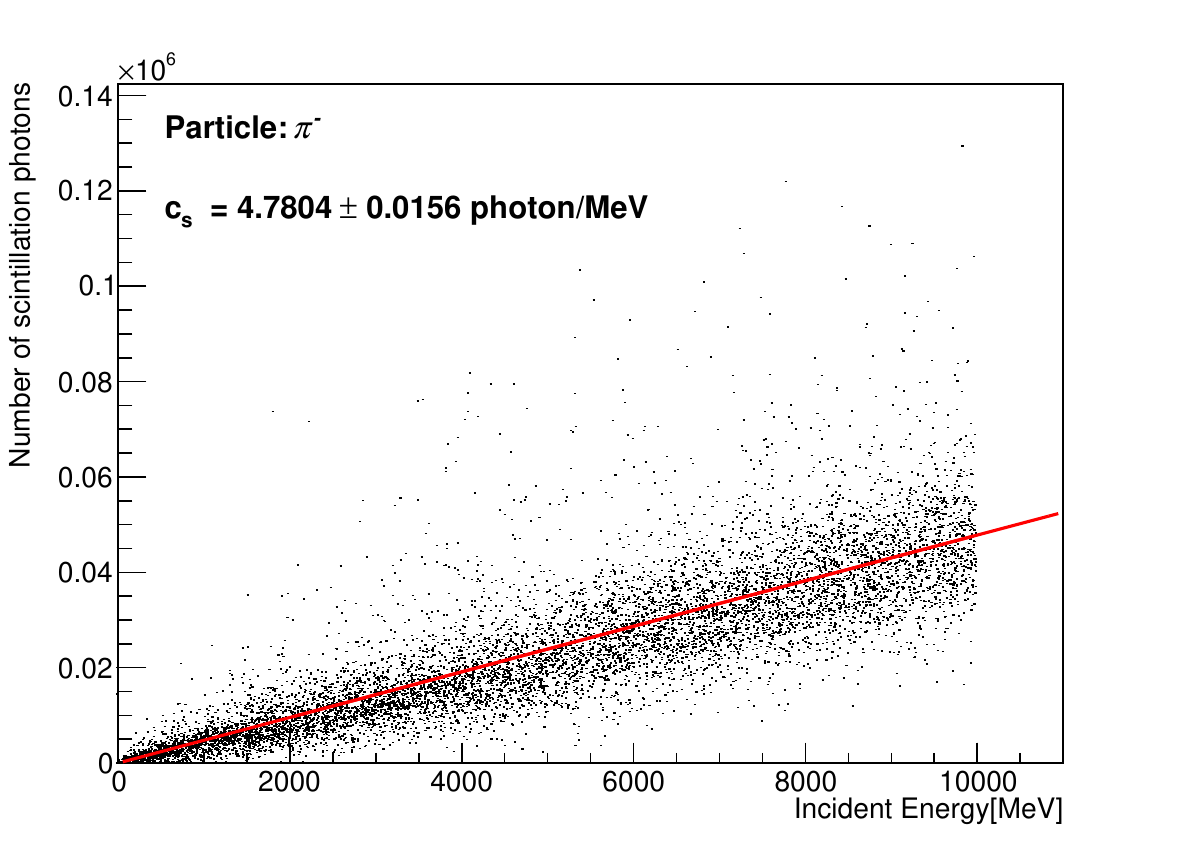}}
\subfigure[]{
\label{Sampling_SamePhotons}
\includegraphics[width=0.4\textwidth]{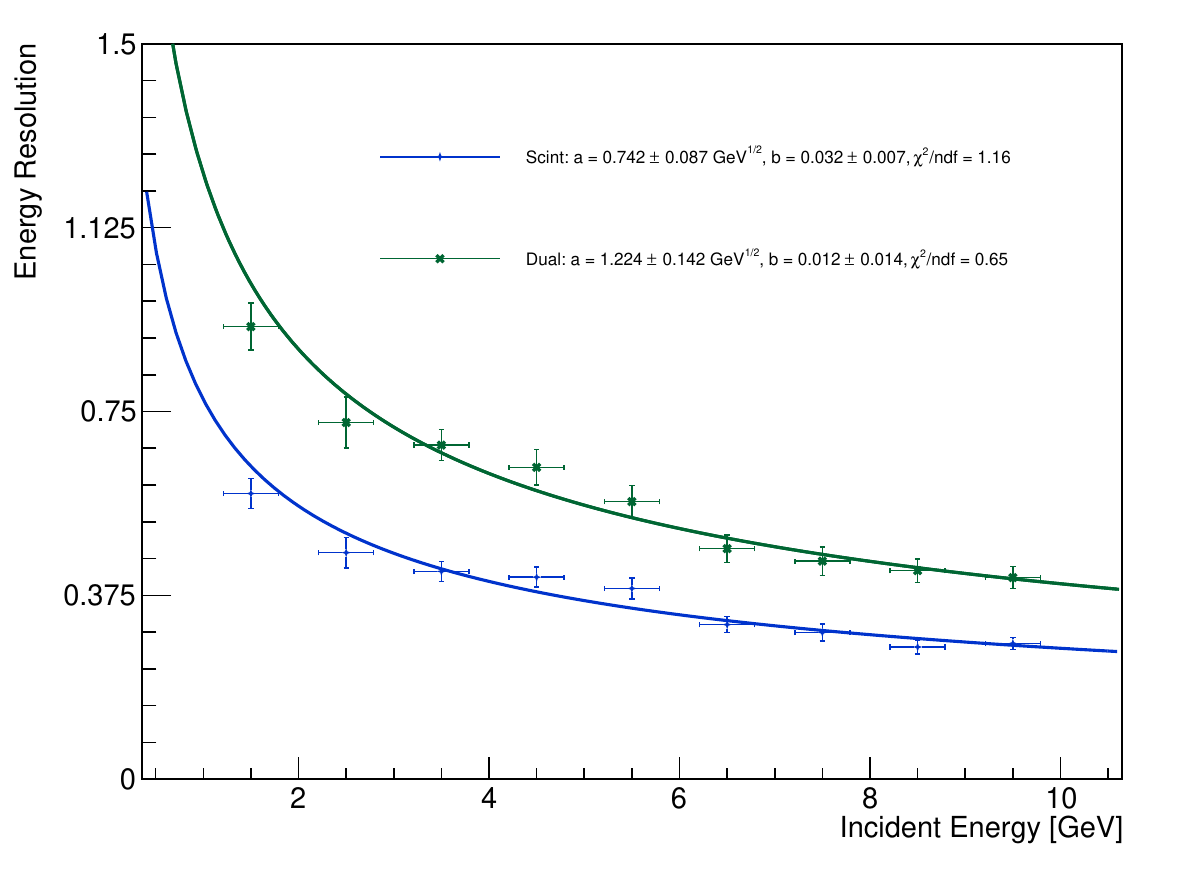}}
\caption{Photon counts and energy resolution for the sampling calorimeter with a 3\% sampling ratio and 50~mm segmentation. Plot (a) shows the number of scintillation photon counts versus incident particle energy. The red line is the linear fit. Plot (b) shows energy resolution for single scintillation and dual-readout channels. The fitting is performed in the range of $1.75$–$9.75 \text{ GeV}$.}
\label{Detected_Sampling}
\end{figure}

It can be observed that after accounting for the detection process, the light yield decreases from 121 photon/MeV to 4.7 photon/MeV. Despite this reduction, the energy resolution reconstructed via the conventional algorithm shifts from $(17.2\pm2.4)\%/\sqrt{E/GeV}$ to $(19.2\pm2.0)\%/\sqrt{E/GeV}$ for the single scintillation channel, and from $(16.5\pm2.0)\%/\sqrt{E/GeV}$ to $(17.1\pm2.0)\%/\sqrt{E/GeV}$ for the dual-readout channels. Although the energy resolution degrades, the impact is relatively limited, suggesting that the fluctuations introduced by detection effects are not the dominant factors compared to the intrinsic fluctuations of hadronic showers. In addition to conventional algorithms, we also tested the detected photon distributions using the machine learning model previously trained on initial generated photons. To ensure the model's applicability, the detected photon counts were scaled by the factors shown in fig.~\ref{PWO_Scint} to maintain the same order of magnitude. The machine learning results similarly indicate that despite a partial degradation in energy resolution, the overall performance is consistent with the results obtained from generated photons. For further comparison, an equivalent photon count was achieved by reducing the sampling ratio and increasing the sampling segmentation instead of implementing the detector effects. Fig.~\ref{SamplingScint} illustrates the photon count distribution obtained for a configuration with a 3\% sampling ratio and a 50~mm sampling segmentation.

It can be observed from fig.~\ref{Sampling_SamePhotons} that with the adoption of the sampling structure,  the energy resolution for the single scintillation channel degrades to $(74.2\pm6.7)\%/\sqrt{E/GeV}$, while the dual-readout resolution also degrades to to $(122.4\pm14.2)\%/\sqrt{E/GeV}$. This degradation relative to the homogeneous crystal indicates that the sampling error contributes more to the overall energy resolution, reaching or exceeding the level of intrinsic fluctuations. By extension, the homogeneous calorimeter can be viewed as a sampling calorimeter with infinitesimal segmentation; in this case, the reduction in photon counts corresponds to statistical fluctuations associated with a lower sampling ratio. Under the condition of infinitesimal segmentation, the energy resolution remains relatively stable despite a reduction in photon counts due to the decreased sampling ratio.

\begin{figure}[htbp]
\centering
\subfigure[]{
\label{90cut_Photon_Generate}
\includegraphics[width=0.4\textwidth]{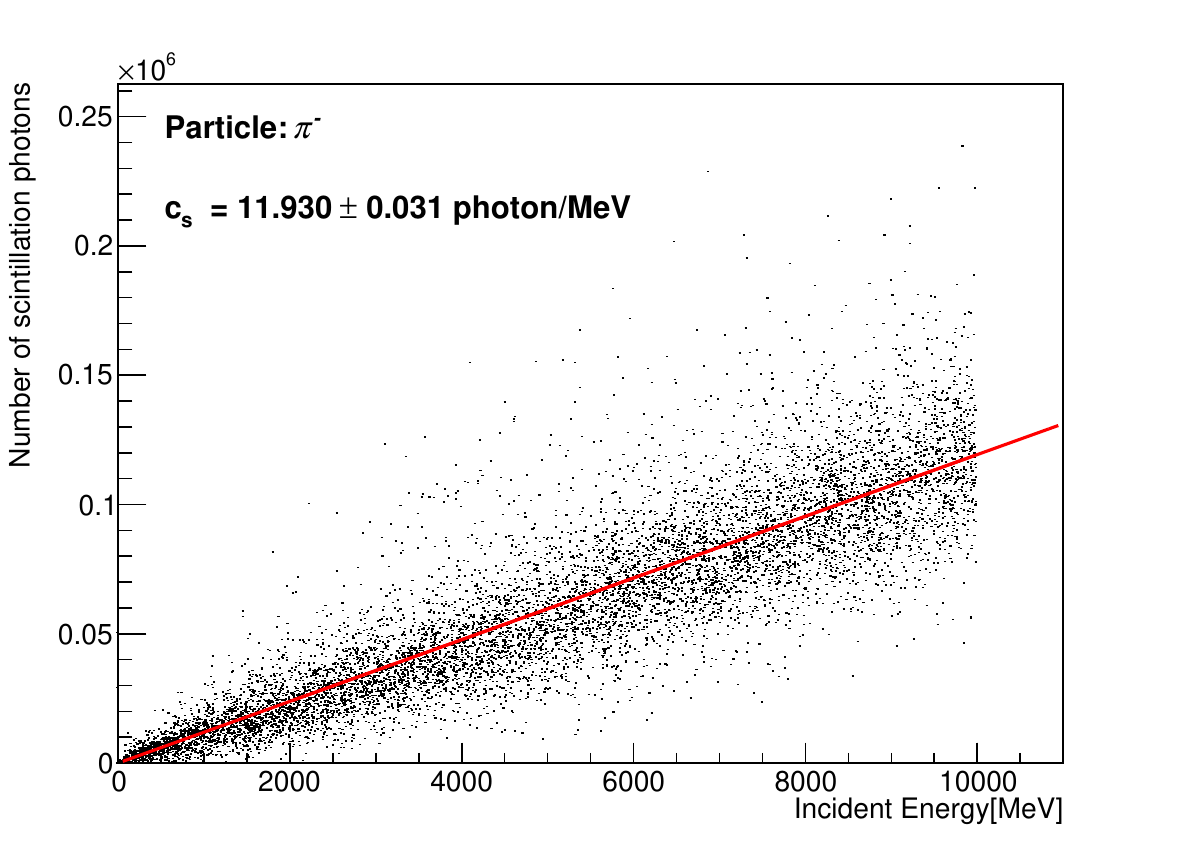}}
\subfigure[]{
\label{90cut_Photo_Detected2}
\includegraphics[width=0.4\textwidth]{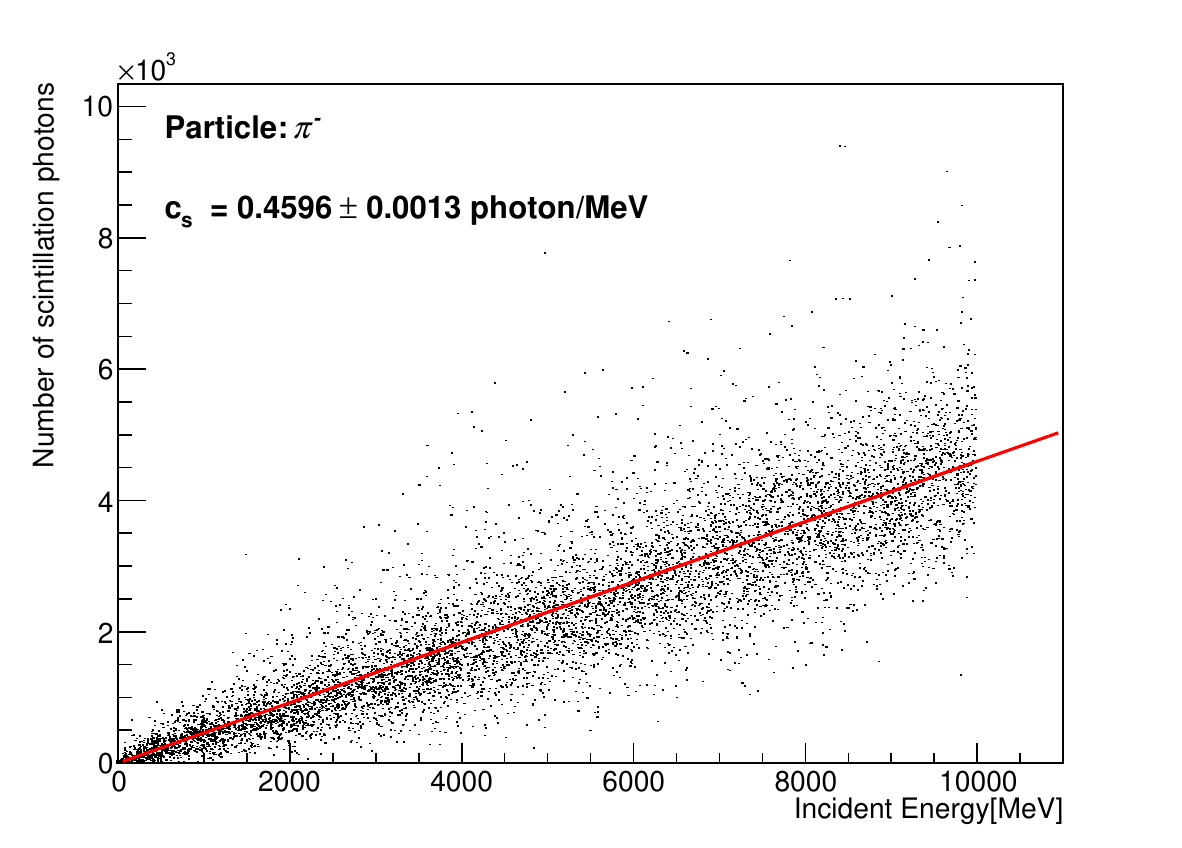}}
\caption{Comparison of scintillation photon counts in the sampling calorimeter with a 10\% sampling ratio and 50~mm sampling segmentation before and after the detection process, where (a) represents the initial generated photons and (b) represents the final detected photons. The red line is the linear fit.} 
\label{90cut_Photo}
\end{figure}

\begin{figure}[htbp]
\centering
\subfigure[]{
\label{90cut_Generate}
\includegraphics[width=0.4\textwidth]{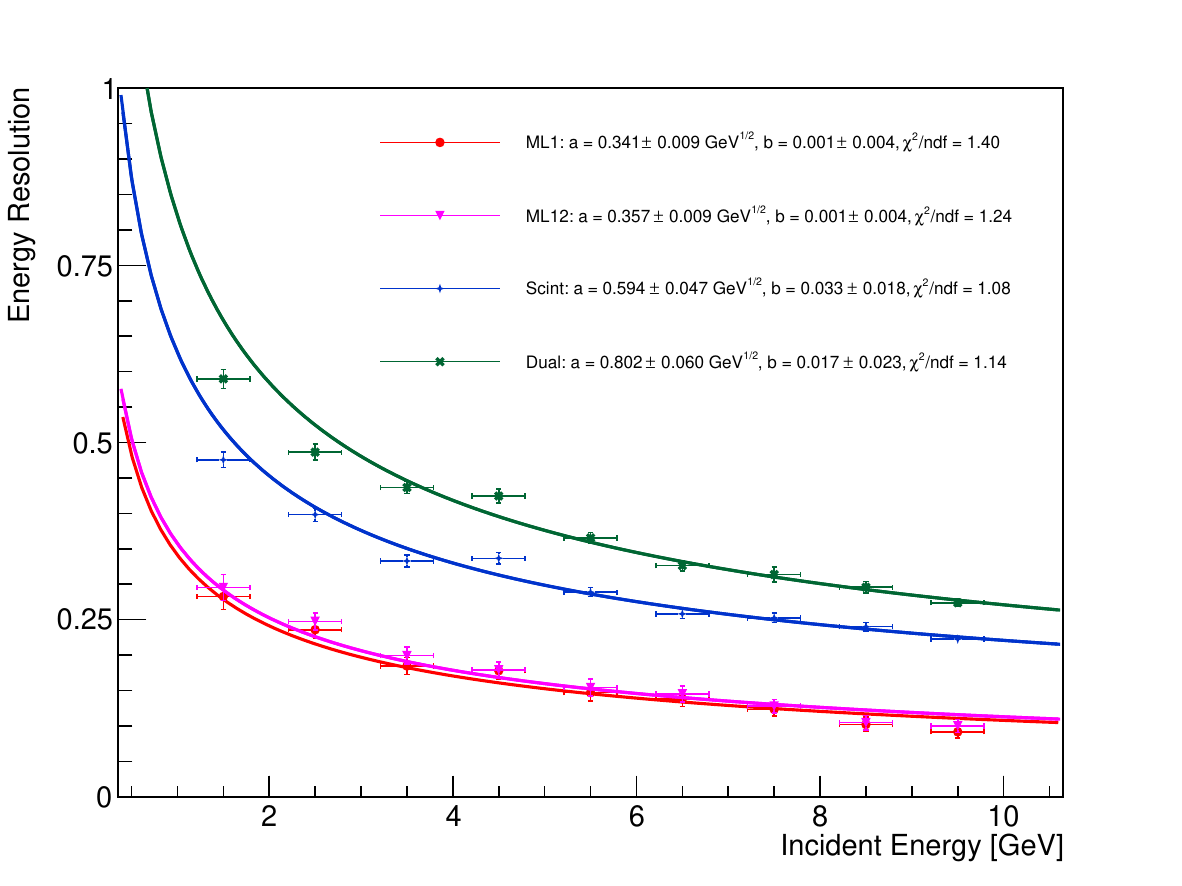}}
\subfigure[]{
\label{90cut_Detected2}
\includegraphics[width=0.4\textwidth]{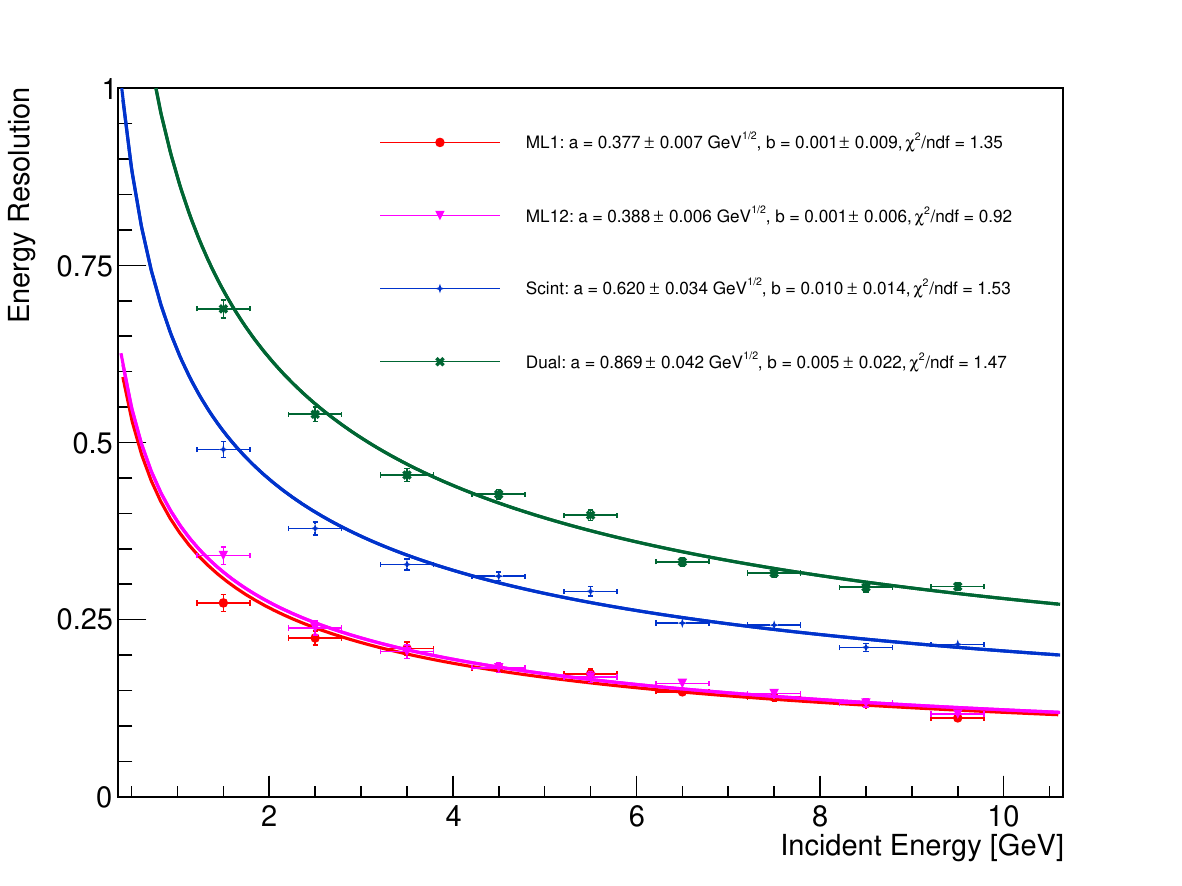}}
\caption{Comparison of energy resolution in the sampling calorimeter with a 10\% sampling ratio and 50~mm sampling segmentation before and after the detection process, where (a) represents the resolution obtained using initial generated photons and (b) represents the resolution obtained using final detected photons. The fitting is performed in the range of $1.75$–$9.75 \text{ GeV}$.} 
\label{90cut_Detected}
\end{figure}

Similarly, the performance of a sampling calorimeter with a 10\% sampling ratio and a 50~mm sampling segmentation was investigated using the same methodology as the homogeneous calorimeter. Regarding the detection of photons, those generated in the absorber layers were manually excluded, meaning the detection simulation was applied only to photons generated in the active layers. The resulting photon count distributions and energy resolution variations are shown in fig.~\ref{90cut_Photo} and fig.~\ref{90cut_Detected}.

It can be observed that after accounting for the detection efficiency, the light yield shifts from 11.9 photon/MeV to 0.46 photon/MeV. Correspondingly, the energy resolution for the single scintillation channel changes from $(59.4\pm4.7)\%/\sqrt{E/GeV}$ to $(62.0\pm3.4)\%/\sqrt{E/GeV}$, while the dual-readout resolution changes from $(80.2\pm6.0)\%/\sqrt{E/GeV}$ to $(86.9\pm4.2)\%/\sqrt{E/GeV}$. Compared to the fluctuations introduced by the sampling structure, the impact of detection effects on the energy resolution is not the dominant factor. The analysis indicates that the contribution of photon detection efficiency to the energy resolution is smaller than those of intrinsic hadronic fluctuations and sampling-induced fluctuations. Given that this study aims to employ machine learning to address these two principal sources of fluctuations, the inclusion of the photon detection process does not change the trends observed. Consequently, to evaluate the physically optimal energy resolution, the photon detection effect is not modeled in the subsequent research.

\section{Results}
\label{sec:results}
Although homogeneous calorimeters based on crystals such as $\mathrm{PbWO_4}$ are rarely used in practical hadronic calorimetry due to their prohibitively high cost, studying their performance remains highly instructive. Such an analysis provides a benchmark for the ultimate energy resolution achievable in hadronic showers and allows us to quantitatively assess the contributions of different signal channels, including scintillation photons, Cherenkov photons, and charged particles, as well as the three-dimensional spatial information of the shower. By understanding these intrinsic limits, we gain valuable insight into the physical constraints on energy reconstruction and the potential benefits of exploiting multiple observables and detailed shower topology.

\subsection{Homogeneous Calorimeter}
\begin{figure}[htbp]
\subfigure[]{
\label{Pure_ML}
\includegraphics[width=0.4\textwidth]{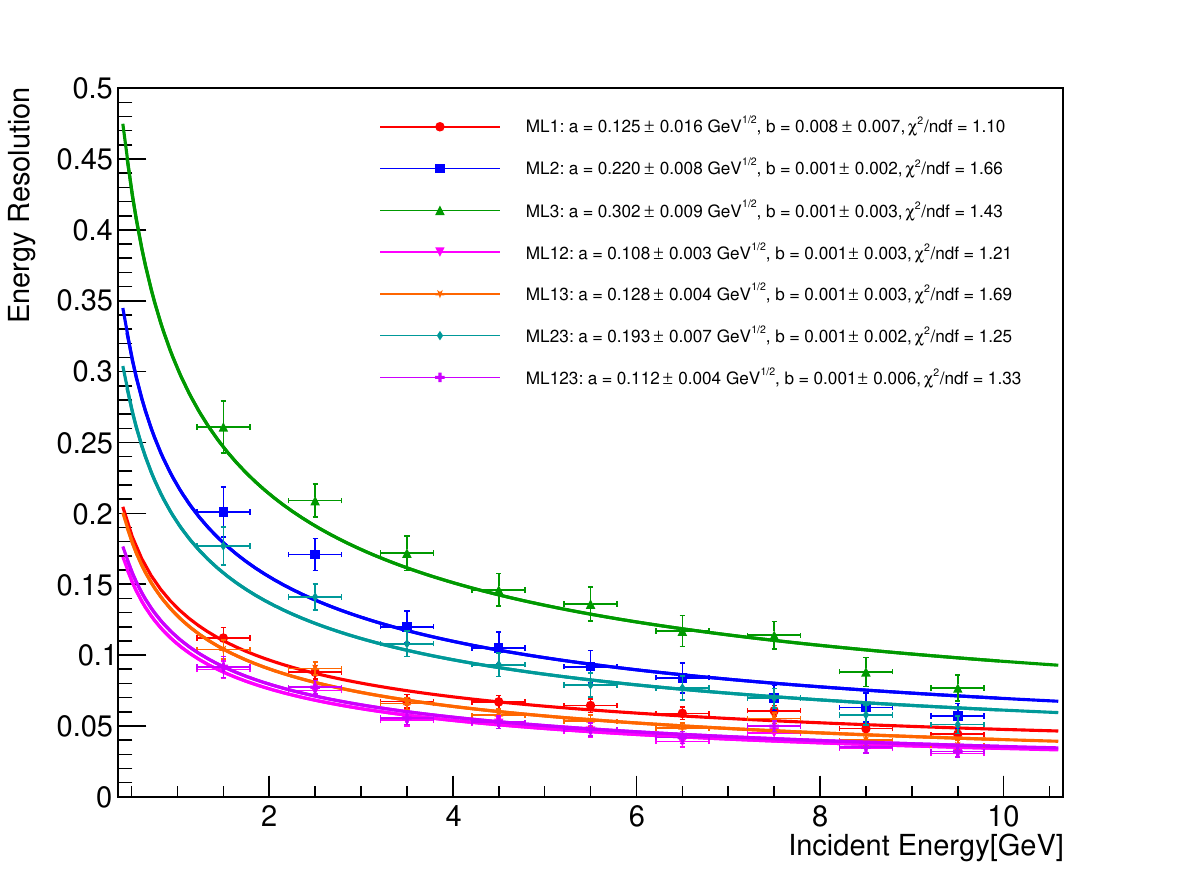}}
\subfigure[]{
\label{Pure_ML_Dual}
\includegraphics[width=0.4\textwidth]{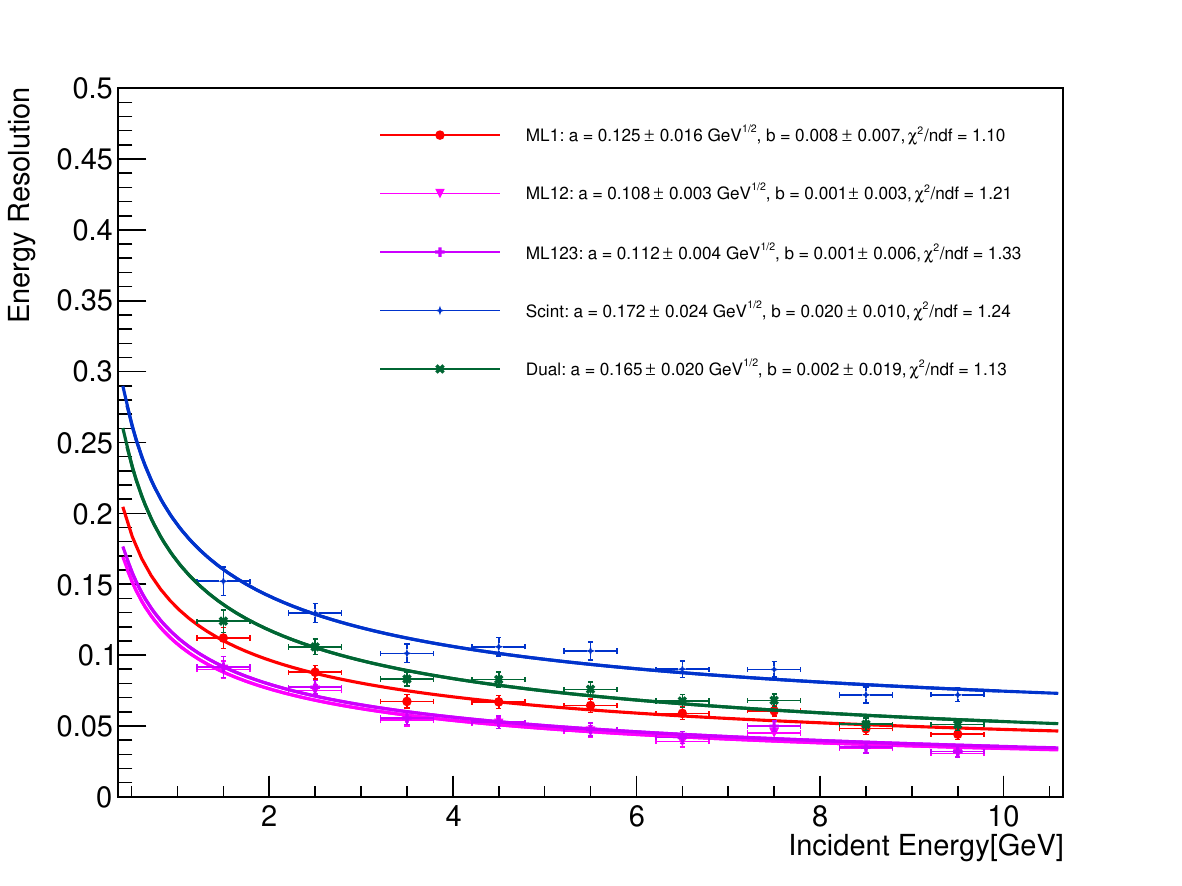}}
\caption{Energy resolution of the homogeneous calorimeter. Plot (a) compares the machine-learning–based energy reconstruction using different input-channel configurations, while plot (b) presents a comparison between machine-learning approaches and conventional reconstruction methods. The fitting is performed in the range of $1.75$–$9.75 \text{ GeV}$.} 
\label{Pure}
\end{figure}

For the homogeneous $\mathrm{PbWO_4}$ calorimeter, we evaluate the energy resolution achieved with the conventional single-channel reconstruction, the dual-readout method, and the machine-learning-based approach. The corresponding results are presented in fig.~\ref{Pure_ML}. To further quantify the contribution of individual observables, we perform a systematic study of the machine-learning performance using different input configurations, including each single channel, all possible two-channel combinations, and the full three-channel input. The resulting energy resolutions are summarized in fig.~\ref{Pure_ML_Dual}.

The energy resolutions obtained using the three individual channels, scintillation, Cherenkov, and charged particles, are $(12.5 \pm 1.6)\%/\sqrt{E/GeV}$, $(22.0 \pm 0.8)\%/\sqrt{E/GeV}$, and $(30.2 \pm 0.9)\%/\sqrt{E/GeV}$, respectively. This trend reflects the relative signal strengths: the scintillation channel produces the largest number of photons, the Cherenkov channel fewer, and the charged-particle channel the least, resulting in progressively degraded resolutions.

For the two-channel combinations, because the single-channel resolutions of the Cherenkov and charged-particle signals are already relatively poor, the combined two-channel performance only reaches $(19.3 \pm 0.7)\%/\sqrt{E/GeV}$. In contrast, for the two-channel combinations that include the scintillation signal, adding either the Cherenkov or the charged-particle channel the resolutions become $(10.8\pm0.3)\%/\sqrt{E/GeV}$ and $(12.8\pm0.4)\%/\sqrt{E/GeV}$, respectively. When using all three channels simultaneously, the resulting energy resolution is $(11.2\pm0.4)\%/\sqrt{E/GeV}$. Considering the fit uncertainty, this indicates that adding the charged-particle channel does not significantly improve the energy prediction.

A comparison between the resolutions obtained from the photon counts and those reconstructed via machine learning indicates a significant improvement: the single-channel resolution is enhanced from $(17.2\pm2.4)\%/\sqrt{E/GeV}$ to $(12.5\pm1.6)\%/\sqrt{E/GeV}$, while the two-channel resolution improves from $(16.5\pm0.2)\%/\sqrt{E/GeV}$ to $(10.8\pm0.3)\%/\sqrt{E/GeV}$. Although machine learning clearly outperforms the traditional algorithm, it is important to note that the dominant fluctuations in hadronic showers arise from intrinsic physical processes. Consequently, even when incorporating spatial information, the achievable improvement in energy resolution via machine learning remains fundamentally limited.

The energy resolutions obtained for the homogeneous $\mathrm{PbWO_4}$ calorimeter under different signal-channel configurations and reconstruction strategies are summarized in Table~\ref{tab:PureML_resolution}. 

\begin{table}[htbp]
\centering
\caption{Summary of energy resolutions for the homogeneous $\mathrm{PbWO_4}$ calorimeter using different signal-channel configurations and reconstruction methods.}
\label{tab:PureML_resolution}
\begin{tabular}{l c}
\hline
\hline
Input configuration / Method & Energy resolution $/\sqrt{E/GeV}$ \\
\hline
\multicolumn{2}{c}{\textbf{Machine-learning-based reconstruction}} \\
\hline
ML1       & $(12.5\pm1.6)\%$ \\
ML2            & $(22.0\pm0.8)\%$ \\
ML3   & $(30.2\pm0.9)\%$ \\
ML12      & $(10.8\pm0.3)\%$ \\
ML13 & $(12.8\pm0.4)\%$ \\
ML23  & $(19.3\pm0.7)\%$ \\
ML123 & $(11.2\pm0.4)\%$ \\
\hline
\multicolumn{2}{c}{\textbf{Conventional reconstruction (photon counting)}} \\
\hline
Scint       & $(17.2\pm2.4)\%$ \\
Dual & $(16.5\pm0.2)\%$ \\
\hline
\hline
\end{tabular}
\end{table}

The intrinsic physical limit of the hadronic energy resolution is approximately $(10.8\pm0.3)\%/\sqrt{E/GeV}$. This limit is achieved when simultaneously exploiting scintillation light, Cherenkov light, and detailed 3D shower topology information, indicating that this combination provides the maximal recoverable information on the shower development. Further inclusion of additional channels yields no statistically significant improvement, underscoring that the ultimate performance is constrained by intrinsic hadronic shower fluctuations rather than by the reconstruction methodology itself.

\subsection{Sampling Calorimeter}

We consider the hadronic calorimeter configuration employed by the ATLAS Collaboration, which consists of a 14 mm steel absorber and a 3 mm plastic scintillator active layer , achieving a stochastic term of approximately $(56.4 \pm 0.4)\%/\sqrt{E/\text{GeV}}$\cite{TheATLASCollaboration2008}.

\begin{figure}[htbp]
    \centering
    \includegraphics[width=0.6\textwidth]{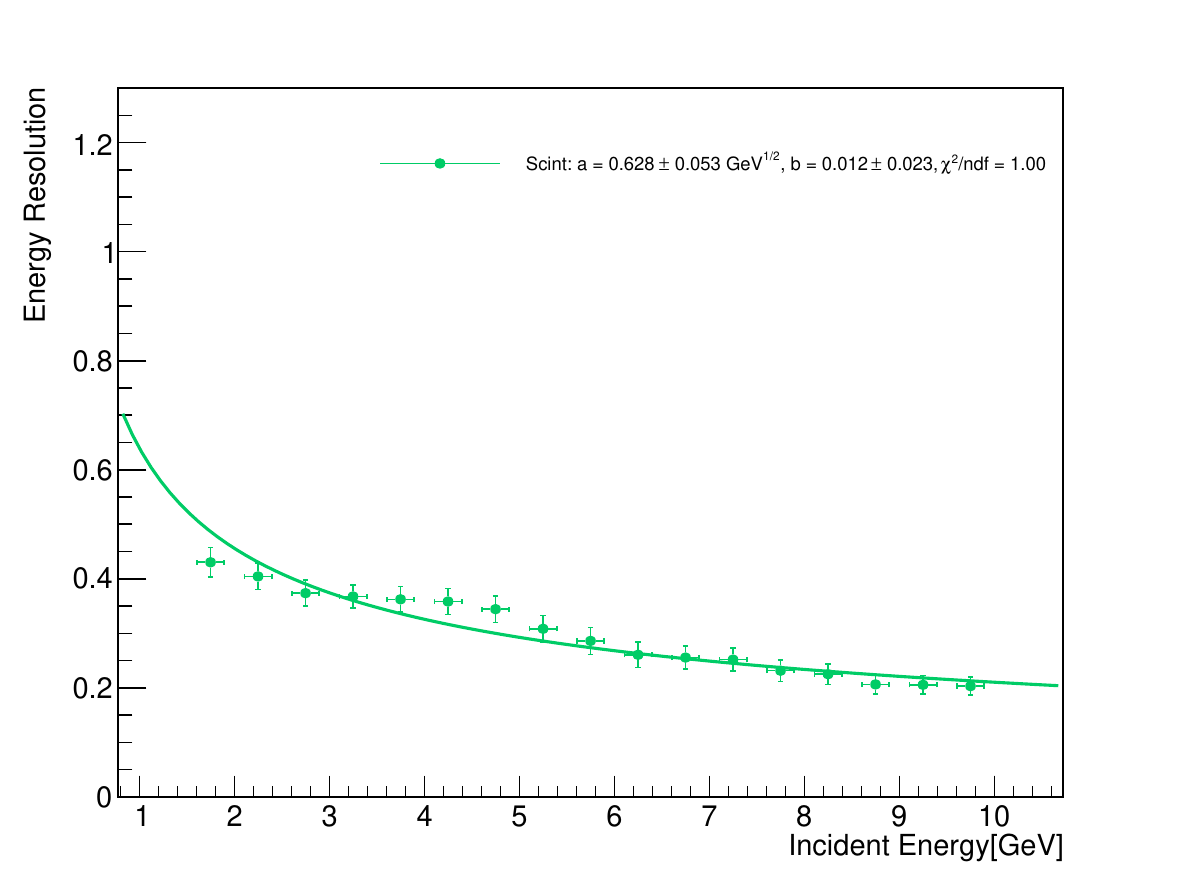} 
    \caption{Energy resolution obtained from simulations based on the ATLAS calorimeter structure, see text for details. The fitting is performed in the range of $1.75$–$9.75 \text{ GeV}$.}
    \label{8_164}
\end{figure}

To construct a physically motivated equivalent sampling geometry for simulation studies, the physical thicknesses of the layers are scaled by accounting for material densities and mass pion interaction lengths to ensure consistency in the interaction probability. The scaling relation is defined as:

\begin{equation}
t_{\text{tar}} = t_{\text{ref}} \cdot \left( \frac{\lambda_{\text{w,tar}}}{\lambda_{\text{w,ref}}} \right) \cdot \left( \frac{\rho_{\text{ref}}}{\rho_{\text{tar}}} \right)
\end{equation}

where $t_\mathrm{tar}$ denotes the physical thickness of the target material. $\lambda_\mathrm{w,tar}$ and $\lambda_\mathrm{w,ref}$ represent the mass pion interaction lengths for the target and the reference material, while $\rho_\mathrm{tar}$ and $\rho_\mathrm{ref}$ denote their mass densities.This scaling effectively preserves the number of pion interaction lengths per layer while accounting for differences in material density, thereby providing a consistent basis for comparing different material configurations. Using this prescription and the material properties sourced from the PDG database as summarized in Table~\ref{tab:pi}, the ATLAS-like configuration is mapped to an equivalent geometry consisting of a 0.8~mm active layer and a 16.4~mm absorber layer. The resulting energy resolution of $(62.8 \pm 5.3)\%/\sqrt{E/GeV}$ shown in fig.~\ref{8_164} is in reasonable agreement with $(56.4\pm0.4)\%/\sqrt{E/GeV}$ reported by the ATLAS experiment in the energy range 20-350 GeV.

\begin{table}[htbp]
    \centering
    \caption{Mass pion interaction lengths $\lambda_w$ and density $\rho$ in different materials.}
    \begin{tabular}{ccc}
        \toprule
        Materials &  $\lambda_w~$[$\mathrm{g\cdot cm^{-2}}$] &  $\rho~$[$\mathrm{g\cdot cm^{-3}}$] \\
        \midrule
        Plastic scintillator & 	113.7 & 1.06\\
        Steel & 	160.8 & 	7.87\\
        $\mathrm{PbWO_4}$ & 199.5 & 8.30\\
        \bottomrule
    \end{tabular}
    \label{tab:pi}
\end{table}

Since most modern hadronic calorimeters adopt a sampling structure with relatively low sampling ratios, typically below 10\%, we first simulate a configuration with a 5 mm active layer and a 45 mm absorber layer, corresponding to a 10\% sampling ratio. To mitigate statistical fluctuations, two independent simulation datasets of 60,000 events each are generated. The resulting single-channel scintillation-light energy resolution is shown in fig.~\ref{60,000}.

\begin{figure}[htbp]
\centering
\includegraphics[width=0.6\textwidth]{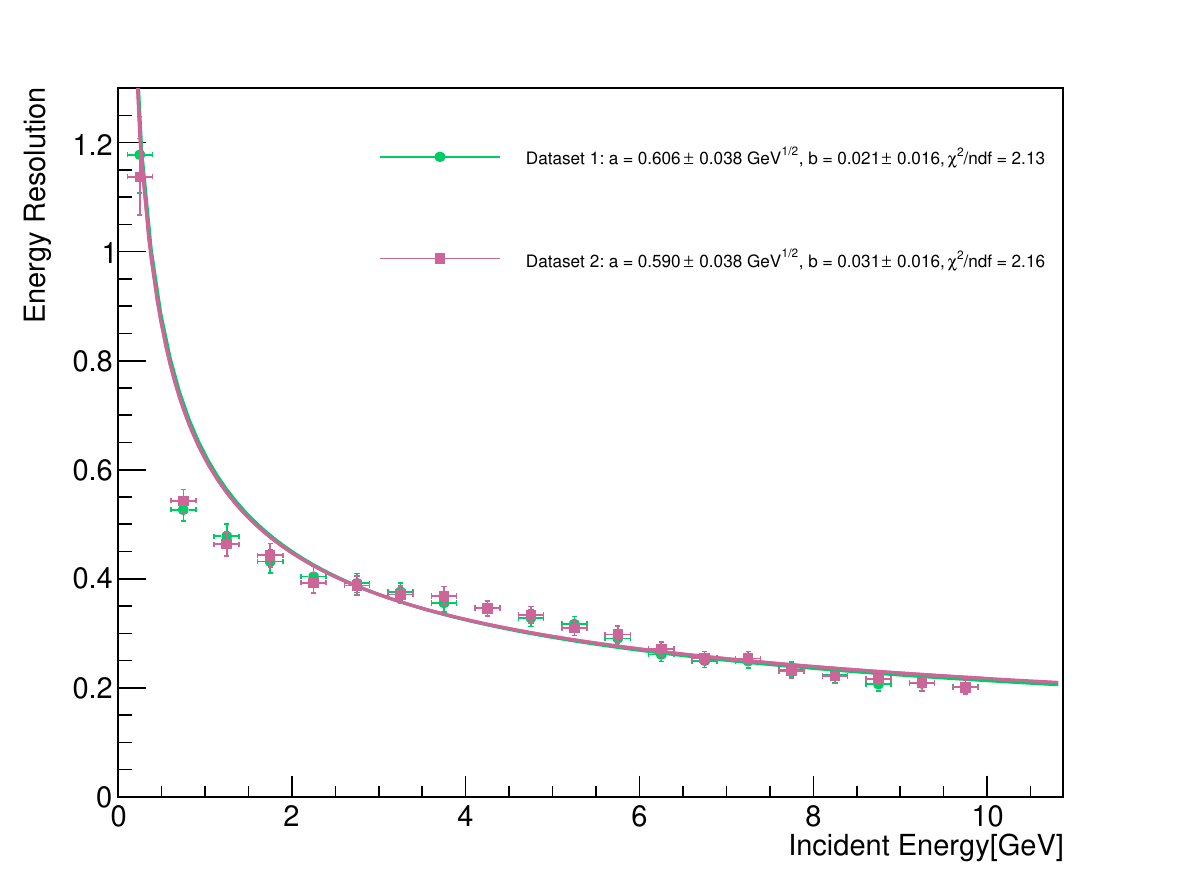} 
\caption{Energy resolution obtained from the scintillation channel for the configuration with a 5 mm active layer and a 45 mm absorber layer. The fitting is performed in the range of $0.25$–$9.75 \text{ GeV}$.}
\label{60,000}
\end{figure}

By comparing the two datasets, we observe fluctuations in the 0.25--1.25~$\mathrm{GeV}$ energy range due to the low incident particle energy. To avoid the influence of these statistical variations on the fitting procedure, only the energy points within the 1.25--9.75~$\mathrm{GeV}$ range are used for subsequent fits.

\begin{figure}[htbp]
\subfigure[]{
\label{0.25}
\includegraphics[width=0.22\textwidth]{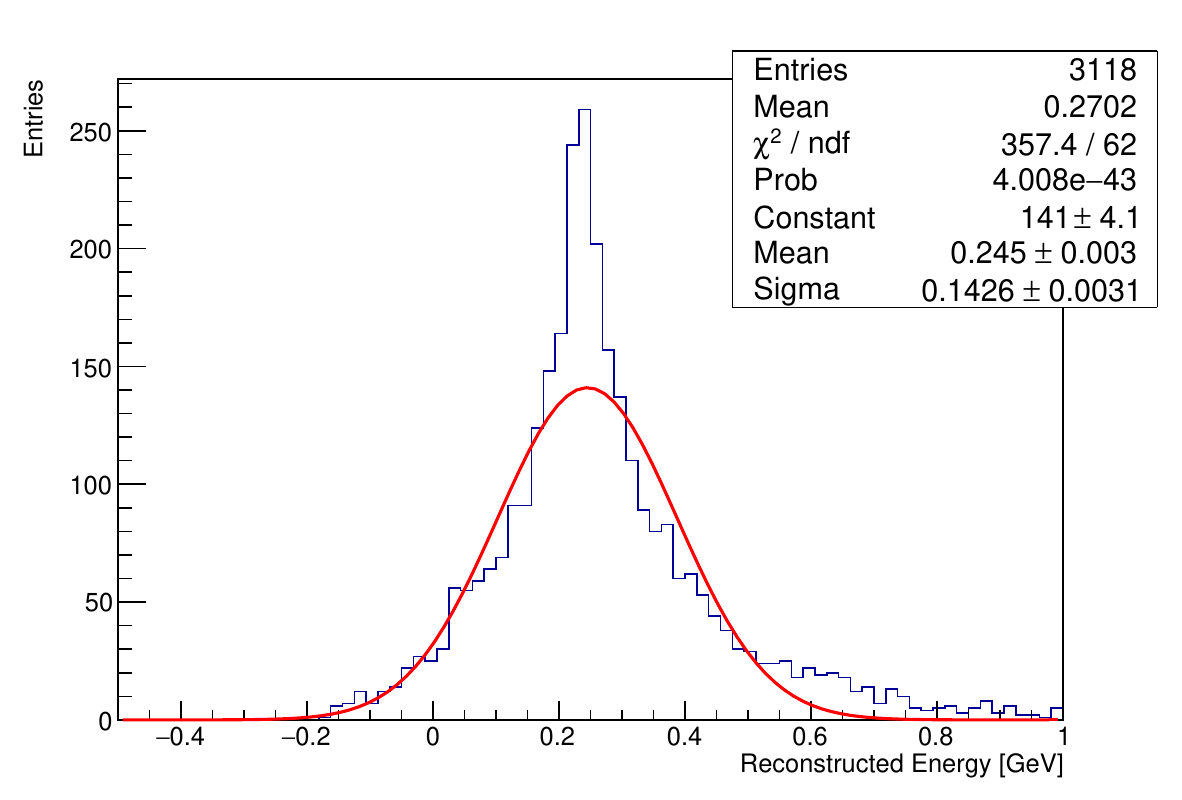}}
\subfigure[]{
\label{0.75}
\includegraphics[width=0.22\textwidth]{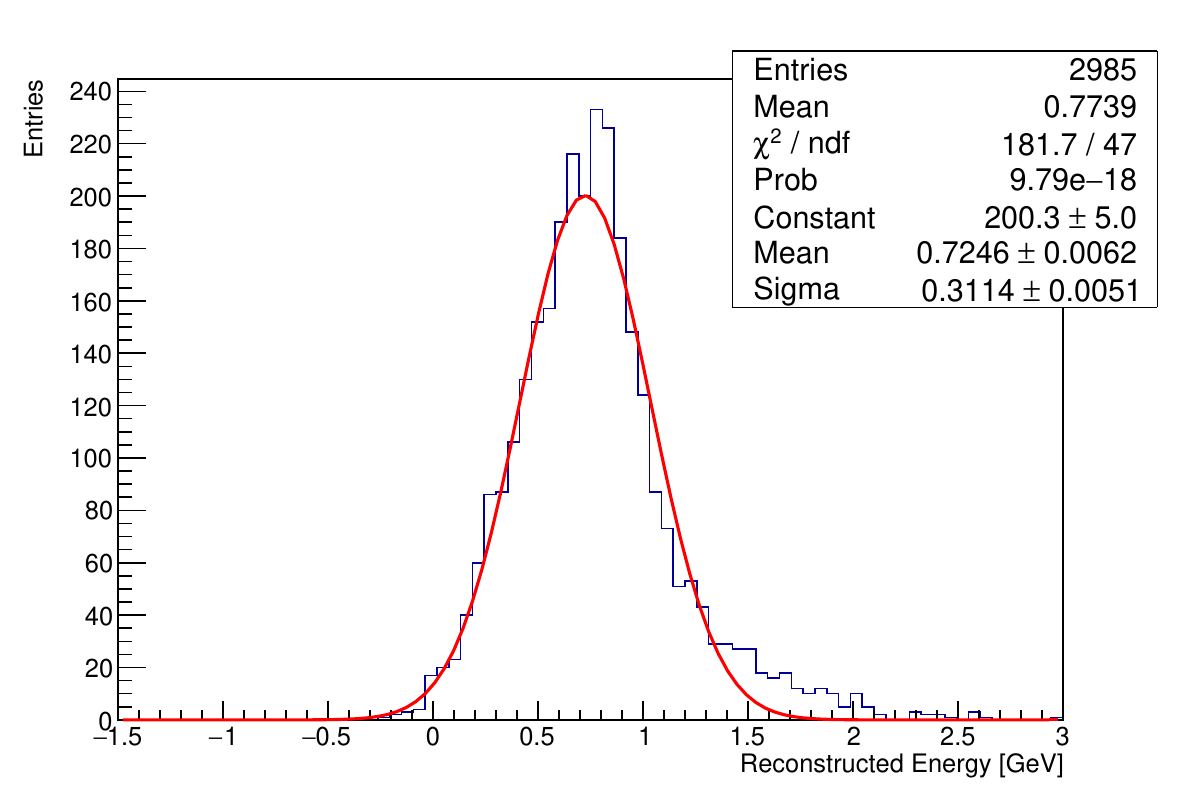}}
\subfigure[]{
\label{1.25}
\includegraphics[width=0.22\textwidth]{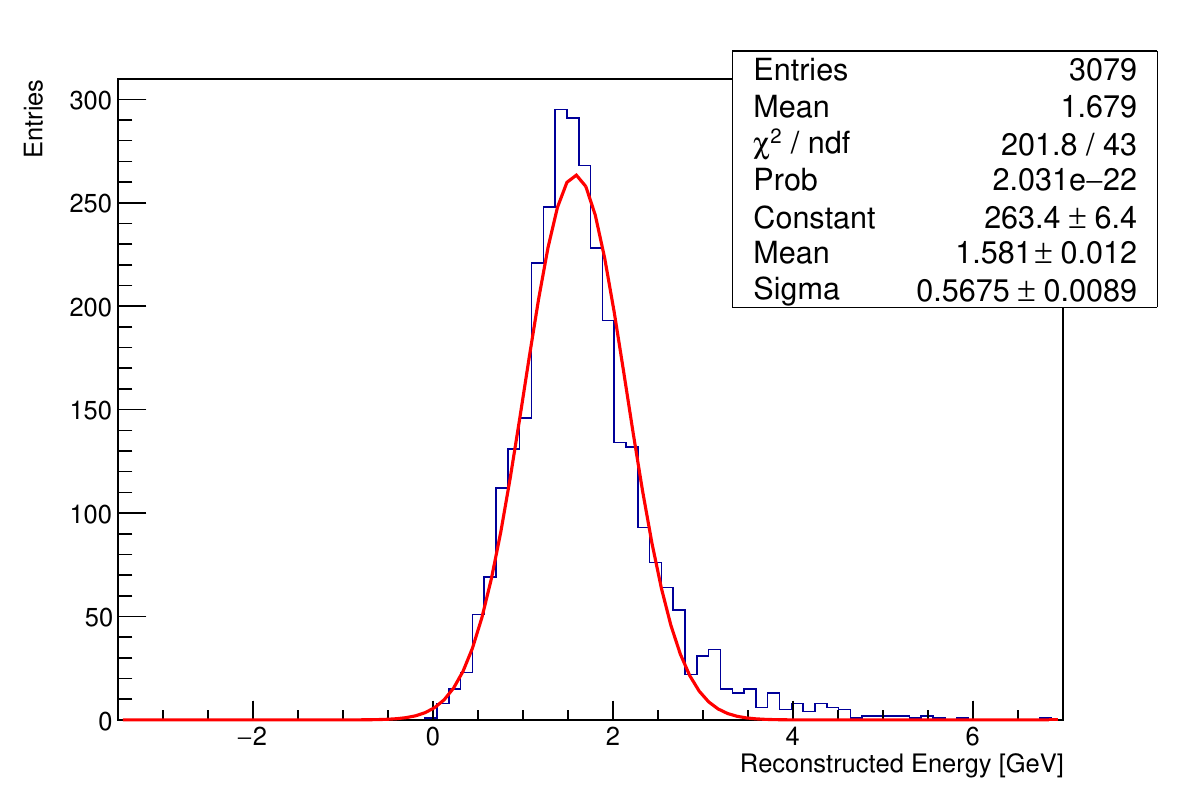}}
\subfigure[]{
\label{1.75}
\includegraphics[width=0.22\textwidth]{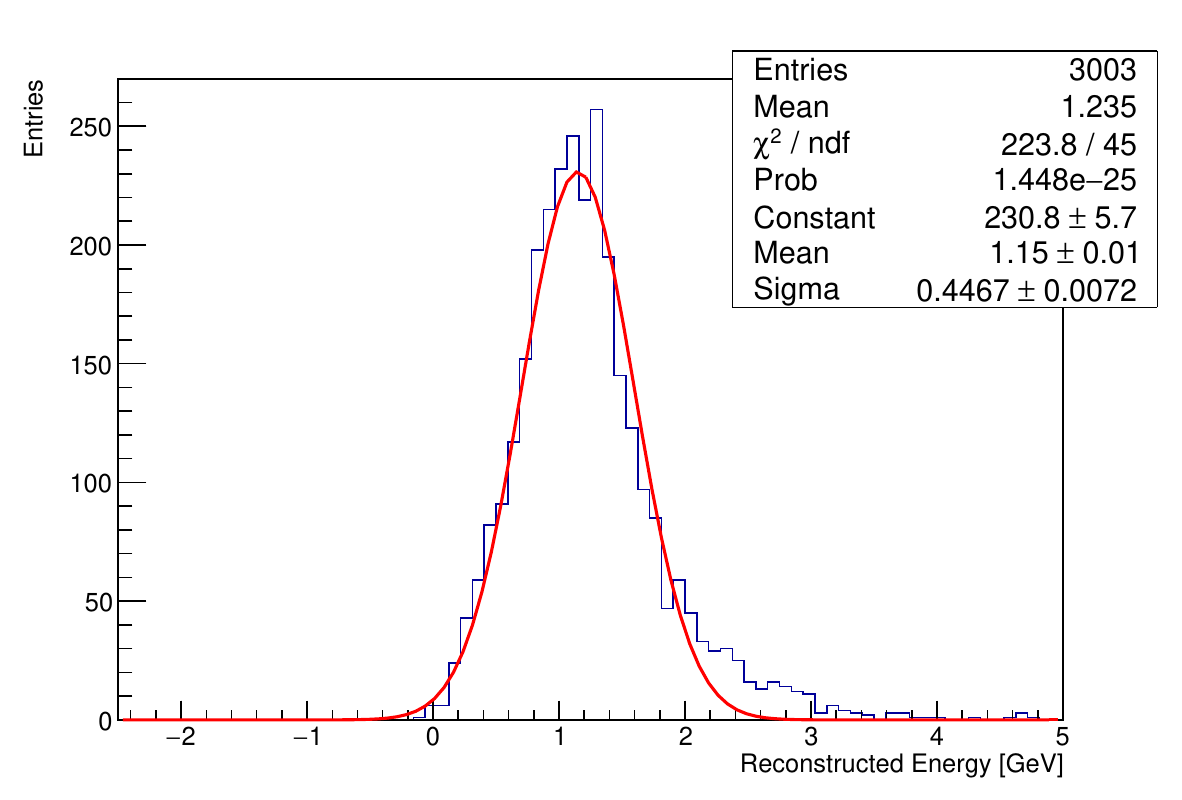}}
\caption{Reconstructed energy distributions for four representative energy intervals: (a) 0–0.5 GeV, (b) 0.5–1 GeV, (c) 1–1.5 GeV and (d) 1.5–2 GeV. See text for details.} 
\label{below2.25}
\end{figure}

It is also observed that for this calorimeter configuration, the energy resolution tends to lie below the fitted curve for incident energies below 2.25 GeV, while it is slightly above the fitted curve in the range of 3.25–-5.75 GeV. The Gaussian distributions of the first four energy points were therefore examined, and the corresponding results are shown in the fig.~\ref{below2.25}.

As shown in the figure, all four reconstructed energy distributions exhibit long non-Gaussian tails to different extents, with the most pronounced feature observed in the 0-–0.5 GeV range, where a deviation between the Gaussian fit and the original distribution is present. To exclude the influence of this point and to compare the Gaussian fitting and RMS methods, the same dataset was analyzed in the 0.75-–9.75 GeV range using both approaches. The results are shown in fig.~\ref{60,000_Gauss}.

\begin{figure}[htbp]
\centering
\includegraphics[width=0.6\textwidth]{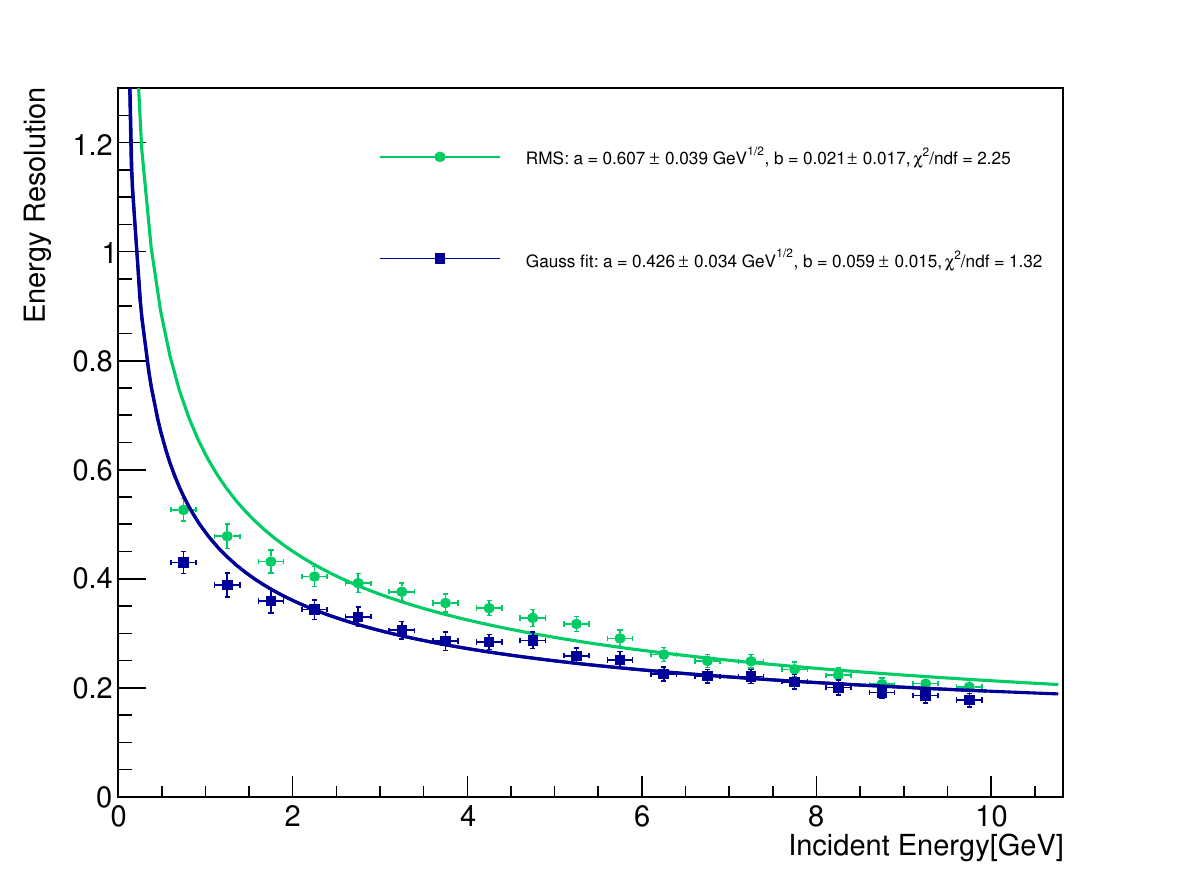} 
\caption{Energy resolution obtained from the scintillation channel for the configuration with a 5 mm active layer and a 45 mm absorber layer. The fitting is performed in the range of $0.75$–$9.75 \text{ GeV}$.}
\label{60,000_Gauss}
\end{figure}

First, for the RMS result, removing the 0.25 GeV data point shows only a minor change in the fitted energy resolution, indicating that this point has a limited impact on the overall fit. Second, the comparison between the Gaussian fitting and RMS methods shows that the energy resolution obtained from the Gaussian fit is lower than that from the RMS method, as the Gaussian fitting procedure suppresses the contribution of non-Gaussian tails. For the Gaussian-fit results, the deviations become smaller and the data points are consistent with the fitting curve from eq. \ref{fiteq} within uncertainties. Since these tail components represent fluctuations in the calorimeter response, the RMS method is adopted for subsequent performance evaluations to provide a more conservative assessment.

\begin{figure}[htbp]
\subfigure[]{
\label{4-6bias2}
\includegraphics[width=0.4\textwidth]{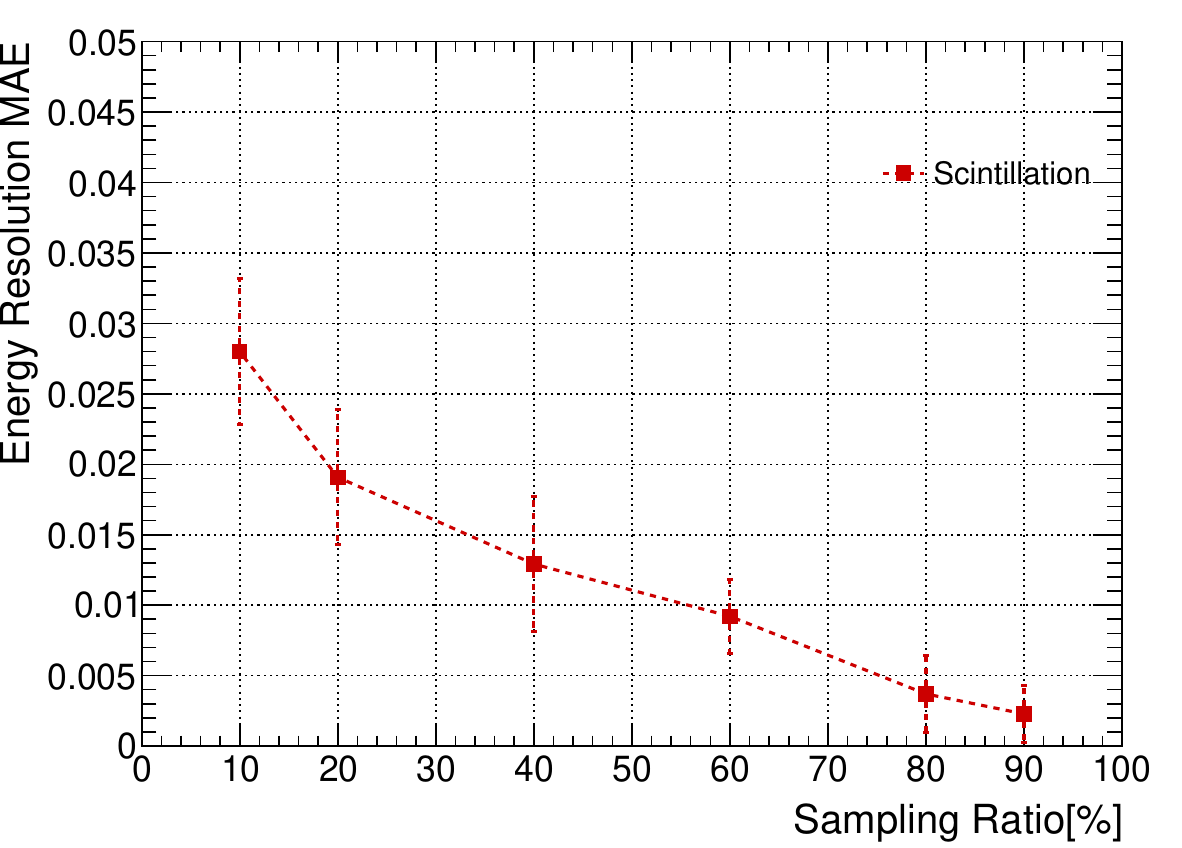}}
\subfigure[]{
\label{fig:biasSamplerlayer}
\includegraphics[width=0.4\textwidth]{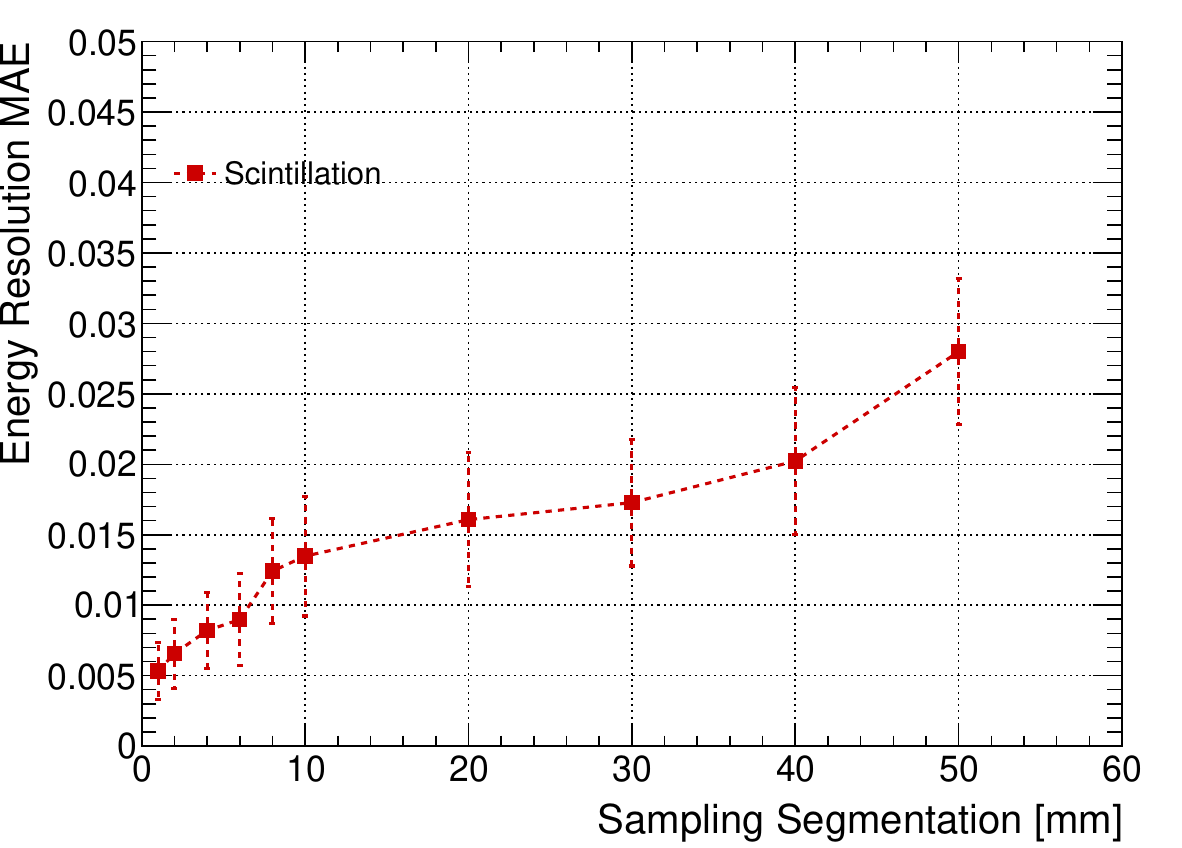}}
\caption{Variation of the energy-resolution MAE of the sampling calorimeter in the 3.25--5.75~$\mathrm{GeV}$ range. (a) MAE as a function of the sampling ratio (with a fixed sampling segmentation of 50 mm); (b) MAE as a function of the sampling segmentation (with a fixed sampling ratio of 10\%)} 
\label{biasSample}
\end{figure}

The impact of the sampling geometry on the deviations within the 3.25–5.75~$\mathrm{GeV}$ range is evaluated through two distinct sets of simulation configurations. First, the sampling segmentation is fixed at 50~mm while the sampling ratio varies; second, the sampling ratio is fixed at 10\% while the sampling segmentation varies. We then calculate the mean absolute error (MAE) of the resolution points in the 3.25--5.75~$\mathrm{GeV}$ range relative to the fitted curve to quantify the deviations. The results are shown in fig.~\ref{biasSample}.

As shown in fig.~\ref{biasSample}, increasing the sampling ratio reduces the MAE from approximately 0.028 to 0.002. Increasing the sampling segmentation increases the MAE from 0.006 to 0.028. Overall, the energy-resolution deviation decreases with both an increasing sampling ratio and a decreasing sampling segmentation. This deviation is primarily measured in the low-energy region of 3.25--5.75~$\mathrm{GeV}$ and becomes smaller at higher incident energies.

In addition, regarding the observed deviations in the fitting procedure, it is worth noting the characteristics of the utilized Geant4 physics list, FTFP--BERT. For the inelastic scattering of incident $\pi^{-}$ mesons, this physics list employs a transition between two distinct hadronic models. The Bertini (BERT) model is primarily active from 0 to 6~$\mathrm{GeV}$, while the Fritiof (FTF) parton model operates from 3~$\mathrm{GeV}$ to 100~$\mathrm{TeV}$. In the overlapping region, the BERT model is invoked with a probability that decreases linearly from 1.0 to 0.0, with the FTF model invoked using the complementary probability. Furthermore, these two models employ different nuclear de-excitation mechanisms: when the FTF model is used, the Precompound model (P) is also invoked to de-excite the remnant nucleus after the initial high energy interaction. When the Bertini model is used, its own, simpler precompound and de-excitation models are invoked. The performance of the FTFP--BERT physics list in the low-energy region has been evaluated by the CALICE collaboration using Geant4 version 9 series \cite{BILKI2015240}. Their findings indicated that the transition between FTFP and BERT models introduces discontinuities, particularly in the radial distributions of the hadronic showers within the $2\text{--}10\text{ GeV}$ energy range.

\begin{figure}[htbp]
\subfigure[]{
\label{SampleRatio}
\includegraphics[width=0.4\textwidth]{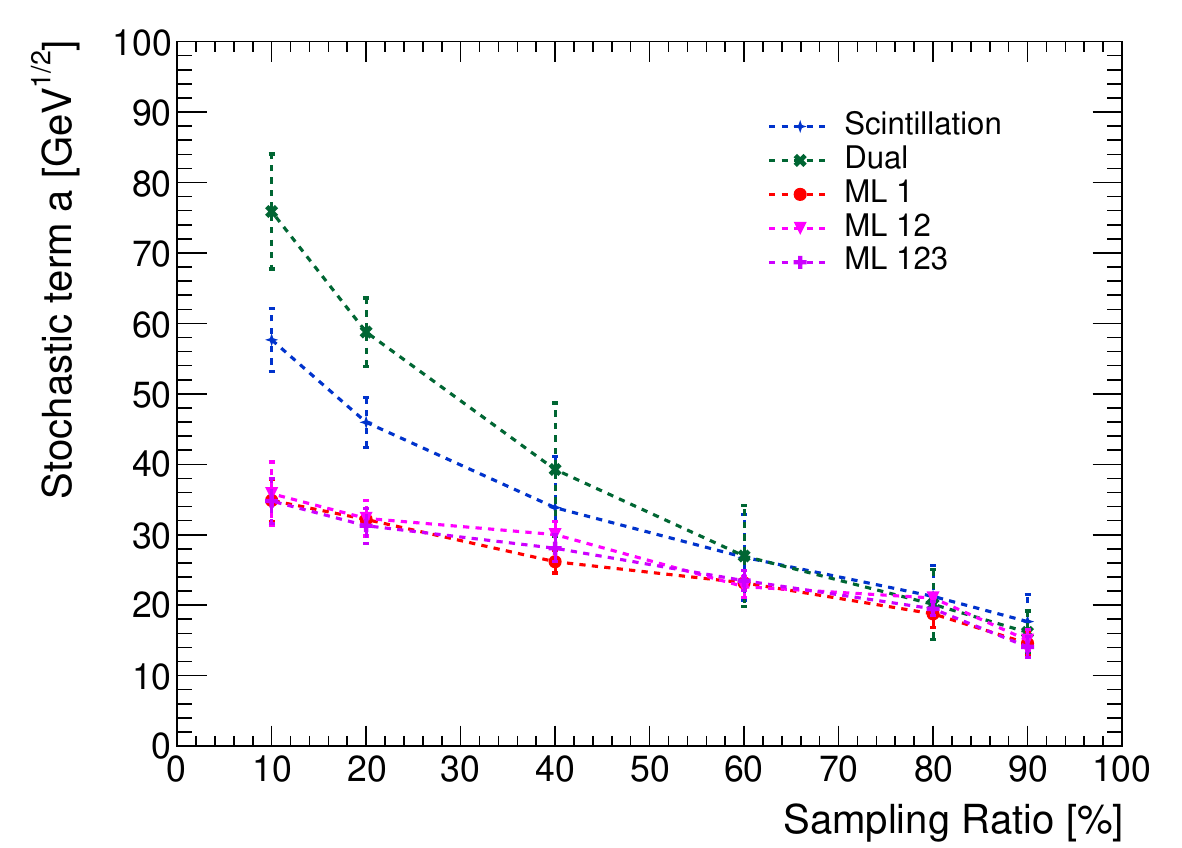}}
\subfigure[]{
\label{Samplelayer}
\includegraphics[width=0.45\textwidth]{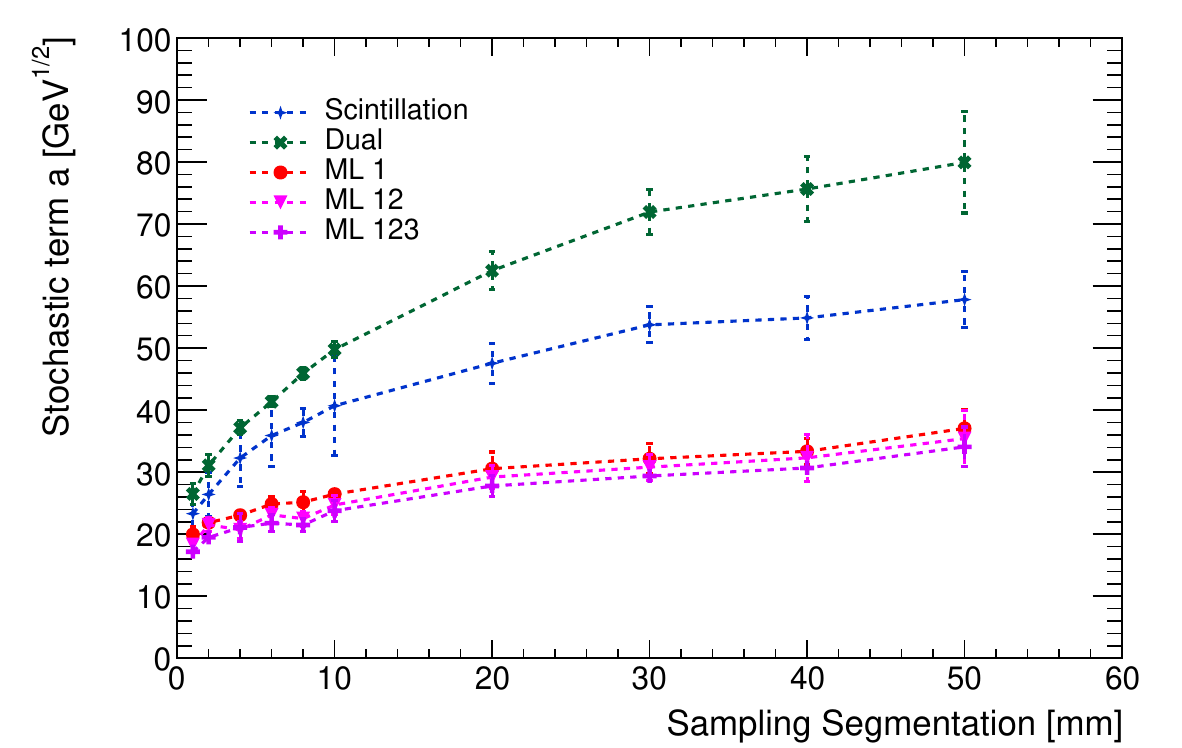}}
\caption{Trends of the energy resolutions obtained from the Scint, Dual, and ML methods. Plot (a) shows the energy resolution as a function of sampling ratio(with a fixed sampling segmentation of 50 mm), and Plot (b) shows the energy resolution as a function of sampling segmentation(with a fixed sampling ratio of 10\%).} 
\label{SampleResolution}
\end{figure}

To study the effects of sampling ratio and sampling segmentation on energy resolution, as well as the correction capability of machine learning under different detector configurations, we simulated sampling calorimeters under two conditions:(1) the sampling segmentation is fixed at 50 mm while the sampling ratio varies;(2) the sampling ratio is fixed at 10\% while the sampling segmentation varies. The resulting scintillation-channel, dual-readout, and machine-learning-based energy resolutions are shown in fig.~\ref{SampleResolution}. For consistency in comparison, all energy resolutions presented here refer to the stochastic term $a$.

From fig.~\ref{SampleResolution}, it can be seen that the machine-learning method significantly corrects the resolution deviations introduced by sampling calorimeters, and the correction becomes more pronounced at lower sampling ratios. For example, for a calorimeter with a sampling ratio of $10\%$ and a sampling segmentation of 50 mm, the scintillation-channel resolution is $(57.6\pm3.7)\%/\sqrt{E/GeV}$, whereas the machine-learning reconstructed result is $(34.1\pm2.8)\%/\sqrt{E/GeV}$, demonstrating a substantial improvement.

Similarly, as the sampling segmentation increases, the improvement achieved by the machine-learning model becomes more evident. This indicates that the machine-learning method effectively exploits the three-dimensional spatial distribution of the shower, enabling accurate particle-energy reconstruction even when the sampling structure reduces the number of collected photons. In other words, although the sampling structure limits photon statistics, it does not destroy the spatial features of the shower; the machine-learning model can recover the energy information by learning these nonlinear spatial characteristics.

A further comparison of the machine-learning results using different input channels shows that the resolutions obtained with three-channel, two-channel, and single-channel inputs are all consistent within uncertainties. For instance, for a calorimeter with a sampling ratio of $10\%$ and a sampling segmentation of 50 mm, the three-channel machine-learning result is $(34.1\pm2.8)\%/\sqrt{E/GeV}$, the two-channel result is $(35.7\pm3.8)\%/\sqrt{E/GeV}$, and the single-channel result is $(35.0\pm2.5)\%/\sqrt{E/GeV}$, showing no significant difference. This suggests that the scintillation signal alone already contains sufficient shower information for accurate energy reconstruction, and additional channels provide only marginal improvement.

Meanwhile, comparison between the dual-readout and single-channel results shows that, for sampling calorimeters, the dual-readout algorithm can even worsen the resolution under conditions of low sampling ratio or large sampling segmentation due to sampling-structure–induced fluctuations. For example, for a calorimeter with a sampling ratio of $10\%$ and a sampling segmentation of 50 mm, the scintillation-channel resolution is $(57.6\pm3.7)\%/\sqrt{E/GeV}$, whereas the dual-readout resolution reaches $(75.8\pm8.1)\%/\sqrt{E/GeV}$. This indicates that the advantage of the dual-readout method over the single-channel mode is highly sensitive to the geometric configuration, and the dual-readout performance may even deteriorate in structures with a low sampling ratio and large segmentation.

\section{Summary and conclusions}
\label{sec:summary}

This work investigates the effects of different calorimeter structures, additional readout channels, and machine-learning reconstruction methods on energy resolution through comprehensive simulations of both homogeneous and sampling calorimeters, providing guidance for the optimization of future calorimeter designs.

For the homogeneous $\mathrm{PbWO_4}$ calorimeter, nearly all of the particle energy is deposited within the detector, making it a physical reference for the limit of calorimetric measurement. Simulation results indicate that the intrinsic physical limit of the hadronic energy resolution is approximately $(10.8\pm0.3)\%/\sqrt{E/GeV}$. This limit is achieved when simultaneously exploiting scintillation light, Cherenkov light, and detailed 3D shower topology information. The scintillation channel alone can achieve an energy resolution of approximately $(17.2\pm2.4)\%/\sqrt{E/GeV}$, while the inclusion of Cherenkov and charged-particle channels provides only limited improvement. Although machine-learning methods can further enhance the resolution to some extent, the intrinsic fluctuations of hadronic showers impose a fundamental limit on achievable improvement. The constant terms in all resolution fits are consistent with zero within the fitting uncertainties across the $0$–$10 \text{ GeV}$ energy range. At $10 \text{ GeV}$, the contribution from the constant term is approximately an order of magnitude smaller than that of the stochastic term. This relationship between the two components indicates that energy leakage in the simulated calorimeter system does not dominate the resolution within the investigated energy range.

Furthermore, the simulations indicate that the traditional dual-readout method does not necessarily improve the resolution in sampling structures. When the sampling error is small, dual-readout can reduce both stochastic and constant terms. However, for thicker sampling modules or lower sampling ratios, the performance of the dual-readout method degrades, resulting in worse low-energy resolution compared to single-channel readout.

For future hadronic calorimeter designs, priority should be given to reducing sampling errors—through higher sampling ratios or thinner individual layers—to ensure that dual-readout techniques can truly be advantageous; otherwise, single-channel readout may be more cost-effective.

Machine-learning methods, by exploiting the nonlinear features of the three-dimensional shower distribution, can significantly recover energy information even in sampling systems with limited photon statistics. For example, at a sampling ratio of only $10\%$, the resolution can be improved from $(57.6\pm3.7)\%/\sqrt{E/GeV}$ to about $(34.1\pm2.8)\%/\sqrt{E/GeV}$. This demonstrates the significant potential of machine learning for future high-granularity calorimeters and PFA-based reconstruction frameworks.

In summary, for the design of calorimeters in future high-energy physics experiments, homogeneous calorimeters should be regarded as the performance upper limit. From an engineering perspective, high sampling ratios, thin sampling modules, and the incorporation of machine-learning reconstruction methods can be used to achieve energy resolutions approaching the physical limit.

\appendix
\section{Single-channel and Dual-readout Parameters}

\begin{table}[htbp]
\centering
\caption{Single-channel response coefficients $c_s$ and $c_c$ under different sampling segmentation}
\begin{tabular}{ccc} 
\hline
sampling segmentation (mm) & $c_s$ (photon/MeV) & $c_c$ (photon/MeV) \\
\hline
1 & 12.080 & 2.1279 \\
2 & 12.083 & 2.1273 \\
4 & 12.076 & 2.1275 \\
6 & 12.084 & 2.1265 \\
8 & 12.050 & 2.1222 \\
10 & 12.047 & 2.1283 \\
20 & 12.021 & 2.1220 \\
30 & 11.972 & 2.1174 \\
40 & 11.965 & 2.1138 \\
50 & 11.925 & 2.0934 \\
\hline
\label{layer_coeff}
\end{tabular}
\end{table}

\begin{table}[htbp]
\centering
\caption{Single-channel response coefficients $c_s$ and $c_c$ under different sampling ratios}
\begin{tabular}{ccc}
\hline
sampling ratio & $c_s$ (photon/MeV) & $c_c$ (photon/MeV) \\
\hline
10 &  11.925 & 2.0934 \\
20 &  23.876 & 4.1956 \\
40 &  48.024 & 8.4299 \\
60 &  72.252 & 12.691 \\
80 &  96.604 & 17.012 \\
90 &  108.73 & 19.104 \\
100 & 121.01 & 21.262 \\
\hline
\label{Sample_coeff}
\end{tabular}
\end{table}

\begin{table}[htbp]
\centering
\caption{Single-channel scintillation $c_s$ (photon/MeV) under different sampling ratios and segmentation}
\begin{tabular}{ccccccccccc}
\hline
 & 1mm & 2mm & 3mm & 4mm & 5mm & 6mm & 7mm & 8mm & 9mm & 10mm \\
\hline
10 & 12.080 & 12.083 & 12.083 & 12.076 & 12.078 & 12.084 & 12.061 & 12.050 & 12.087 & 12.047 \\
20 & 24.169 & 24.156 & 24.189 & 24.170 & 24.157 & 24.157 & 24.124 & 24.102 & 24.152 & 24.083 \\
30 & 36.247 & 36.248 & 36.283 & 36.242 & 36.240 & 36.217 & 36.183 & 36.167 & 36.232 & 36.149 \\
40 & 48.334 & 48.336 & 48.369 & 48.341 & 48.333 & 48.289 & 48.282 & 48.237 & 48.314 & 48.252 \\
50 & 60.426 & 60.414 & 60.448 & 60.426 & 60.426 & 60.425 & 60.365 & 60.322 & 60.407 & 60.358 \\
60 & 72.522 & 72.500 & 72.521 & 72.506 & 72.499 & 72.473 & 72.454 & 72.442 & 72.469 & 72.462 \\
70 & 84.613 & 84.591 & 84.602 & 84.592 & 84.589 & 84.580 & 84.559 & 84.562 & 84.563 & 84.540 \\
80 & 96.702 & 96.687 & 96.691 & 96.684 & 96.705 & 96.671 & 96.684 & 96.660 & 96.675 & 96.614 \\
90 & 114.83 & 114.82 & 114.86 & 114.85 & 114.86 & 114.85 & 114.85 & 114.85 & 114.84 & 114.79 \\
\hline
\label{compare_coeff_s}
\end{tabular}
\end{table}

\begin{table}[htbp]
\centering
\caption{Single-channel Cherenkov $c_c$ (photon/MeV) under different sampling ratios and segmentation}
\begin{tabular}{ccccccccccc}
\hline
 & 1mm & 2mm & 3mm & 4mm & 5mm & 6mm & 7mm & 8mm & 9mm & 10mm \\
\hline
10 & 2.1279 & 2.1273 & 2.1252 & 2.1275 & 2.1258 & 2.1265 & 2.1232 & 2.1222 & 2.1229 & 2.1283 \\
20 & 4.2554 & 4.2555 & 4.2522 & 4.2544 & 4.2530 & 4.2571 & 4.2450 & 4.2466 & 4.2486 & 4.2553 \\
30 & 6.3835 & 6.3820 & 6.3808 & 6.3814 & 6.3792 & 6.3853 & 6.3677 & 6.3715 & 6.3743 & 6.3843 \\
40 & 8.5108 & 8.5104 & 8.5098 & 8.5082 & 8.5130 & 8.4938 & 8.4971 & 8.5018 & 8.5019 & 8.5169 \\
50 & 10.638 & 10.638 & 10.639 & 10.636 & 10.634 & 10.638 & 10.620 & 10.624 & 10.631 & 10.646 \\
60 & 12.766 & 12.767 & 12.768 & 12.764 & 12.762 & 12.765 & 12.750 & 12.756 & 12.758 & 12.779 \\
70 & 14.893 & 14.894 & 14.895 & 14.892 & 14.894 & 14.892 & 14.881 & 14.887 & 14.886 & 14.908 \\
80 & 17.022 & 17.023 & 17.024 & 17.022 & 17.023 & 17.021 & 17.017 & 17.017 & 17.018 & 17.039 \\
90 & 20.214 & 20.215 & 20.218 & 20.217 & 20.218 & 20.217 & 20.216 & 20.216 & 20.217 & 20.223 \\
\hline
\label{compare_coeff_c}
\end{tabular}
\end{table}

\begin{table}[htbp]
\centering
\caption{Dual-readout $e_s$, $e_c$, and $k$ under different sampling segmentation }
\begin{tabular}{cccc}
\hline
sampling segmentation (mm) & $e_s$ (photon/MeV) & $e_c$ (photon/MeV) & $k$ \\
\hline
1 & 13.942 & 3.477 & 2.94 \\
2 & 13.940 & 3.492 & 2.91 \\
4 & 13.948 & 3.464 & 2.82 \\
6 & 13.938 & 3.494 & 2.85 \\
8 & 13.908 & 3.467 & 2.78 \\
10 & 13.914 & 3.469 & 2.79 \\
20 & 13.939 & 3.479 & 2.60 \\
30 & 13.944 & 3.482 & 2.50 \\
40 & 14.022 & 3.498 & 2.32 \\
50 & 14.035 & 3.501 & 2.17 \\
\hline
\label{dual_layer}
\end{tabular}
\end{table}

\begin{table}[htbp]
\centering
\caption{Dual-readout $e_s$, $e_c$, and $k$ under different sampling ratios}
\begin{tabular}{cccc}
\hline
sampling ratios & $e_s$ (photon/MeV) & $e_c$ (photon/MeV) & $k$ \\
\hline
10 & 14.035 & 3.501 & 2.17 \\
20 & 28.432 & 7.908 & 2.26 \\
40 & 57.384 & 14.35 & 2.48 \\
60 & 85.349 & 21.34 & 2.70 \\
80 & 112.18 & 28.01 & 2.96 \\
90 & 125.63 & 31.35 & 3.03 \\
100 & 139.40 & 34.77 & 3.12 \\
\hline
\label{dual_sample}
\end{tabular}
\end{table}

\begin{table}[htbp]
\centering
\caption{Dual-readout scintillation EM response $e_s$ (photon/MeV) under different sampling ratios and segmentation}
\begin{tabular}{ccccccccccc}
\hline
 & 1mm & 2mm & 3mm & 4mm & 5mm & 6mm & 7mm & 8mm & 9mm & 10mm \\
\hline
10 & 13.942 & 13.940 & 13.952 & 13.948 & 13.937 & 13.938 & 13.923 & 13.908 & 13.950 & 13.914 \\
20 & 27.889 & 27.897 & 27.822 & 27.864 & 27.850 & 27.875 & 27.891 & 27.846 & 27.904 & 27.838 \\
30 & 41.838 & 41.821 & 41.847 & 41.796 & 41.785 & 41.825 & 41.793 & 41.761 & 41.832 & 41.780 \\
40 & 55.780 & 55.746 & 55.783 & 55.746 & 55.732 & 55.760 & 55.712 & 55.698 & 55.783 & 55.770 \\
50 & 69.727 & 69.687 & 69.730 & 69.699 & 69.670 & 69.693 & 69.668 & 69.680 & 69.692 & 69.705 \\
60 & 83.702 & 83.637 & 83.665 & 83.659 & 83.629 & 83.671 & 83.589 & 83.621 & 83.648 & 83.630 \\
70 & 97.650 & 97.600 & 97.598 & 97.605 & 97.601 & 97.615 & 97.577 & 97.598 & 97.617 & 97.601 \\
80 & 111.61 & 111.58 & 111.56 & 111.57 & 111.56 & 111.56 & 111.56 & 111.56 & 111.59 & 111.55 \\
90 & 125.56 & 125.56 & 125.55 & 125.55 & 125.52 & 125.52 & 125.56 & 125.55 & 125.53 & 125.52 \\
\hline
\label{compare_es}
\end{tabular}
\end{table}

\begin{table}[htbp]
\centering
\caption{Dual-readout Cherenkov EM response $e_c$ (photon/MeV) under different sampling ratios and segmentation}
\begin{tabular}{ccccccccccc}
\hline
 & 1mm & 2mm & 3mm & 4mm & 5mm & 6mm & 7mm & 8mm & 9mm & 10mm \\
\hline
10 & 3.4774 & 3.4792 & 3.4825 & 3.4764 & 3.4754 & 3.4794 & 3.4743 & 3.4668 & 3.4804 & 3.4690 \\
20 & 6.9598 & 6.9559 & 6.9496 & 6.9479 & 6.9481 & 6.9518 & 6.9556 & 6.9434 & 6.9599 & 6.9454 \\
30 & 10.436 & 10.434 & 10.440 & 10.427 & 10.422 & 10.431 & 10.427 & 10.415 & 10.437 & 10.423 \\
40 & 13.915 & 13.911 & 13.918 & 13.908 & 13.905 & 13.911 & 13.904 & 13.891 & 13.917 & 13.909 \\
50 & 17.400 & 17.389 & 17.397 & 17.395 & 17.383 & 17.392 & 17.375 & 17.379 & 17.386 & 17.388 \\
60 & 20.887 & 20.877 & 20.875 & 20.876 & 20.862 & 20.870 & 20.860 & 20.860 & 20.873 & 20.871 \\
70 & 24.371 & 24.359 & 24.355 & 24.363 & 24.355 & 24.358 & 24.348 & 24.352 & 24.358 & 24.356 \\
80 & 27.856 & 27.847 & 27.841 & 27.849 & 27.843 & 27.841 & 27.840 & 27.840 & 27.845 & 27.832 \\
90 & 31.338 & 31.336 & 31.335 & 31.338 & 31.332 & 31.327 & 31.339 & 31.337 & 31.328 & 31.328 \\
\hline
\label{compare_ec}
\end{tabular}
\end{table}

\begin{table}[htbp]
\centering
\caption{Dual-readout slope $k$ under different sampling ratios and segmentation}
\label{tab:detector_data}
\begin{tabular}{ccccccccccc}
\hline
 & 1mm & 2mm & 3mm & 4mm & 5mm & 6mm & 7mm & 8mm & 9mm & 10mm \\ \hline
10 & 2.94 & 2.91 & 2.86 & 2.82 & 2.85 & 2.85 & 2.78 & 2.78 & 2.76 & 2.79 \\ 
20 & 3.00 & 2.97 & 3.07 & 2.93 & 3.00 & 2.99 & 2.91 & 2.87 & 2.87 & 2.90 \\ 
30 & 3.04 & 2.99 & 3.06 & 2.99 & 3.01 & 2.96 & 2.95 & 2.96 & 2.91 & 2.94 \\ 
40 & 3.06 & 3.06 & 3.06 & 3.02 & 3.02 & 2.98 & 3.01 & 2.99 & 2.97 & 2.94 \\ 
50 & 3.06 & 3.07 & 3.05 & 3.04 & 3.04 & 2.98 & 3.03 & 2.99 & 3.00 & 2.98 \\ 
60 & 3.06 & 3.07 & 3.03 & 3.06 & 3.03 & 3.00 & 3.02 & 3.08 & 3.02 & 3.01 \\ 
70 & 3.07 & 3.08 & 3.03 & 3.08 & 3.06 & 3.01 & 3.05 & 3.05 & 3.01 & 3.03 \\ 
80 & 3.06 & 3.09 & 3.07 & 3.07 & 3.07 & 3.00 & 3.09 & 3.06 & 3.04 & 3.06 \\ 
90 & 3.10 & 3.10 & 3.08 & 3.09 & 3.08 & 3.05 & 3.09 & 3.08 & 3.08 & 3.09 \\ 
\hline
\label{compare_k}
\end{tabular}
\end{table}

\paragraph{Data Availability Statement}

The key simulation codes and underlying data supporting the findings of this study have been deposited in the Science Data Bank (ScienceDB) at https://doi.org/10.57 760/sciencedb.36836.

\acknowledgments

This work was partially supported by the National Key R\&D Program of China (Grant No. 2024YFA1611002) and by the National Natural Science Foundation of China (NSFC) under Grant Nos. 12475187 and 12547102. The computations in this research were performed using the CFFF platform of Fudan University.



\bibliographystyle{JHEP}
\bibliography{biblio.bib}

@book{Fernow_2023, 
place={Cambridge}, 
title={Introduction to Experimental Particle Physics}, 
publisher={Cambridge University Press}, 
author={Fernow, Richard Clinton}, 
year={2023}}

@article{Accardi_2016,
	author = {Accardi, A. and Albacete, J. L. and Anselmino, M. and Armesto, N. and Aschenauer, E. C. and Bacchetta, A. and Boer, D. and Brooks, W. K. and Burton, T. and Chang, N. -B. and Deng, W. -T. and Deshpande, A. and Diehl, M. and Dumitru, A. and Dupr{\'e}, R. and Ent, R. and Fazio, S. and Gao, H. and Guzey, V. and Hakobyan, H. and Hao, Y. and Hasch, D. and Holt, R. and Horn, T. and Huang, M. and Hutton, A. and Hyde, C. and Jalilian-Marian, J. and Klein, S. and Kopeliovich, B. and Kovchegov, Y. and Kumar, K. and Kumeri{\v c}ki, K. and Lamont, M. A. C. and Lappi, T. and Lee, J. -H. and Lee, Y. and Levin, E. M. and Lin, F. -L. and Litvinenko, V. and Ludlam, T. W. and Marquet, C. and Meziani, Z. -E. and McKeown, R. and Metz, A. and Milner, R. and Morozov, V. S. and Mueller, A. H. and M{\"u}ller, B. and M{\"u}ller, D. and Nadel-Turonski, P. and Paukkunen, H. and Prokudin, A. and Ptitsyn, V. and Qian, X. and Qiu, J. -W. and Ramsey-Musolf, M. and Roser, T. and Sabati{\'e}, F. and Sassot, R. and Schnell, G. and Schweitzer, P. and Sichtermann, E. and Stratmann, M. and Strikman, M. and Sullivan, M. and Taneja, S. and Toll, T. and Trbojevic, D. and Ullrich, T. and Venugopalan, R. and Vigdor, S. and Vogelsang, W. and Weiss, C. and Xiao, B. -W. and Yuan, F. and Zhang, Y. -H. and Zheng, L.},
	date = {2016/09/08},
	doi = {10.1140/epja/i2016-16268-9},
	id = {Accardi2016},
	isbn = {1434-601X},
	journal = {The European Physical Journal A},
	number = {9},
	pages = {268},
	title = {Electron-Ion Collider: The next QCD frontier},
	url = {https://doi.org/10.1140/epja/i2016-16268-9},
	volume = {52},
	year = {2016}
}

@article{Benedikt_2025,
	author = {Benedikt, M. and Zimmermann, F. and Auchmann, B. and Bartmann, W. and Burnet, J. P. and Carli, C. and Chanc{\'e}, A. and Craievich, P. and Giovannozzi, M. and Grojean, C. and Gutleber, J. and Hanke, K. and Henriques, A. and Janot, P. and Louren{\c c}o, C. and Mangano, M. and Otto, T. and Poole, J. and Rajagopalan, S. and Raubenheimer, T. and Todesco, E. and Ulrici, L. and Watson, T. and Wilkinson, G. and Abada, A. and Abbrescia, M. and Abdolmaleki, H. and Abidi, S. H. and Abramov, A. and Adam, C. and Ady, M. and Ad\v{z}i\'{c}, P. R. and Agapov, I. and Aguglia, D. and Ahmed, I. and Aiba, M. and Aielli, G. and Akan, T. and Akchurin, N. and Akturk, D. and Al-Thakeel, M. and Alberghi, G. L. and Alcaraz Maestre, J. and Aleksa, M. and Aleksan, R. and Alharthi, F. and Alimena, J. and Alimenti, A. and Alioli, S. and Alix, L. and Allanach, B. C. and Allwicher, L. and Altintas, A. A. and Altınlı, M. and Alviggi, M. and Ambrosio, G. and Amhis, Y. and Amiri, A. and Ammirabile, G. and Andeen, T. and Andr{\'e}, K. D. J. and Andrea, J. and Andreazza, A. and Andreini, M. and Andriollo, T. and Angel, L. and Angelucci, M. and Antusch, S. and Anwar, M. N. and Apolin{\'a}rio, L. and Apollinari, G. and Appleby, R. B. and Apresyan, A. and Apyan, Aram and Apyan, Armen and Arbey, A. and Argiento, B. and Ari, V. and Arias, S. and Arias Alonso, B. and Arnaez, O. and Arnaldi, R. and Arneodo, F. and Arnold, H. and Arrutia Sota, P. and Ascioti, M. E. and Assamagan, K. A. and Aumiller, S. and Aydın, G. and Azizi, K. and Azzi, P. and Bacchetta, N. and Bacci, A. and Bai, B. and Bai, Y. and Balconi, L. and Baldinelli, G. and Balhan, B. and Ball, A. H. and Ballarino, A. and Banerjee, S. and Banik, S. and Barber, D. P. and Barbero, M. B. and Barducci, D. and Barna, D. and Barnaf{\"o}ldi, G. G. and Barnes, M. J. and Barr, A. J. and Bartek, R. and Bartosik, H. and Bass, S. A. and Bassler, U. and Basso, M. J. and Bastianin, A. and Bataillard, P. and Battistin, M. and Bauche, J. and Baudin, L. and Baudot, J. and Baudouy, B. and Bauerdick, L. and Bayındır, C. and Beck, H. P. and Bedeschi, F. and Bee, C. and Begel, M. and Behtouei, M. and Bellagamba, L. and Bellegarde, N. and Belli, E. and Bellingeri, E. and Belomestnykh, S. and Benaglia, A. D. and Bencivenni, G. and Bendavid, J. and Benmergui, M. and Benoit, M. and Benvenuti, D. and Bergauer, T. and Bernachot, N. and Bernardi, G. and Bernardi, J. and Berthet, Q. and Bertoni, S. and Bertulani, C. and Besana, M. I. and Besson, A. and Bettelini, M. and Bettoni, S. and Beuvier, S. and Bhat, P. C. and Bhattacharya, S. and Bhom, J. and Biagini, M. E. and Bibet-Chevalier, A. and Bicrel, M. and Biglietti, M. and Bilei, G. M. and Bilki, B. and Bisgaard Christensen, K. and Biswas, T. and Blanc, F. and Blekman, F. and Blondel, A. and Bl{\"u}mlein, J. and Boccanfuso, D. and Bogomyagkov, A. and Boillon, P. and Boivin, P. and Boland, M. J. and Bologna, S. and Bolukbasi, O. and Bonnet, R. and Borburgh, J. and Bordry, F. and Borges de Sousa, P. and Borghello, G. and Borriello, L. and Bortoletto, D. and Boscolo, M. and Bottura, L. and Boudoul, G. and Boudry, V. and Boughezal, R. and Bourilkov, D. and Boyd, M. and Boye, D. and Bozzi, G. and Braccini, V. and Bracco, C. and Bradu, B. and Braghieri, A. and Braibant, S. and Bramante, J. and Branco, G. C. and Brenner, R. and Brisa, N. and Britzger, D. and Broggi, G. and Bromiley, L. and Brost, E. and Bruant, Q. and Bruce, R. and Br{\"u}ndermann, E. and Brunetti, L. and Br{\"u}ning, O. and Brunner, O. and Buffat, X. and Bulyak, E. and Burdyko, A. and Burkhardt, H. and Burrows, P. N. and Busatto, S. and Buschaert, S. and Buttazzo, D. and Butterworth, A. and Butti, D. and Cacciapaglia, G. and Cai, Y. and Caiffi, B. and Cairo, V. and Cakir, O. and Calafiura, P. and Calaga, R. and Calatroni, S. and Caldwell, D. G. and {\c C}alı{\c s}kan, A. and Calpini, C. and Calviani, M. and Camacho-P{\'e}rez, E. and Camarri, P. and Caminada, L. and Campajola, M. and Canbay, A. C. and Canderan, K. and Candido, S. and Canelli, F. and Canepa, A. and Cantarella, S. and Cant{\'u}n-Avila, K. B. and Capriotti, L. and Caram, A. and Carbone, A. and Carceller, J. M. and Carini, G. and Carlier, F. and Carloni Calame, C. M. and Carra, F. and Cartannaz, C. and Casenove, S. and Catalano, G. and Cavaliere, V. and Cazzaniga, C. and Cecchi, C. and Celiberto, F. G. and Cepeda, M. and Cerutti, F. and Cetorelli, F. and Chachamis, G. and Chae, Y. and Chagnet, F. and Chaikovska, I. and Chalhoub, M. and Chamizo-Llatas, M. and Champagne, M. and Chanal, H. and Chapelier, G. and Charitos, P. and Charles, C. and Charles, T. K. and Charlot, C. and Chatterjee, S. and Chaudhuri, A. and Chehab, R. and Chekanov, S. V. and Chen, H. and Chesne, T. and Chiapponi, F. and Chiarello, G. and Chiesa, M. and Chiggiato, P. and Chomaz, P. and Chorowski, M. and Chou, J. P. and Chrzaszcz, M. and Chung, W. and Ciarlantini, S. and Ciarma, A. and Cieri, D. and Ciftci, A. K. and Ciftci, R. and Cimino, R. and Cirotto, F. and Ciuchini, M. and Cobal, M. and Coccaro, A. and Coelho Lopes De Sa, R. and Coleman-Smith, J. A. and Collamati, F.},
	date = {2025/11/01},
	doi = {10.1140/epjs/s11734-025-01958-5},
	id = {Benedikt2025},
	isbn = {1951-6401},
	journal = {The European Physical Journal Special Topics},
	number = {17},
	pages = {5113--5383},
	title = {Future Circular Collider Feasibility Study Report},
	url = {https://doi.org/10.1140/epjs/s11734-025-01958-5},
	volume = {234},
	year = {2025}
}

@article{Brau:2010zz,
    author = "Brau, James E. and Jaros, John A. and Ma, Hong",
    title = "{Advances in calorimetry}",
    reportNumber = "SLAC-REPRINT-2013-845",
    doi = "10.1146/annurev.nucl.012809.104449",
    journal = "Ann. Rev. Nucl. Part. Sci.",
    volume = "60",
    pages = "615--644",
    year = "2010"
}

@article{WELLISCH200165,
title = {Hadronic shower models in Geant4 — the frameworks},
journal = {Computer Physics Communications},
volume = {140},
number = {1},
pages = {65-75},
year = {2001},
note = {CHEP2000},
issn = {0010-4655},
doi = {https://doi.org/10.1016/S0010-4655(01)00256-9},
url = {https://www.sciencedirect.com/science/article/pii/S0010465501002569},
author = {J.P. Wellisch}
}

@article{Lai_2024,
doi = {10.1088/1748-0221/19/04/P04037},
url = {https://doi.org/10.1088/1748-0221/19/04/P04037},
year = {2024},
month = {apr},
publisher = {IOP Publishing},
volume = {19},
number = {04},
pages = {P04037},
author = {Lai, S. and Utehs, J. and Wilhahn, A. and Bach, O. and Brianne, E. and Ebrahimi, A. and Gadow, K. and Göttlicher, P. and Hartbrich, O. and Heuchel, D. and Irles, A. and Krüger, K. and Kvasnicka, J. and Lu, S. and Neubüser, C. and Provenza, A. and Reinecke, M. and Sefkow, F. and Schuwalow, S. and De Silva, M. and Sudo, Y. and Tran, H.L. and Buhmann, E. and Garutti, E. and \.{H}uck, S. and Kasieczka, G. and Martens, S. and Rolph, J. and Wellhausen, J. and Blazey, G.C. and Dyshkant, A. and Francis, K. and Zutshi, V. and Bilki, B. and Northacker, D. and Onel, Y. and Hummer, F. and Simon, F. and Kawagoe, K. and Onoe, T. and Suehara, T. and Tsumura, S. and Yoshioka, T. and Fouz, M.C. and Emberger, L. and Graf, C. and Wagner, M. and Pöschl, R. and Richard, F. and Zerwas, D. and Boudry, V. and Brient, J-C. and Nanni, J. and Videau, H. and Liu, L. and Masuda, R. and Murata, T. and Ootani, W. and Takatsu, T. and Tsuji, N. and Chadeeva, M. and Danilov, M. and Korpachev, S. and Rusinov, V. and The CALICE collaboration},
title = {Software compensation for highly granular calorimeters using machine learning},
journal = {Journal of Instrumentation}
}

@article{HILGER1987488,
title = {The ZEUS uranium-scintillator calorimeter for HERA},
journal = {Nuclear Instruments and Methods in Physics Research Section A: Accelerators, Spectrometers, Detectors and Associated Equipment},
volume = {257},
number = {3},
pages = {488-498},
year = {1987},
issn = {0168-9002},
doi = {https://doi.org/10.1016/0168-9002(87)90952-1},
url = {https://www.sciencedirect.com/science/article/pii/0168900287909521},
author = {Erwin Hilger}
}

@article{CELLETTI1984493,
title = {Test of prototype hadron calorimeter for the L3 experiment},
journal = {Nuclear Instruments and Methods in Physics Research},
volume = {225},
number = {3},
pages = {493-497},
year = {1984},
issn = {0167-5087},
doi = {https://doi.org/10.1016/0167-5087(84)90091-7},
url = {https://www.sciencedirect.com/science/article/pii/0167508784900917},
author = {F. Celletti and A. Marchionni and P. Spillantini and Yu. Kamyshkov and V. Pojidaev and M. Cerrada and H. Zeidler and F. Ferroni and M. Steuer and K. Deiters and G. Gianolli and P. Lecomte and P. {Le Coultre} and H. Suter}
}

@article{RevModPhys.90.025002,
  title = {Dual-readout calorimetry},
  author = {Lee, Sehwook and Livan, Michele and Wigmans, Richard},
  journal = {Rev. Mod. Phys.},
  volume = {90},
  issue = {2},
  pages = {025002},
  numpages = {40},
  year = {2018},
  month = {Apr},
  publisher = {American Physical Society},
  doi = {10.1103/RevModPhys.90.025002},
  url = {https://link.aps.org/doi/10.1103/RevModPhys.90.025002}
}

@Article{instruments6030036,
AUTHOR = {Wigmans, Richard},
TITLE = {25 Years of Dual-Readout Calorimetry},
JOURNAL = {Instruments},
VOLUME = {6},
YEAR = {2022},
NUMBER = {3},
ARTICLE-NUMBER = {36},
URL = {https://www.mdpi.com/2410-390X/6/3/36},
ISSN = {2410-390X},
DOI = {10.3390/instruments6030036}
}

@article{Pareti_2024,
doi = {10.1088/1748-0221/19/04/C04029},
url = {https://doi.org/10.1088/1748-0221/19/04/C04029},
year = {2024},
month = {apr},
publisher = {IOP Publishing},
volume = {19},
number = {04},
pages = {C04029},
author = {Pareti, A. and on behalf of the IDEA dual-readout calorimeter group},
title = {Status of dual-readout calorimetry for future high-energy physics experiments},
journal = {Journal of Instrumentation}
}

@article{VANDERKOLK2020162059,
title = {FoCal: A highly granular digital calorimeter},
journal = {Nuclear Instruments and Methods in Physics Research Section A: Accelerators, Spectrometers, Detectors and Associated Equipment},
volume = {958},
pages = {162059},
year = {2020},
note = {Proceedings of the Vienna Conference on Instrumentation 2019},
issn = {0168-9002},
doi = {https://doi.org/10.1016/j.nima.2019.04.013},
url = {https://www.sciencedirect.com/science/article/pii/S0168900219304619},
author = {N. {van der Kolk}}
}

@article{Amendola_2025,
doi = {10.1088/1748-0221/20/01/C01031},
url = {https://doi.org/10.1088/1748-0221/20/01/C01031},
year = {2025},
month = {jan},
publisher = {IOP Publishing},
volume = {20},
number = {01},
pages = {C01031},
author = {Amendola, Chiara and on behalf of the CMS collaboration},
title = {An overview of the CMS High-Granularity Calorimeter},
journal = {Journal of Instrumentation}
}

@Article{particles9010001,
AUTHOR = {Wei, Yide and Wang, Hui},
TITLE = {Review of the Performance of the CMS Hadron Calorimeter},
JOURNAL = {Particles},
VOLUME = {9},
YEAR = {2026},
NUMBER = {1},
ARTICLE-NUMBER = {1},
URL = {https://www.mdpi.com/2571-712X/9/1/1},
ISSN = {2571-712X},
DOI = {10.3390/particles9010001}
}

@article{Solovyanov_2009,
doi = {10.1088/1742-6596/160/1/012053},
url = {https://doi.org/10.1088/1742-6596/160/1/012053},
year = {2009},
month = {apr},
publisher = {},
volume = {160},
number = {1},
pages = {012053},
author = {O Solovyanov and (on behalf ofAtlas Tile Community)},
title = {The ATLAS tile calorimeter: Commissioning and preparation for collisions},
journal = {Journal of Physics: Conference Series}
}

@article{achasov_2023, 
title={STCF conceptual design report (Volume 1): Physics \& detector}, 
volume={19}, 
ISSN={2095-0470}, 
url={http://dx.doi.org/10.1007/s11467-023-1333-z},
DOI={10.1007/s11467-023-1333-z},
number={1}, 
journal={Frontiers of Physics}, 
publisher={China Engineering Science Press Co. Ltd.}, 
author={Achasov, M. and Ai, X. C. and An, L. P. and Aliberti, R. and An, Q. and Bai, X. Z. and Bai, Y. and Bakina, O. and Barnyakov, A. and Blinov, V. and Bobrovnikov, V. and Bodrov, D. and Bogomyagkov, A. and Bondar, A. and Boyko, I. and Bu, Z. H. and Cai, F. M. and Cai, H. and Cao, J. J. and Cao, Q. H. and Cao, X. and Cao, Z. and Chang, Q. and Chao, K. T. and Chen, D. Y. and Chen, H. and Chen, H. X. and Chen, J. F. and Chen, K. and Chen, L. L. and Chen, P. and Chen, S. L. and Chen, S. M. and Chen, S. and Chen, S. P. and Chen, W. and Chen, X. and Chen, X. F. and Chen, X. R. and Chen, Y. and Chen, Y. Q. and Cheng, H. Y. and Cheng, J. and Cheng, S. and Cheng, T. G. and Dai, J. P. and Dai, L. Y. and Dai, X. C. and Dedovich, D. and Denig, A. and Denisenko, I. and Dias, J. M. and Ding, D. Z. and Dong, L. Y. and Dong, W. H. and Druzhinin, V. and Du, D. S. and Du, Y. J. and Du, Z. G. and Duan, L. M. and Epifanov, D. and Fan, Y. L. and Fang, S. S. and Fang, Z. J. and Fedotovich, G. and Feng, C. Q. and Feng, X. and Feng, Y. T. and Fu, J. L. and Gao, J. and Gao, Y. N. and Ge, P. S. and Geng, C. Q. and Geng, L. S. and Gilman, A. and Gong, L. and Gong, T. and Gou, B. and Gradl, W. and Gu, J. L. and Guevara, A. and Gui, L. C. and Guo, A. Q. and Guo, F. K. and Guo, J. C. and Guo, J. and Guo, Y. P. and Guo, Z. H. and Guskov, A. and Han, K. L. and Han, L. and Han, M. and Hao, X. Q. and He, J. B. and He, S. Q. and He, X. G. and He, Y. L. and He, Z. B. and Heng, Z. X. and Hou, B. L. and Hou, T. J. and Hou, Y. R. and Hu, C. Y. and Hu, H. M. and Hu, K. and Hu, R. J. and Hu, W. H. and Hu, X. H. and Hu, Y. C. and Hua, J. and Huang, G. S. and Huang, J. S. and Huang, M. and Huang, Q. Y. and Huang, W. Q. and Huang, X. T. and Huang, X. J. and Huang, Y. B. and Huang, Y. S. and H\"{u}sken, N. and Ivanov, V. and Ji, Q. P. and Jia, J. J. and Jia, S. and Jia, Z. K. and Jiang, H. B. and Jiang, J. and Jiang, S. Z. and Jiao, J. B. and Jiao, Z. and Jing, H. J. and Kang, X. L. and Kang, X. S. and Ke, B. C. and Kenzie, M. and Khoukaz, A. and Koop, I. and Kravchenko, E. and Kuzmin, A. and Lei, Y. and Levichev, E. and Li, C. H. and Li, C. and Li, D. Y. and Li, F. and Li, G. and Li, G. and Li, H. B. and Li, H. and Li, H. N. and Li, H. J. and Li, H. L. and Li, J. M. and Li, J. and Li, L. and Li, L. and Li, L. Y. and Li, N. and Li, P. R. and Li, R. H. and Li, S. and Li, T. and Li, W. J. and Li, X. and Li, X. H. and Li, X. Q. and Li, X. H. and Li, Y. and Li, Y. Y. and Li, Z. J. and Liang, H. and Liang, J. H. and Liang, Y. T. and Liao, G. R. and Liao, L. Z. and Liao, Y. and Lin, C. X. and Lin, D. X. and Lin, X. S. and Liu, B. J. and Liu, C. W. and Liu, D. and Liu, F. and Liu, G. M. and Liu, H. B. and Liu, J. and Liu, J. J. and Liu, J. B. and Liu, K. and Liu, K. Y. and Liu, K. and Liu, L. and Liu, Q. and Liu, S. B. and Liu, T. and Liu, X. and Liu, Y. W. and Liu, Y. and Liu, Y. L. and Liu, Z. Q. and Liu, Z. Y. and Liu, Z. W. and Logashenko, I. and Long, Y. and Lu, C. G. and Lu, J. X. and Lu, N. and L\"{u}, Q. F. and Lu, Y. and Lu, Y. and Lu, Z. and Lukin, P. and Luo, F. J. and Luo, T. and Luo, X. F. and Lyu, H. J. and Lyu, X. R. and Ma, J. P. and Ma, P. and Ma, Y. and Ma, Y. M. and Maas, F. and Malde, S. and Matvienko, D. and Meng, Z. X. and Mitchell, R. and Nefediev, A. and Nefedov, Y. and Olsen, S. L. and Ouyang, Q. and Pakhlov, P. and Pakhlova, G. and Pan, X. and Pan, Y. and Passemar, E. and Pei, Y. P. and Peng, H. P. and Peng, L. and Peng, X. Y. and Peng, X. J. and Peters, K. and Pivovarov, S. and Pyata, E. and Qi, B. B. and Qi, Y. Q. and Qian, W. B. and Qian, Y. and Qiao, C. F. and Qin, J. J. and Qin, J. J. and Qin, L. Q. and Qin, X. S. and Qiu, T. L. and Rademacker, J. and Redmer, C. F. and Sang, H. Y. and Saur, M. and Shan, W. and Shan, X. Y. and Shang, L. L. and Shao, M. and Shekhtman, L. and Shen, C. P. and Shen, J. M. and Shen, Z. T. and Shi, H. C. and Shi, X. D. and Shwartz, B. and Sokolov, A. and Song, J. J. and Song, W. M. and Song, Y. and Song, Y. X. and Sukharev, A. and Sun, J. F. and Sun, L. and Sun, X. M. and Sun, Y. J. and Sun, Z. P. and Tang, J. and Tang, S. S. and Tang, Z. B. and Tian, C. H. and Tian, J. S. and Tian, Y. and Tikhonov, Y. and Todyshev, K. and Uglov, T. and Vorobyev, V. and Wan, B. D. and Wang, B. L. and Wang, B. and Wang, D. Y. and Wang, G. Y. and Wang, G. L. and Wang, H. L. and Wang, J. and Wang, J. H. and Wang, J. C. and Wang, M. L. and Wang, R. and Wang, R. and Wang, S. B. and Wang, W. and Wang, W. P. and Wang, X. C. and Wang, X. D. and Wang, X. L. and Wang, X. L. and Wang, X. P. and Wang, X. F. and Wang, Y. D. and Wang, Y. P. and Wang, Y. Q. and Wang, Y. L. and Wang, Y. G. and Wang, Z. Y. and Wang, Z. Y. and Wang, Z. L. and Wang, Z. G. and Wei, D. H. and Wei, X. L. and Wei, X. M. and Wen, Q. G. and Wen, X. J. and Wilkinson, G. and Wu, B. and Wu, J. J. and Wu, L. and Wu, P. and Wu, T. W. and Wu, Y. S. and Xia, L. and Xiang, T. and Xiao, C. W. and Xiao, D. and Xiao, M. and Xie, K. P. and Xie, Y. H. and Xing, Y. and Xing, Z. Z. and Xiong, X. N. and Xu, F. R. and Xu, J. and Xu, L. L. and Xu, Q. N. and Xu, X. C. and Xu, X. P. and Xu, Y. C. and Xu, Y. P. and Xu, Y. and Xu, Z. Z. and Xuan, D. W. and Xue, F. F. and Yan, L. and Yan, M. J. and Yan, W. B. and Yan, W. C. and Yan, X. S. and Yang, B. F. and Yang, C. and Yang, H. J. and Yang, H. R. and Yang, H. T. and Yang, J. F. and Yang, S. L. and Yang, Y. D. and Yang, Y. H. and Yang, Y. S. and Yang, Y. L. and Yang, Z. W. and Yang, Z. Y. and Yao, D. L. and Yin, H. and Yin, X. H. and Yokozaki, N. and You, S. Y. and You, Z. Y. and Yu, C. X. and Yu, F. S. and Yu, G. L. and Yu, H. L. and Yu, J. S. and Yu, J. Q. and Yuan, L. and Yuan, X. B. and Yuan, Z. Y. and Yue, Y. F. and Zeng, M. and Zeng, S. and Zhang, A. L. and Zhang, B. W. and Zhang, G. Y. and Zhang, G. Q. and Zhang, H. J. and Zhang, H. B. and Zhang, J. Y. and Zhang, J. L. and Zhang, J. and Zhang, L. and Zhang, L. M. and Zhang, Q. A. and Zhang, R. and Zhang, S. L. and Zhang, T. and Zhang, X. and Zhang, Y. and Zhang, Y. J. and Zhang, Y. X. and Zhang, Y. T. and Zhang, Y. F. and Zhang, Y. C. and Zhang, Y. and Zhang, Y. and Zhang, Y. M. and Zhang, Y. L. and Zhang, Z. H. and Zhang, Z. Y. and Zhang, Z. Y. and Zhao, H. Y. and Zhao, J. and Zhao, L. and Zhao, M. G. and Zhao, Q. and Zhao, R. G. and Zhao, R. P. and Zhao, Y. X. and Zhao, Z. G. and Zhao, Z. X. and Zhemchugov, A. and Zheng, B. and Zheng, L. and Zheng, Q. B. and Zheng, R. and Zheng, Y. H. and Zhong, X. H. and Zhou, H. J. and Zhou, H. Q. and Zhou, H. and Zhou, S. H. and Zhou, X. and Zhou, X. K. and Zhou, X. P. and Zhou, X. R. and Zhou, Y. L. and Zhou, Y. and Zhou, Y. X. and Zhou, Z. Y. and Zhu, J. Y. and Zhu, K. and Zhu, R. D. and Zhu, R. L. and Zhu, S. H. and Zhu, Y. C. and Zhu, Z. A. and Zhukova, V. and Zhulanov, V. and Zou, B. S. and Zuo, Y. B.}, 
year={2023},
month={Nov} 
}

@article{Lai_2024_1,
doi = {10.1088/1748-0221/19/10/P10027},
url = {https://doi.org/10.1088/1748-0221/19/10/P10027},
year = {2024},
month = {oct},
publisher = {IOP Publishing},
volume = {19},
number = {10},
pages = {P10027},
author = {Lai, S. and Utehs, J. and Wilhahn, A. and Fouz, M.C. and Bach, O. and Brianne, E. and Ebrahimi, A. and Gadow, K. and Göttlicher, P. and Hartbrich, O. and Heuchel, D. and Irles, A. and Krüger, K. and Kvasnicka, J. and Lu, S. and Neubüser, C. and Provenza, A. and Reinecke, M. and Sefkow, F. and Schuwalow, S. and De Silva, M. and Sudo, Y. and Tran, H.L. and Liu, L. and Masuda, R. and Murata, T. and Ootani, W. and Seino, T. and Takatsu, T. and Tsuji, N. and Pöschl, R. and Richard, F. and Zerwas, D. and Hummer, F. and Simon, F. and Boudry, V. and Brient, J-C. and Nanni, J. and Videau, H. and Buhmann, E. and Garutti, E. and Huck, S. and Kasieczka, G. and Martens, S. and Rolph, J. and Wellhausen, J. and Bilki, B. and Northacker, D. and Onel, Y. and Emberger, L. and Graf, C. and The CALICE collaboration},
title = {Shower separation in five dimensions for highly granular calorimeters using machine learning},
journal = {Journal of Instrumentation}
}

@article{WIGMANS2013475,
title = {The dual-readout approach to calorimetry},
journal = {Nuclear Instruments and Methods in Physics Research Section A: Accelerators, Spectrometers, Detectors and Associated Equipment},
volume = {732},
pages = {475-479},
year = {2013},
note = {Vienna Conference on Instrumentation 2013},
issn = {0168-9002},
doi = {https://doi.org/10.1016/j.nima.2013.04.005},
url = {https://www.sciencedirect.com/science/article/pii/S0168900213003847},
author = {Richard Wigmans}
}

@article{RevModPhys.75.1243,
  title = {Calorimetry for particle physics},
  author = {Fabjan, Christian W. and Gianotti, Fabiola},
  journal = {Rev. Mod. Phys.},
  volume = {75},
  issue = {4},
  pages = {1243--1286},
  numpages = {0},
  year = {2003},
  month = {Oct},
  publisher = {American Physical Society},
  doi = {10.1103/RevModPhys.75.1243},
  url = {https://link.aps.org/doi/10.1103/RevModPhys.75.1243}
}

@article{GROOM2007633,
title = {Energy flow in a hadronic cascade: Application to hadron calorimetry},
journal = {Nuclear Instruments and Methods in Physics Research Section A: Accelerators, Spectrometers, Detectors and Associated Equipment},
volume = {572},
number = {2},
pages = {633-653},
year = {2007},
issn = {0168-9002},
doi = {https://doi.org/10.1016/j.nima.2006.11.070},
url = {https://www.sciencedirect.com/science/article/pii/S0168900206023965},
author = {Donald E. Groom}
}

@article{AKCHURIN2008273,
title = {Dual-readout calorimetry with lead tungstate crystals},
journal = {Nuclear Instruments and Methods in Physics Research Section A: Accelerators, Spectrometers, Detectors and Associated Equipment},
volume = {584},
number = {2},
pages = {273-284},
year = {2008},
issn = {0168-9002},
doi = {https://doi.org/10.1016/j.nima.2007.09.035},
url = {https://www.sciencedirect.com/science/article/pii/S016890020702075X},
author = {N. Akchurin and L. Berntzon and A. Cardini and R. Ferrari and G. Gaudio and J. Hauptman and H. Kim and L. {La Rotonda} and M. Livan and E. Meoni and H. Paar and A. Penzo and D. Pinci and A. Policicchio and S. Popescu and G. Susinno and Y. Roh and W. Vandelli and R. Wigmans}
}

@article{Thomson_2011,
doi = {10.1088/1742-6596/293/1/012021},
url = {https://doi.org/10.1088/1742-6596/293/1/012021},
year = {2011},
month = {apr},
publisher = {},
volume = {293},
number = {1},
pages = {012021},
author = {Thomson, Mark A},
title = {Particle Flow Calorimetry},
journal = {Journal of Physics: Conference Series}
}

@article{THOMSON200925,
title = {Particle flow calorimetry and the PandoraPFA algorithm},
journal = {Nuclear Instruments and Methods in Physics Research Section A: Accelerators, Spectrometers, Detectors and Associated Equipment},
volume = {611},
number = {1},
pages = {25-40},
year = {2009},
issn = {0168-9002},
doi = {https://doi.org/10.1016/j.nima.2009.09.009},
url = {https://www.sciencedirect.com/science/article/pii/S0168900209017264},
author = {M.A. Thomson}
}

@article{Aamir_2024,
doi = {10.1088/1748-0221/19/11/P11025},
url = {https://doi.org/10.1088/1748-0221/19/11/P11025},
year = {2024},
month = {nov},
publisher = {IOP Publishing},
volume = {19},
number = {11},
pages = {P11025},
author = {Aamir, M. and Adamov, G. and Adams, T. and Adloff, C. and Afanasiev, S. and Agrawal, C. and Agrawal, C. and Ahmad, A. and Ahmed, H.A. and Akbar, S. and Akchurin, N. and Akgul, B. and Akgun, B. and Akpinar, R.O. and Aktas, E. and Al Kadhim, A. and Alexakhin, V. and Alimena, J. and Alison, J. and Alpana, A. and Alshehri, W. and Alvarez Dominguez, P. and Alyari, M. and Amendola, C. and Amir, R.B. and Andersen, S.B. and Andreev, Y. and Antoszczuk, P.D. and Aras, U. and Ardila, L. and Aspell, P. and Avila, M. and Awad, I. and Aydilek, O. and Azimi, Z. and Aznar Pretel, A. and Bach, O.A. and Bainbridge, R. and Bakshi, A. and Bam, B. and Banerjee, S. and Barney, D. and Bayraktar, O. and Beaudette, F. and Beaujean, F. and Becheva, E. and Behera, P.K. and Belloni, A. and Bergauer, T. and Besancon, M. and Bessidskaia Bylund, O. and Bhatt, L. and Bhattacharya, S. and Bhowmil, D. and Blekman, F. and Blinov, P. and Bloch, P. and Bodek, A. and Boger, a. and Bonnemaison, A. and Bouyjou, F. and Brennan, L. and Brondolin, E. and Brusamolino, A. and Bubanja, I. and Buchot Perraguin, A. and Bunin, P. and Burazin Misura, A. and Butler-nalin, A. and Cakir, A. and Callier, S. and Campbell, S. and Candemir, Y.B. and Canderan, K. and Cankocak, K. and Cappati, A. and Caregari, S. and Carron, S. and Carty, C. and Cauchois, A. and Ceard, L. and Cerci, S. and Chang, P.J. and Chatterjee, R.M. and Chatterjee, S. and Chattopadhyay, P. and Chatzistavrou, T. and Chaudhary, M.S. and Chen, J.A. and Chen, J. and Chen, Y. and Cheng, K. and Cheung, H. and Chhikara, J. and Chiron, A. and Chiusi, M. and Chokheli, D. and Chudasama, R. and Clement, E. and Coco Mendez, S. and Coko, D. and Coskun, K. and Couderc, F. and Crossman, B. and Cui, Z. and Cuisset, T. and Cummings, G. and Curtis, E.M. and D'Alfonso, M. and Döhler-Ball, J. and Dadazhanova, O. and Damgov, J. and Das, I. and Das Gupta, S. and Dauncey, P. and David Tinoco Mendes, A. and Davies, G. and Davignon, O. and de Barbaro, P. and De La Taille, C. and De Silva, M. and De Wit, A. and Debbins, P. and Defranchis, M.M. and Delagnes, E. and Devouge, P. and Di Guglielmo, G. and Diehl, L. and Dilsiz, K. and Dincer, G.G. and Dittmann, J. and Dragicevic, M. and Du, D. and Dubinchik, B. and Dugad, S. and Dulucq, F. and Dumanoglu, I. and Duran, B. and Dutta, S. and Dutta, V. and Dychkant, A. and Dünser, M. and Edberg, T. and Ehle, I.T. and El Berni, A. and Elias, F. and Eno, S.C. and Erdogan, E.N. and Erkmen, B. and Ershov, Y. and Ertorer, E.Y. and Extier, S. and Eychenne, L. and Fedar, Y.E. and Fedi, G. and Figueiredo De Sá Sousa De Almeida, J.P. and Fontana Santos Alves, B.A. and Frahm, E. and Francis, K. and Freeman, J. and French, T. and Gaede, F. and Gandhi, P.K. and Ganjour, S. and Garcia-Bellido, A. and Gastaldi, F. and Gazi, L. and Gecse, Z. and Gerwig, H. and Gevin, O. and Ghosh, S. and Ghosh, S. and Gill, K. and Gingu, C. and Gleyzer, S. and Godinovic, N. and Goettlicher, P. and Goff, R. and Gok, M. and Golunov, A. and Gonultas, B. and González Martínez, J.D. and Gorbounov, N. and Gouskos, L. and Gray, A. and Gray, L. and Grieco, C. and Groenroos, S. and Groner, D. and Gruber, A. and Grummer, A. and Grönroos, S. and Guerrero, D. and Guilloux, F. and Guler, Y. and Gungordu, A.D. and Guo, J. and Guo, K. and Gurpinar Guler, E. and Gutti, H.K. and Guvenli, A.A. and Gülmez, E. and Hacisahinoglu, B. and Halkin, Y. and Hamilton Ilha Machado, G. and Hare, H.S. and Hatakeyama, K. and Heering, A.H. and Hegde, V. and Heintz, U. and Hinton, N. and Hinzmann, A. and Hirschauer, J. and Hitlin, D. and Hoff, J. and Hos, \.{I}. and Hou, B. and Hou, X. and Howard, A. and Howe, C. and Hsieh, H. and Hsu, T. and Hua, H. and Hummer, F. and Imran, M. and Incandela, J. and Iren, E. and Isildak, B. and Jackson, P.S. and Jackson, W.J. and Jain, S. and Jana, P. and Jaroslavceva, J. and Jena, S. and Jige, A. and Jordano, P.P. and Joshi, U. and Kaadze, K. and Kachanov, V. and Kafizov, A. and Kalipoliti, L. and Kallil Tharayil, A. and Kaluzinska, O. and Kamble, S. and Kaminskiy, A. and Kanemura, M. and Kanso, H. and Kao, Y. and Kapic, A. and Kapsiak, C. and Karjavine, V. and Karmakar, S. and Karneyeu, A. and Kaya, M. and Kayis Topaksu, A. and Kaynak, B. and Kazhykarim, Y. and Khan, F.A. and Khudiakov, A. and Kieseler, J. and Kim, R.S. and Klijnsma, T. and Kloiber, E.G. and Klute, M. and Kocak, Z. and Kodali, K.R. and Koetz, K. and Kolberg, T. and Kolcu, O.B. and Komaragiri, J.R. and Komm, M. and Kopsalis, I. and Krause, H.A. and Krawczyk, M.A. and Krishnaswamy Vinayakam, T.R. and Kristiansen, K. and Kristic, A. and Krohn, M. and Kronheim, B. and Krüger, K. and Kudtarkar, C. and Kulis, S. and Kumar, M. and Kumar, N. and Kumar, S. and Kumar Verma, R. and Kunori, S. and Kunts, A. and Kuo, C. and Kurenkov, A. and Kuryatkov, V. and Kyre, S. and Ladenson, J. and Lamichhane, K. and Landsberg, G. and Langford, J. and Laudrain, A. and Laughlin, R. and Lawhorn, J. and Le Dortz, O. and Lee, S.W. and Lektauers, A. and Lelas, D. and Leon, M. and Levchuk, L. and Li, A.J. and Li, J. and Li, Y. and Liang, Z. and Liao, H. and Lin, K. and Lin, W. and Lin, Z. and Lincoln, D. and Linssen, L. and Litomin, A. and Liu, G. and Liu, Y. and Lobanov, A. and Lohezic, V. and Loiseau, T. and Lu, C. and Lu, R. and Lu, S.Y. and Lukens, P. and Mackenzie, M. and Magnan, A. and Magniette, F. and Mahjoub, A. and Mahon, D. and Majumder, G. and Makarenko, V. and Malakhov, A. and Malgeri, L. and Mallios, S. and Mandloi, C. and Mankel, A. and Mannelli, M. and Mans, J. and Mantilla, C. and Martinez, G. and Massa, C. and Masterson, P. and Matthewman, M. and Matveev, V. and Mayekar, S. and Mazlov, I. and Mehta, A. and Mestvirishvili, A. and Miao, Y. and Milella, G. and Mirza, I.R. and Mitra, P. and Moccia, S. and Mohanty, G.B. and Monti, F. and Moortgat, F. and Murthy, S. and Music, J. and Musienko, Y. and Nabili, S. and Nelson, J.W. and Nema, A. and Neutelings, I. and Niedziela, J. and Nikitenko, A. and Noonan, D. and Noy, M. and Nurdan, K. and Obraztsov, S. and Ochando, C. and Ogul, H. and Olsson, J. and Onel, Y. and Ozkorucuklu, S. and Paganis, E. and Palit, P. and Pan, R. and Pandey, S. and Pantaleo, F. and Papageorgakis, C. and Paramesvaran, S. and Paranjpe, M.M. and Parolia, S. and Parsons, A.G. and Parygin, P. and Pastika, J. and Paulini, M. and Paus, C. and Peñaló Castillo, K. and Pedro, K. and Pekic, V. and Peltola, T. and Peng, B. and Perego, A. and Perini, D. and Petrilli, A. and Pham, H. and Podem, S.K. and Popov, V. and Portales, L. and Potok, O. and Pradeep, P.B. and Pramanik, R. and Prosper, H. and Prvan, M. and Qasim, S.R. and Qu, H. and Quast, T. and Quiroga Trivio, A. and Rabour, L. and Raicevic, N. and Rao, M.A. and Rapacz, K. and Redjeb, W. and Reinecke, M. and Revering, M. and Roberts, A. and Rohlf, J. and Rosado, P. and Rose, A. and Rothman, S. and Rout, P.K. and Rovere, M. and Roy, A. and Rubinov, P. and Rumerio, P. and Rusack, R. and Rygaard, L. and Ryjov, V. and Sadivnycha, S. and Sahin, M.Ö. and Sakarya, U. and Salerno, R. and Saradhy, R. and Saraf, M. and Sarbandi, K. and Sarkisla, M.A. and Satyshev, I. and Saud, N. and Sauvan, J. and Schindler, G. and Schmidt, A. and Schmidt, I. and Schmitt, M.H. and Sculac, A. and Sculac, T. and Sedelnikov, A. and Seez, C. and Sefkow, F. and Selivanova, D. and Selvaggi, M. and Sergeychik, V. and Sert, H. and Shahid, M. and Sharma, P. and Sharma, R. and Sharma, S. and Shelake, M. and Shenai, A. and Shih, C.W. and Shinde, R. and Shmygol, D. and Shukla, R. and Sicking, E. and Silva, P. and Simsek, C. and Simsek, E. and Sirasva, B.K. and Sirois, Y. and Song, S. and Song, Y. and Soudais, G. and Sriram, S. and St Jacques, R.R. and Stahl Leiton, A.G. and Steen, A. and Stein, J. and Strait, J. and Strobbe, N. and Su, X. and Sukhov, E. and Suleiman, A. and Sunar Cerci, D. and Suryadevara, P. and Swain, K. and Syal, C. and Tali, B. and Tanay, K. and Tang, W. and Tanvir, A. and Tao, J. and Tarabini, A. and Tatli, T. and Taylor, R. and Taysi, Z.C. and Teafoe, G. and Tee, C.Z. and Terrill, W. and Thienpont, D. and Thomas, P.E. and Thomas, R. and Titov, M. and Todd, C. and Todd, E. and Toms, M. and Tosun, A. and Troska, J. and Tsai, L. and Tsamalaidze, Z. and Tsionou, D. and Tsipolitis, G. and Tsirigoti, M. and Tu, R. and Tural Polat, S.N. and Undleeb, S. and Usai, E. and Uslan, E. and Ustinov, V. and Uzunian, A. and Vernazza, E. and Viahin, O. and Viazlo, O. and Vichoudis, P. and Vijay, A. and Virdee, T. and Voirin, E. and Vojinovic, M. and Vámi, T.Á. and Wade, A. and Walter, D. and Wang, C. and Wang, F. and Wang, J. and Wang, K. and Wang, X. and Wang, X. and Wang, Y. and Wang, Z. and Wanlin, E. and Wayne, M. and Wetzel, J. and Whitbeck, A. and Wickwire, R. and Wilmot, D. and Wilson, J. and Wu, H. and Xiao, M. and Yang, J. and Yazici, B. and Ye, Y. and Yerli, B. and Yetkin, T. and Yi, R. and Yohay, R. and Yu, T. and Yuan, C. and Yuan, X. and Yuksel, O. and YushmanoV, I. and Yusuff, I. and Zabi, A. and Zareckis, D. and Zehetner, P. and Zghiche, A. and Zhang, C. and Zhang, D. and Zhang, H. and Zhang, J. and Zhang, J. and Zhang, Z. and Zhao, X. and Zhong, J. and Zhou, Y. and Zorbilmez, \c{C}. and The CMS HGCAL collaboration and The CALICE AHCAL collaborations},
title = {Using graph neural networks to reconstruct charged pion showers in the CMS High Granularity Calorimeter},
journal = {Journal of Instrumentation}
}

@article{WangXY2024,
author = {Wang, Xi-Yang and Wang, Shuai-Chun and He, Wan-Bing and Ma, Yu-Gang},
year = {2024},
month = {09},
pages = {169857},
title = {Cosmic ray test of shashlik electromagnetic calorimeter modules for NICA-MPD},
volume = {1069},
journal = {Nuclear Instruments and Methods in Physics Research Section A: Accelerators, Spectrometers, Detectors and Associated Equipment},
doi = {10.1016/j.nima.2024.169857}
}

@article{Wang2024NST,
	author = {Wang, Zhen and Yuan, Rui and Liu, Han-Qing and Chen, Jing and Chen, Xiang and Khaw, Kim Siang and Li, Liang and Li, Shu and Liu, Kun and Liu, Qi-Bin and Song, Si-Yuan and Sun, Tong and Wang, Xiao-Long and Wang, Yu-Feng and Yang, Hai-Jun and Zhang, Jun-Hua and Zhang, Yu-Lei and Zhao, Zhi-Yu and Zhu, Chun-Xiang and Zhu, Xu-Liang and Zhu, Yi-Fan},
	date = {2024/08/28},
	doi = {10.1007/s41365-024-01502-5},
	id = {Wang2024},
	isbn = {2210-3147},
	journal = {Nuclear Science and Techniques},
	number = {9},
	pages = {148},
	title = {Design of hadronic calorimeter for DarkSHINE experiment},
	url = {https://doi.org/10.1007/s41365-024-01502-5},
	volume = {35},
	year = {2024}}

@article{Zhao2025NST,
	author = {Zhao, Zhi-Yu and Liu, Qi-Bin and Chen, Ji-Yuan and Chen, Jing and Chen, Jun-Feng and Chen, Xiang and Fu, Chang-Bo and Guo, Jun and Khaw, Kim Siang and Li, Liang and Li, Shu and Liu, Dan-Ning and Liu, Kun and Song, Si-Yuan and Sun, Tong and Tang, Jian-Nan and Wang, Yu-Feng and Wang, Zhen and Wu, Wei-Hao and Yang, Hai-Jun and Lin, Yu-Ming and Yuan, Rui and Zhang, Yu-Lei and Zhang, Yun-Long and Zhou, Bai-Hong and Zhu, Xu-Liang and Zhu, Yi-Fan},
	date = {2025/01/28},
	doi = {10.1007/s41365-024-01618-8},
	id = {Zhao2025},
	isbn = {2210-3147},
	journal = {Nuclear Science and Techniques},
	number = {3},
	pages = {41},
	title = {Design of a LYSO crystal electromagnetic calorimeter for dark photon detection in the DarkSHINE experiment},
	url = {https://doi.org/10.1007/s41365-024-01618-8},
	volume = {36},
	year = {2025}}

@article{Fei2025NST,
	author = {Fei, Jia-Le and Yuan, Ao and Wei, Ke and Sun, Liang and Wang, Ji-Ke},
	date = {2025/06/28},
	doi = {10.1007/s41365-025-01740-1},
	id = {Fei2025},
	isbn = {2210-3147},
	journal = {Nuclear Science and Techniques},
	number = {9},
	pages = {166},
	title = {Layered reconstruction framework for longitudinal segmented electromagnetic calorimeter},
	url = {https://doi.org/10.1007/s41365-025-01740-1},
	volume = {36},
	year = {2025}}

@article{He2023NST,
	author = {He, Wan-Bing and Ma, Yu-Gang and Pang, Long-Gang and Song, Hui-Chao and Zhou, Kai},
	date = {2023/06/21},
	doi = {10.1007/s41365-023-01233-z},
	id = {He2023},
	isbn = {2210-3147},
	journal = {Nuclear Science and Techniques},
	number = {6},
	pages = {88},
	title = {High-energy nuclear physics meets machine learning},
	url = {https://doi.org/10.1007/s41365-023-01233-z},
	volume = {34},
	year = {2023}}

@article{Zubankov2025NST,
	author = {Zubankov, A. and Afanasiev, S. and Golubeva, M. and Guber, F. and Ivashkin, A. and Karpushkin, N. and Kutinova, O. and Lyapin, D. and Makhnev, A. and Morozov, S. and Parfenov, P. and Pshenichnov, I. and Sakulin, D. and Savenkov, S. and Shabanov, A. and Sukhov, E. and Svetlichnyi, A. and Taer, G. and Ustinov, V.},
	date = {2025/09/20},
	doi = {10.1007/s41365-025-01817-x},
	id = {Zubankov2025},
	isbn = {2210-3147},
	journal = {Nuclear Science and Techniques},
	number = {11},
	pages = {226},
	title = {Performance study of the Highly Granular Neutron Detector prototype in the BM{\char64}N experiment},
	url = {https://doi.org/10.1007/s41365-025-01817-x},
	volume = {36},
	year = {2025}}

@article{Liao2024,
	author = {Liao, C. L. and Quan, Z. and Dong, Y. W. and Xu. , M. and Zhang, C. and Wang, J. J. and Yang, X. G. and Wu, Q. and Sun, J. Y. and Liu. , X. and Wang. , Z. G. and Wang. , R. J.},
	date = {2024/10/19},
	doi = {10.1007/s10686-024-09957-5},
	id = {Liao2024},
	isbn = {1572-9508},
	journal = {Experimental Astronomy},
	number = {3},
	pages = {12},
	title = {Application of machine learning method for energy reconstruction on space based high granularity calorimeter},
	url = {https://doi.org/10.1007/s10686-024-09957-5},
	volume = {58},
	year = {2024}}

@article{Giannelli2024,
	author = {Giannelli, Michele Faucci and Zhang, Rui},
	date = {2024/07/08},
	doi = {10.1140/epjp/s13360-024-05397-4},
	id = {Giannelli2024},
	isbn = {2190-5444},
	journal = {The European Physical Journal Plus},
	number = {7},
	pages = {597},
	title = {CaloShowerGAN, a generative adversarial network model for fast calorimeter shower simulation},
	url = {https://doi.org/10.1140/epjp/s13360-024-05397-4},
	volume = {139},
	year = {2024}}

@article{Wu2024NST,
	author = {Wu, Huang-Kai and Wang, Xi-Yang and Wang, Yu-Miao and Wang, You-Jing and Fang, De-Qing and He, Wan-Bing and Ma, Wei-Hu and Cao, Xi-Guang and Fu, Chang-Bo and Deng, Xian-Gai and Ma, Yu-Gang},
	date = {2024/10/13},
	doi = {10.1007/s41365-024-01576-1},
	id = {Wu2024},
	isbn = {2210-3147},
	journal = {Nuclear Science and Techniques},
	number = {11},
	pages = {200},
	title = {Fudan multi-purpose active target time projection chamber (fMeta-TPC) for photonuclear reaction experiments},
	url = {https://doi.org/10.1007/s41365-024-01576-1},
	volume = {35},
	year = {2024}}

@article{Yu2024NST,
	author = {Yu, Xiao-Zhou and Wang, Xi-Yang and Ma, Wei-Hu and Fu, Shi-Hong and Sun, Peng-Fei and Song, Jin-Xing and He, Wan-Bing and Shen, Yang and Ma, Long and Chen, Jin-Hui and Huang, Huan-Zhong and Wang, Si-Guang and Zhou, Jing and Li, Xiao-Mei},
	date = {2024/08/22},
	doi = {10.1007/s41365-024-01517-y},
	id = {Yu2024},
	isbn = {2210-3147},
	journal = {Nuclear Science and Techniques},
	number = {8},
	pages = {145},
	title = {Production and test of sPHENIX W/SciFiber electromagnetic calorimeter blocks in China},
	url = {https://doi.org/10.1007/s41365-024-01517-y},
	volume = {35},
	year = {2024}}

@article{ABDULKHALEK2022122447,
title = {Science Requirements and Detector Concepts for the Electron-Ion Collider: EIC Yellow Report},
journal = {Nuclear Physics A},
volume = {1026},
pages = {122447},
year = {2022},
issn = {0375-9474},
doi = {https://doi.org/10.1016/j.nuclphysa.2022.122447},
url = {https://www.sciencedirect.com/science/article/pii/S0375947422000677},
author = {R. {Abdul Khalek} and A. Accardi and J. Adam and D. Adamiak and W. Akers and M. Albaladejo and A. Al-bataineh and M.G. Alexeev and F. Ameli and P. Antonioli and N. Armesto and W.R. Armstrong and M. Arratia and J. Arrington and A. Asaturyan and M. Asai and E.C. Aschenauer and S. Aune and H. Avagyan and C. {Ayerbe Gayoso} and B. Azmoun and A. Bacchetta and M.D. Baker and F. Barbosa and L. Barion and K.N. Barish and P.C. Barry and M. Battaglieri and A. Bazilevsky and N.K. Behera and F. Benmokhtar and V.V. Berdnikov and J.C. Bernauer and V. Bertone and S. Bhattacharya and C. Bissolotti and D. Boer and M. Boglione and M. Bondì and P. Boora and I. Borsa and F. Bossù and G. Bozzi and J.D. Brandenburg and N. Brei and A. Bressan and W.K. Brooks and S. Bufalino and M.H.S. Bukhari and V. Burkert and N.H. Buttimore and A. Camsonne and A. Celentano and F.G. Celiberto and W. Chang and C. Chatterjee and K. Chen and T. Chetry and T. Chiarusi and Y.-T. Chien and M. Chiosso and X. Chu and E. Chudakov and G. Cicala and E. Cisbani and I.C. Cloet and C. Cocuzza and P.L. Cole and D. Colella and J.L. Collins and M. Constantinou and M. Contalbrigo and G. Contin and R. Corliss and W. Cosyn and A. Courtoy and J. Crafts and R. Cruz-Torres and R.C. Cuevas and U. D'Alesio and S. {Dalla Torre} and D. Das and S.S. Dasgupta and C. {Da Silva} and W. Deconinck and M. Defurne and W. DeGraw and K. Dehmelt and A. {Del Dotto} and F. Delcarro and A. Deshpande and W. Detmold and R. {De Vita} and M. Diefenthaler and C. Dilks and D.U. Dixit and S. Dulat and A. Dumitru and R. Dupré and J.M. Durham and M.G. Echevarria and L. {El Fassi} and D. Elia and R. Ent and R. Esha and J.J. Ethier and O. Evdokimov and K.O. Eyser and C. Fanelli and R. Fatemi and S. Fazio and C. Fernandez-Ramirez and M. Finger and M. Finger and D. Fitzgerald and C. Flore and T. Frederico and I. Friščić and S. Fucini and S. Furletov and Y. Furletova and C. Gal and L. Gamberg and H. Gao and P. Garg and D. Gaskell and K. Gates and M.B. {Gay Ducati} and M. Gericke and G. {Gil Da Silveira} and F.-X. Girod and D.I. Glazier and K. Gnanvo and V.P. Goncalves and L. Gonella and J.O. {Gonzalez Hernandez} and Y. Goto and F. Grancagnolo and L.C. Greiner and W. Guryn and V. Guzey and Y. Hatta and M. Hattawy and F. Hauenstein and X. He and T.K. Hemmick and O. Hen and G. Heyes and D.W. Higinbotham and A.N. {Hiller Blin} and T.J. Hobbs and M. Hohlmann and T. Horn and T.-J. Hou and J. Huang and Q. Huang and G.M. Huber and C.E. Hyde and G. Iakovidis and Y. Ilieva and B.V. Jacak and P.M. Jacobs and M. Jadhav and Z. Janoska and A. Jentsch and T. Jezo and X. Jing and P.G. Jones and K. Joo and S. Joosten and V. Kafka and N. Kalantarians and G. Kalicy and D. Kang and Z.B. Kang and K. Kauder and S.J.D. Kay and C.E. Keppel and J. Kim and A. Kiselev and M. Klasen and S. Klein and H.T. Klest and O. Korchak and A. Kostina and P. Kotko and Y.V. Kovchegov and M. Krelina and S. Kuleshov and S. Kumano and K.S. Kumar and R. Kumar and L. Kumar and K. Kumerički and A. Kusina and K. Kutak and Y.S. Lai and K. Lalwani and T. Lappi and J. Lauret and M. Lavinsky and D. Lawrence and D. Lednicky and C. Lee and K. Lee and S.H. Lee and S. Levorato and H. Li and S. Li and W. Li and X. Li and X. Li and W.B. Li and T. Ligonzo and H. Liu and M.X. Liu and X. Liu and S. Liuti and N. Liyanage and C. Lorcé and Z. Lu and G. Lucero and N.S. Lukow and E. Lunghi and R. Majka and Y. Makris and I. Mandjavidze and S. Mantry and H. Mäntysaari and F. Marhauser and P. Markowitz and L. Marsicano and A. Mastroserio and V. Mathieu and Y. Mehtar-Tani and W. Melnitchouk and L. Mendez and A. Metz and Z.-E. Meziani and C. Mezrag and M. Mihovilovič and R. Milner and M. Mirazita and H. Mkrtchyan and A. Mkrtchyan and V. Mochalov and V. Moiseev and M.M. Mondal and A. Morreale and D. Morrison and L. Motyka and H. Moutarde and C. {Muñoz Camacho} and F. Murgia and M.J. Murray and P. Musico and P. Nadel-Turonski and P.M. Nadolsky and J. Nam and P.R. Newman and D. Neyret and D. Nguyen and E.R. Nocera and F. Noferini and F. Noto and A.S. Nunes and V.A. Okorokov and F. Olness and J.D. Osborn and B.S. Page and S. Park and A. Parker and K. Paschke and B. Pasquini and H. Paukkunen and S. Paul and C. Pecar and I.L. Pegg and C. Pellegrino and C. Peng and L. Pentchev and R. Perrino and F. Petriello and R. Petti and A. Pilloni and C. Pinkenburg and B. Pire and C. Pisano and D. Pitonyak and A.A. Poblaguev and T. Polakovic and M. Posik and M. Potekhin and R. Preghenella and S. Preins and A. Prokudin and P. Pujahari and M.L. Purschke and J.R. Pybus and M. Radici and R. Rajput-Ghoshal and P.E. Reimer and M. Rinaldi and F. Ringer and C.D. Roberts and S. Rodini and J. Rojo and D. Romanov and P. Rossi and E. Santopinto and M. Sarsour and R. Sassot and N. Sato and B. Schenke and W.B. Schmidke and I. Schmidt and A. Schmidt and B. Schmookler and G. Schnell and P. Schweitzer and J. Schwiening and I. Scimemi and S. Scopetta and J. Segovia and R. Seidl and S. Sekula and K. Semenov-Tian-Shanskiy and D.Y. Shao and N. Sherrill and E. Sichtermann and M. Siddikov and A. Signori and B.K. Singh and S. Širca and K. Slifer and W. Slominski and D. Sokhan and W.E. Sondheim and Y. Song and O. Soto and H. Spiesberger and A.M. Stasto and P. Stepanov and G. Sterman and J.R. Stevens and I.W. Stewart and I. Strakovsky and M. Strikman and M. Sturm and M.L. Stutzman and M. Sullivan and B. Surrow and P. Svihra and S. Syritsyn and A. Szczepaniak and P. Sznajder and H. Szumila-Vance and L. Szymanowski and A.S. Tadepalli and J.D. {Tapia Takaki} and G.F. Tassielli and J. Terry and F. Tessarotto and K. Tezgin and L. Tomasek and F. {Torales Acosta} and P. Tribedy and A. Tricoli and  Triloki and S. Tripathi and R.L. Trotta and O.D. Tsai and Z. Tu and C. Tuvè and T. Ullrich and M. Ungaro and G.M. Urciuoli and A. Valentini and P. Vancura and M. Vandenbroucke and C. {Van Hulse} and G. Varner and R. Venugopalan and I. Vitev and A. Vladimirov and G. Volpe and A. Vossen and E. Voutier and J. Wagner and S. Wallon and H. Wang and Q. Wang and X. Wang and S.Y. Wei and C. Weiss and T. Wenaus and H. Wennlöf and N. Wickramaarachchi and A. Wikramanayake and D. Winney and C.P. Wong and C. Woody and L. Xia and B.W. Xiao and J. Xie and H. Xing and Q.H. Xu and J. Zhang and S. Zhang and Z. Zhang and Z.W. Zhao and Y.X. Zhao and L. Zheng and Y. Zhou and P. Zurita}
}

@article{MELIKYAN2025170604,
title = {Characterization of the large-size NDL EQR20 silicon photomultipliers},
journal = {Nuclear Instruments and Methods in Physics Research Section A: Accelerators, Spectrometers, Detectors and Associated Equipment},
volume = {1078},
pages = {170604},
year = {2025},
issn = {0168-9002},
doi = {https://doi.org/10.1016/j.nima.2025.170604},
url = {https://www.sciencedirect.com/science/article/pii/S016890022500405X},
author = {Yu A. Melikyan and I.G. Bearden and V. Buchakchiev and S. Jia and V. Kozhuharov and I.P. Møller}
}

@article{BILKI2015240,
title = {Testing hadronic interaction models using a highly granular silicon–tungsten calorimeter},
journal = {Nuclear Instruments and Methods in Physics Research Section A: Accelerators, Spectrometers, Detectors and Associated Equipment},
volume = {794},
pages = {240-254},
year = {2015},
issn = {0168-9002},
doi = {https://doi.org/10.1016/j.nima.2015.05.009},
url = {https://www.sciencedirect.com/science/article/pii/S0168900215006191},
author = {B. Bilki and J. Repond and J. Schlereth and L. Xia and Z. Deng and Y. Li and Y. Wang and Q. Yue and Z. Yang and G. Eigen and Y. Mikami and T. Price and N.K. Watson and M.A. Thomson and D.R. Ward and D. Benchekroun and A. Hoummada and Y. Khoulaki and C. Cârloganu and S. Chang and A. Khan and D.H. Kim and D.J. Kong and Y.D. Oh and G.C. Blazey and A. Dyshkant and K. Francis and J.G.R. Lima and P. Salcido and V. Zutshi and V. Boisvert and B. Green and A. Misiejuk and F. Salvatore and K. Kawagoe and Y. Miyazaki and Y. Sudo and T. Suehara and T. Tomita and H. Ueno and T. Yoshioka and J. Apostolakis and G. Folger and V. Ivantchenko and A. Ribon and V. Uzhinskiy and S. Cauwenbergh and M. Tytgat and N. Zaganidis and J.-Y. Hostachy and L. Morin and K. Gadow and P. Göttlicher and C. Günter and K. Krüger and B. Lutz and M. Reinecke and F. Sefkow and N. Feege and E. Garutti and S. Laurien and S. Lu and I. Marchesini and M. Matysek and M. Ramilli and A. Kaplan and E. Norbeck and D. Northacker and Y. Onel and E.J. Kim and B. {van Doren} and G.W. Wilson and M. Wing and B. Bobchenko and M. Chadeeva and R. Chistov and M. Danilov and A. Drutskoy and A. Epifantsev and O. Markin and R. Mizuk and E. Novikov and V. Popov and V. Rusinov and E. Tarkovsky and D. Besson and E. Popova and M. Gabriel and C. Kiesling and F. Simon and C. Soldner and M. Szalay and M. Tesar and L. Weuste and M.S. Amjad and J. Bonis and S. Callier and S. {Conforti di Lorenzo} and P. Cornebise and Ph. Doublet and F. Dulucq and M. Faucci-Giannelli and J. Fleury and T. Frisson and B. Kégl and N. {van der Kolk} and H. Li and G. Martin-Chassard and F. Richard and Ch. {de la Taille} and R. Pöschl and L. Raux and J. Rouëné and N. Seguin-Moreau and M. Anduze and V. Balagura and E. Becheva and V. Boudry and J-C. Brient and R. Cornat and M. Frotin and F. Gastaldi and F. Magniette and A. Matthieu and P. {Mora de Freitas} and H. Videau and J-E. Augustin and J. David and P. Ghislain and D. Lacour and L. Lavergne and J. Zacek and J. Cvach and P. Gallus and M. Havranek and M. Janata and J. Kvasnicka and D. Lednicky and M. Marcisovsky and I. Polak and J. Popule and L. Tomasek and M. Tomasek and P. Ruzicka and P. Sicho and J. Smolik and V. Vrba and J. Zalesak and D. Jeans and M. Götze}
}







\end{document}